\documentclass[usenatbib]{mn2e}
\usepackage{graphicx}
\usepackage{natbib}
\usepackage{amssymb}
\usepackage{aas_macros}

\def\kms{\hbox{\,km\,s$^{-1}$}}

\title[Compact groups in SDSS DR6]{Compact groups in theory and
practice - III. Compact groups of galaxies in the Sixth Data Release
of the Sloan Digital Sky Survey}

\author[McConnachie et al.] {Alan W. McConnachie$^{1,2}$, David R. Patton$^{3,4}$, Sara L. Ellison$^1$, Luc Simard$^2$\\
$^1$Department of Physics and Astronomy, University of Victoria, Victoria, B.C., V8P 1A1, Canada\\
$^2$NRC Herzberg Institute of Astrophysics, 5071 West Saanich Road, Victoria, B.C., V9E 2E7, Canada\\
$^3$Department of Physics and Astronomy, Trent University, 1600 West Bank Drive, Peterborough, ON, K9J 7B8, Canada\\
$^4$Visiting Researcher, Department of Physics and Astronomy, University of Victoria, Victoria, B.C., V8P 1A1, Canada\\
}

\begin{document}

\maketitle

\begin{abstract}

We present the largest publicly available catalogue of compact groups
of galaxies identified using the original selection criteria of
Hickson, selected from the Sixth Data Release (DR6) of the Sloan
Digital Sky Survey (SDSS). We identify 2297 compact groups down to a
limiting magnitude of $r = 18$ ($\sim0.24$\,groups\,degree$^{-2}$),
and 74791 compact groups down to a limiting magnitude of $r = 21$
($\sim6.7$\,groups\,degree$^{-2}$). This represents $0.9\%$ of all
galaxies in the SDSS~DR6 at these magnitude levels. Contamination due
to gross photometric errors has been removed from the bright sample of
groups, and we estimate it is present in the large sample at the
$14\%$ level. Spectroscopic information is available for 4131 galaxies
in the bright catalogue (43\,\% completeness), and we find that the
median redshift of these groups is ${z_{med}} = 0.09$. The median
line-of-sight velocity dispersion within the compact groups from the
bright catalogue is $\sigma_{LOS} \simeq 230$\,\kms and their typical
inter-galactic separations are of order $50 - 100$\,kpc. We show that
the fraction of groups with interloping galaxies identified as members
is in good agreement with the predictions from our previous study of a
mock galaxy catalogue, and we demonstrate how to select compact groups
such that the interloper fraction is well defined and minimized. This
observational dataset is ideal for large statistical studies of
compact groups, the role of environment on galaxy evolution, and the
effect of galaxy interactions in determining galaxy morphology.
\end{abstract}

\begin{keywords}
catalogues - surveys - galaxies: general - galaxies: interactions
\end{keywords}

\section{Introduction}

Galaxies cluster on all scales, and a basic tenet of the $\Lambda-$CDM
cosmological paradigm is that structure formation is
hierarchical. Consequently, both weak and strong galaxy interactions
are expected to be part of the typical life-cycle of a galaxy, and are
believed to be among the main drivers in determining a galaxy's star
formation history, chemical evolution and morphology.

The majority of $L_\star$ galaxies at $z = 0$ are not found in rich
environments and, while many are members of loose groups with a low
velocity dispersion, few have nearby large companions with which they
are interacting.  In contrast, galaxy clusters have a high galaxy
number density. However, the considerable mass of these structures
means that their velocity dispersion is of order $1000$\kms, and thus
individual galaxy-galaxy mergers are rare. On the other hand, compact
groups (CGs) of galaxies generally have the low velocity dispersions
typical of galaxy groups, but with galaxy-galaxy separations small
enough for interactions to be significant (e.g., \citealt{hickson1997}
and references therein).

Compact groups of galaxies were first quantitatively defined by
\cite{hickson1982}, and there now exist several catalogues of compact
groups identified using a variety of criteria (e.g., the Digitized
POSS CG catalog, \citealt{iovino2003}; the Southern CG catalogue,
\citealt{prandoni1994,iovino2002}; CGs in the UZC galaxy catalogue,
\citealt{focardi2002}; the CfA2 redshift survey, \citealt{barton1996};
the Las Campanas redshift survey, \citealt{allam2000}; the SDSS
Commissioning Data, \citealt{lee2004}). The identification of genuine
groups with only a few members is a difficult problem, and
considerable effort has been spent on determining the three
dimensional properties of the identified systems (particularly the
Hickson Compact Groups) and the effect of interlopers (e.g.,
\citealt{mamon1986,hickson1988b,walke1989,hernquist1995,ponman1996}).

In this series of papers, we are conducting a homogeneous and
systematic study of compact groups using a combination of theoretical
and observational galaxy catalogues to provide a
statistically-powerful platform from which to probe the effect of
galaxy interactions on all aspects of galaxy evolution. In the first
paper of this series (\citealt{mcconnachie2008a}, hereafter Paper~I),
we identified compact groups in a mock catalogue constructed from the
Millennium Simulation (\citealt{springel2005,delucia2007}) and
developed a robust understanding of their spatial and dynamical
properties. This allowed us to refine the compact group selection
criteria to more efficiently identify these systems and understand the
sources of contamination in any observed sample. In the second paper
of this series (\citealt{brasseur2008}, hereafter Paper~II), we
compared the physical properties of compact group galaxies in our mock
catalogue to previous observational studies. We demonstrated
consistency with these earlier studies, and we concluded that
interloping galaxies misidentified as compact group members have
potentially affected our understanding of the influence of
interactions on galaxy evolution and morphology.

In this paper, the third in the series, we create the largest
observational catalogue of compact group galaxies to date by
identifying all such groups in the Sloan Digital Sky Survey Data
Release 6, consisting of more than 29\,million galaxies to a limiting
magnitude of $r = 21$. We make the resulting catalogue of compact
groups publicly available through online tables associated with this
paper. This catalogue is ideal for large statistical studies of
compact groups, the role of environment on galaxy evolution, and the
effect of galaxy interactions in determining galaxy
morphology. Section 2 details our procedure for the identification of
compact groups, Section 3 presents some of the basic observable
properties of these systems, and Section 4 summarises our results.

\section{Compact Group Finding in SDSS DR6}

\subsection{Compact group selection criteria}

\cite{hickson1982} define a CG as a group of galaxies with projected
properties such that

\begin{enumerate}
\item{$N\left(\Delta\,m = 3\right) \geq 4$;}
\item{$\theta_N \geq 3\,\theta_G$;}
\item{$\mu_e \leq 26.0$ mags\,arcsec$^{-2}$.}
\end{enumerate}

\noindent $N\left(\Delta\,m = 3\right)$ is the number of galaxies
within 3\,magnitudes of the brightest galaxy and $\mu_e$ is the
effective surface brightness of these galaxies, where the total flux
of the galaxies is averaged over the smallest circle which contains
their geometric centres and has an angular diameter
$\theta_G$. $\theta_N$ is the angular diameter of the largest
concentric circle which contains no additional galaxies in this
magnitude range or brighter. All magnitudes and surface brightnesses
are measured in the $r-$band. We hereafter refer to these criteria
collectively as the `Hickson criteria'.

Since the Hickson criteria were first introduced, there has been
considerable debate regarding the physical nature of the galaxy
associations identified (e.g.,
\citealt{mamon1986,hickson1988b,walke1989}), and determining the three
dimensional reality of systems identified as compact groups has
remained a focus of theoretical and observational study (e.g.,
\citealt{hernquist1995,ponman1996}). To address this issue, in Paper~I
of this series, we applied the Hickson criteria to an all-sky mock
galaxy catalogue, with a limiting magnitude of $r = 18$, to determine
the physical reality of the systems that these criteria identify. We
found that the identified galaxy associations preferentially had a
three-dimensional linking length, $l \le 200 h^{-1}$\,kpc (comoving,
where $H_o = 100\,h\,$km\,s$^{-1}$Mpc$^{-1}$). Defined in this way,
$\sim 30$\,\% of all the identified galaxy associations were
``genuine'' compact groups, while the remaining systems were found to
contain one or more interloping galaxies and were treated as
contamination. We found that this physically well-defined sample of
genuine compact groups could be identified with greatly reduced levels
of contamination by modifying the original Hickson criteria to select
more isolated groups and higher surface brightness groups: by changing
the latter from $\mu_e \leq 26.0$ to $\mu_e \leq 25.0, 24.0$ and
$23.0$\,mags\,arcsec$^{-2}$, the contamination rates decrease from
$\sim 71$\% to 57\%, 44\% and 33\%, respectively. This is in general
agreement with the search strategies adopted by some earlier surveys
for compact groups (e.g., \citealt{iovino2003,lee2004}) which modified
these criteria to try to reduce contamination.

In what follows, we apply the Hickson criteria in their original form
to create our main compact group catalogues.  Sub-catalogues can be
created from this publicly-available resource, with different cuts
applied in, for example, surface brightness, isolation or brightest
members, depending upon the requirements of the specific project being
undertaken (e.g., statistical size of dataset versus expected
contamination rates).

\subsection{The SDSS DR6 galaxy catalogue}

The SDSS DR6 (\citealt{adelmanmccarthy2008}) catalogue contains
photometric parameters for \mbox{$\sim 287$\,million} unique objects
(stellar and galactic) over \mbox{$\sim 9583$\,degrees$^{2}$} ($\sim
23$\% of the sky), of which approximately 1.27 million objects have
associated spectra. As such, SDSS DR6 is presently the largest
publically available photometric and spectroscopic dataset of
galaxies.

Some previous searches for compact groups of galaxies have identified
compact groups in spectroscopic galaxy catalogues through
linking-length analyses in projection and redshift space (following
\citealt{huchra1982}; e.g., the CfA2 redshift survey,
\citealt{barton1996}; the Las Campanas redshift survey,
\citealt{allam2000}; the UZC galaxy catalogue, \citealt{focardi2002};
the SDSS DR5 and DR6, \citealt{deng2007,deng2008}). However,
spectroscopic galaxy catalogues are necessarily much smaller than
corresponding photometric catalogues, and one of the main motivations
for this work is to obtain a statistically large sample of compact
groups for subsequent study. Further, compact groups generally have
small angular extent ($\lesssim 1$\,arcmin), and the SDSS suffers from
fiber collisions for objects which are very close together
(\citealt{strauss2002}), meaning that not all members of a compact
group can be expected to possess redshift information.  We therefore
choose not to use the spectroscopic catalogue for our compact group
search, and instead identify compact groups based upon their projected
photometric properties.

We extract all objects identified as a galaxy from the SDSS DR6
catalogue, imposing a bright-end limit on the $r$-band magnitude at $r
= 14.5$, since the automated de-blending of galaxies is known to become
unreliable above this limit (\citealt{strauss2002}).  We do not
consider any objects which have been flagged as ``SATURATED'' and/or
``DEBLENDED\_AS\_PSF''.  The former removes all objects which appear
saturated, implying that their photometric parameters are
unreliable. The latter removes any point sources initially identified
as part of larger photometric structures; they are therefore unlikely
to be galactic in nature. The exclusion of galaxies with either of
these two flags decreases the size of our dataset by $3.9\%$.

We construct two different datasets from SDSS DR6, corresponding to
two different faint-end magnitude limits of $r = 18$ and $r = 21$. The
former is identical to the magnitude limit of the mock galaxy
catalogue discussed in Papers I and II, and ensures that the
observational and mock catalogues can be robustly compared. The latter
corresponds to a level above which photometric completeness and
star-galaxy separation are reliable
(\citealt{lupton2001,abazajian2004}), and ensures a large reliable
dataset for statistical studies. All magnitudes are corrected for
foreground Galactic extinction.

Our final $14.5\le r \le 18.0$ galaxy catalogue (hereafter referred to
as Catalogue A) has 1\,107\,622 members.  Our final $14.5\le r \le 21.0$
galaxy catalogue (hereafter referred to as Catalogue B) has
29\,065\,010 members.

\subsection{Application of the search algorithm}

We apply a search algorithm for compact groups by considering each
galaxy in the magnitude-limited catalogues in turn. The galaxies in
the immediate vicinity of the target galaxy (selected within a very
generous 1\,degree radius from the target galaxy) are initially
considered as possible group members. The geometric centre of the
group, its radius, magnitude range, surface brightness and distance to
the nearest non-member galaxy in the appropriate magnitude range, are
calculated. This procedure is repeated, with the most distant member
galaxy from the target galaxy being eliminated each round, until a group of
galaxies is found which satisfies the Hickson criteria. Failing this,
if fewer than 4 galaxies remain in the group, we start the procedure
afresh by considering the next galaxy in the catalogue (unless that
galaxy has been determined to be a member of a previous compact
group). This is an identical procedure to that which was applied to
the mock catalogue in Paper~I.

To speed up this computationally intensive algorithm, Catalogue B is
split into 34 sub-areas and the algorithm applied separately to each
sub-area. Catalogue A is split into 5 sub-areas. Compact groups which
happen to lie on the borders of these sub-areas could conceivably
elude detection. However, the maximum diameter of our compact groups
is found to be $\sim 2$\,arcmin and so only groups lying within a
negligible area compared to the total survey area will not be
detected.

In each dataset we identify all groups irrespective of the magnitude
of the brightest galaxy; for groups where the brightest member is less
than 3 magnitudes brighter than the magnitude limit of the catalogue,
this means that the isolation criteria may not be met when nearby
galaxies fainter than the magnitude limit of the catalogue are
considered. This relaxation of the original Hickson criteria increases
the number of groups identified significantly and allows for direct
comparisons with the mock catalogue studied in Paper I and II which
uses the same selection criteria.

In total, we initially identify 3108 compact groups of galaxies in
Catalogue~A and 74791 compact groups in Catalogue~B, consisting of a
total of 13233 and 313508 individual objects, respectively.

\subsection{Contamination and Health Warnings}

\begin{figure*}
  \begin{center}
    \includegraphics[angle=0, width=4.cm]{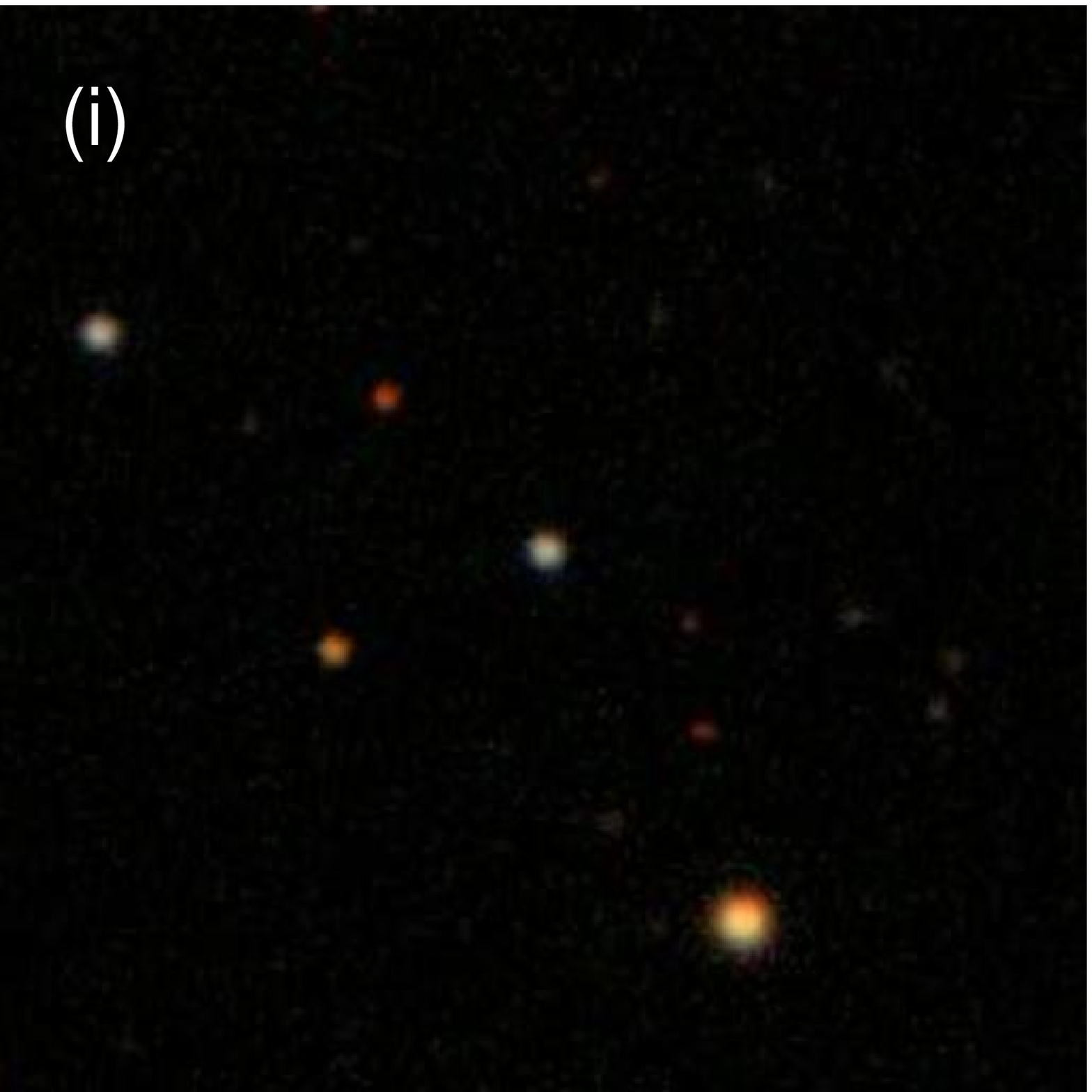}
    \includegraphics[angle=0, width=4.cm]{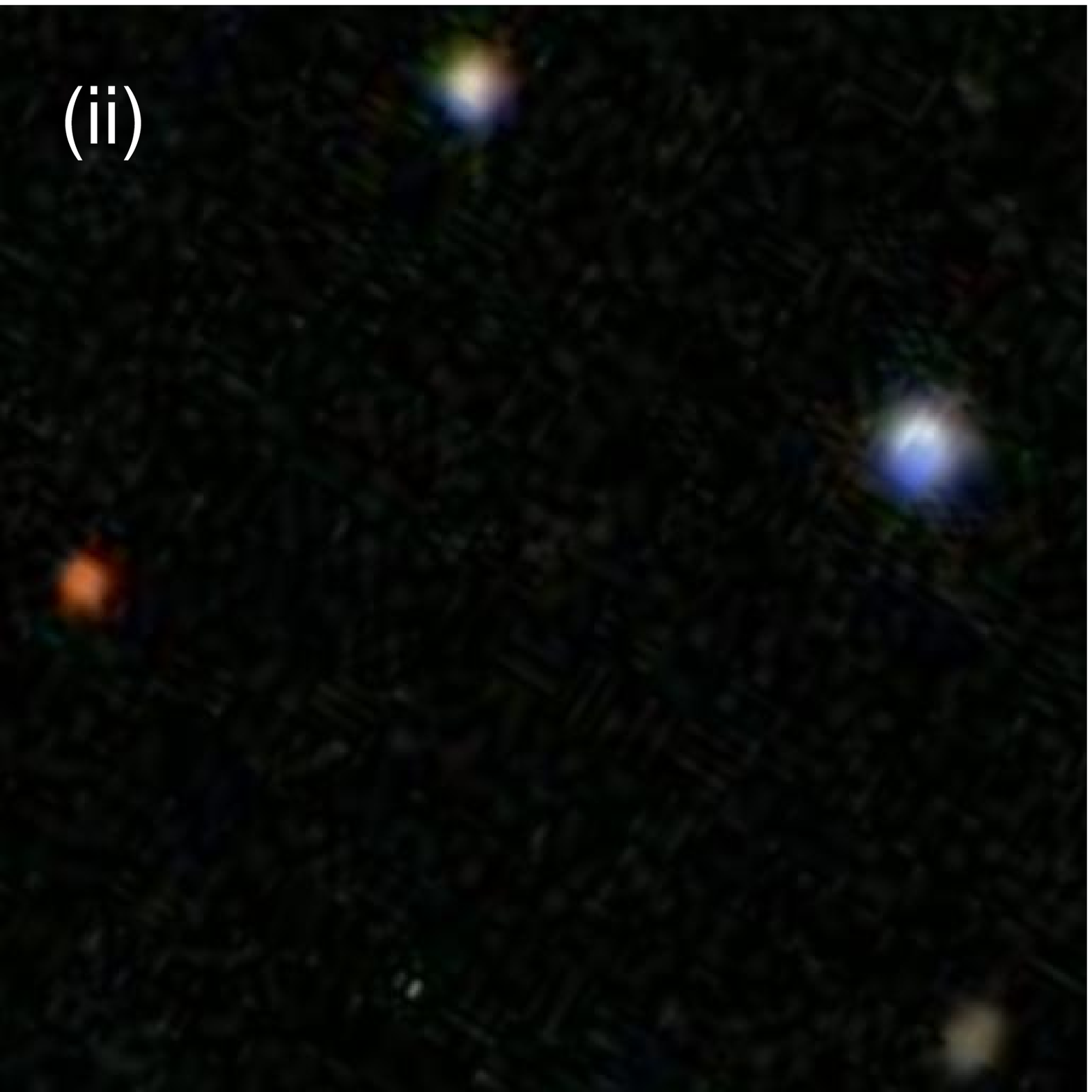}
    \includegraphics[angle=0, width=4.cm]{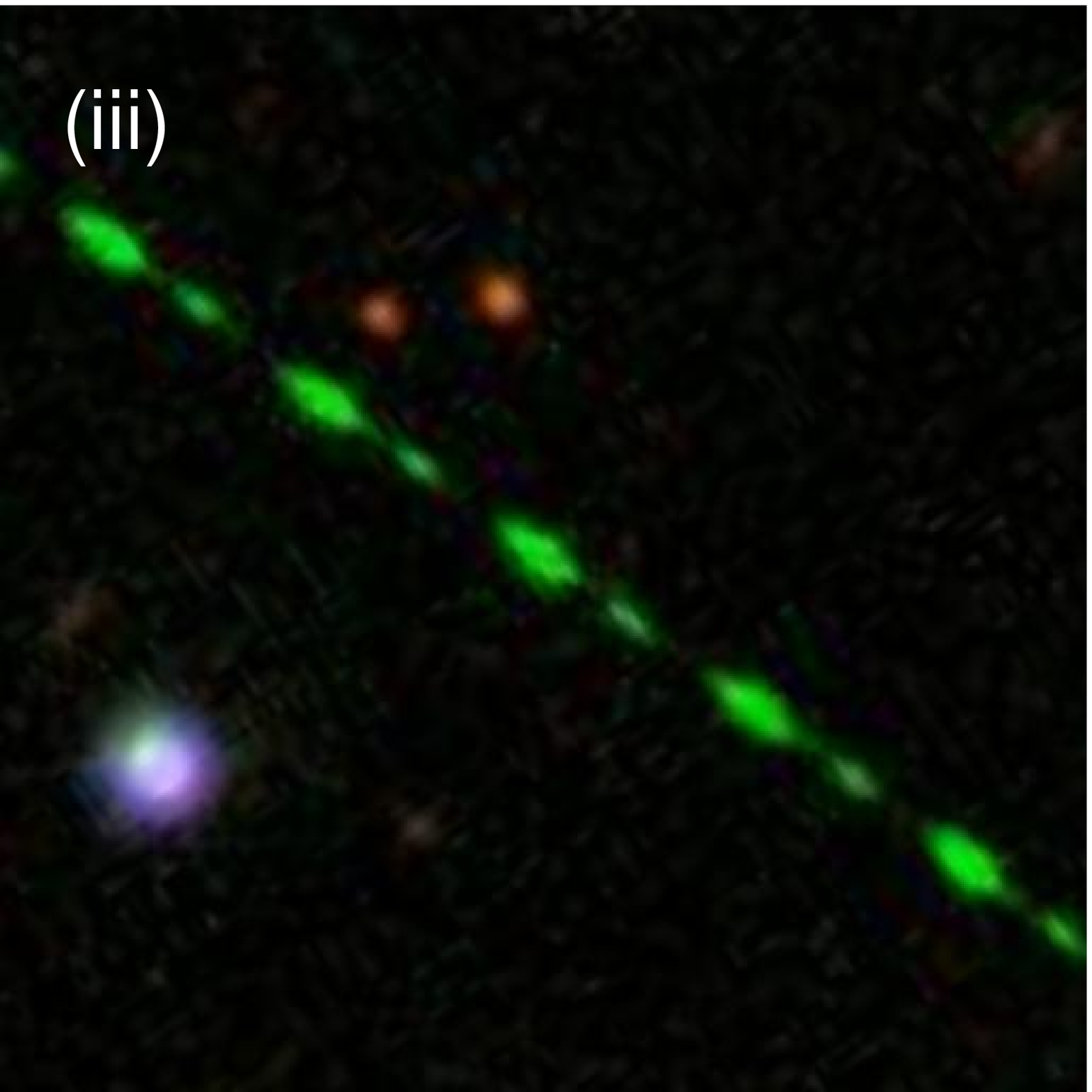}\\
    \includegraphics[angle=0, width=4.cm]{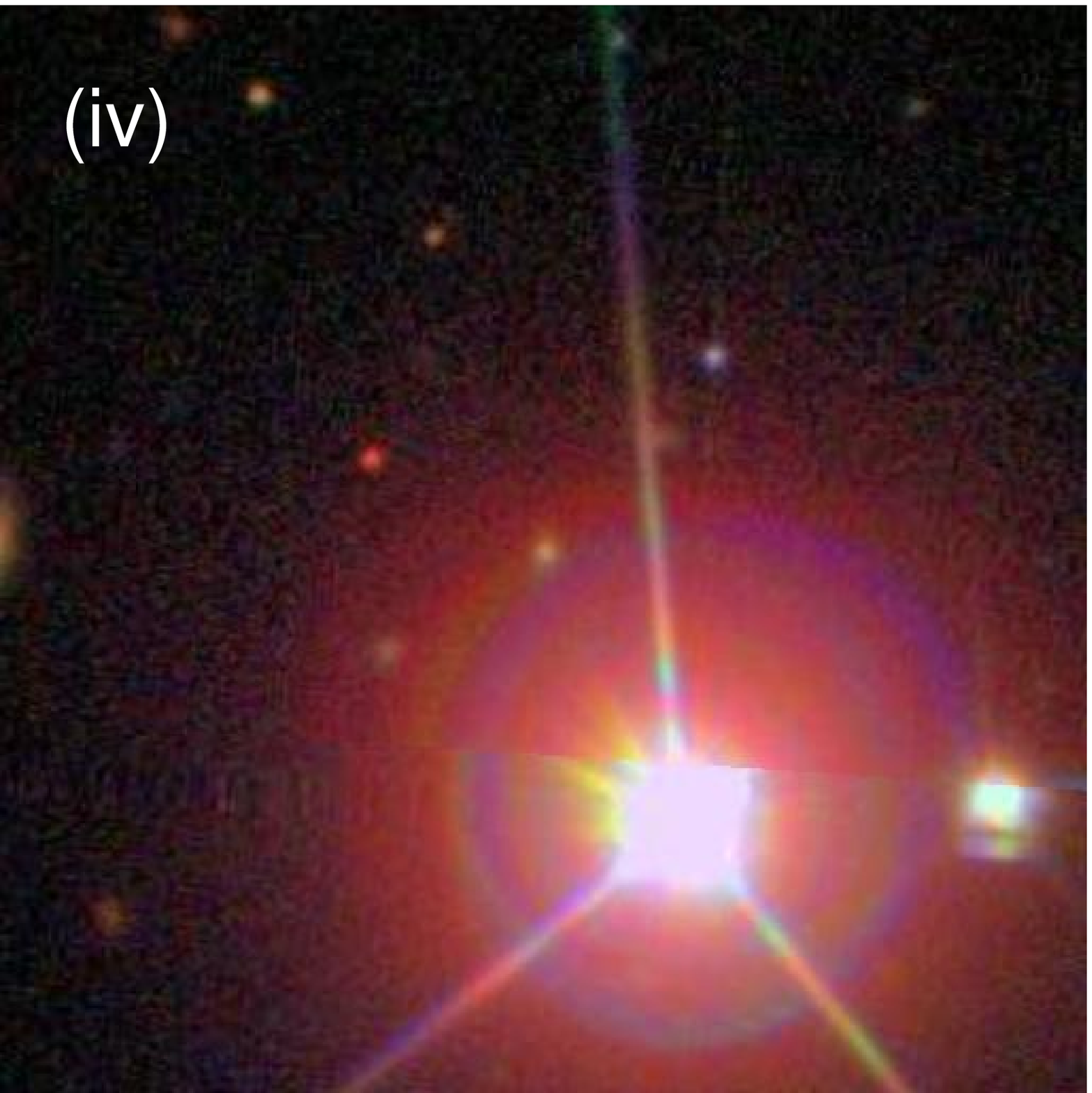}
    \includegraphics[angle=0, width=4.cm]{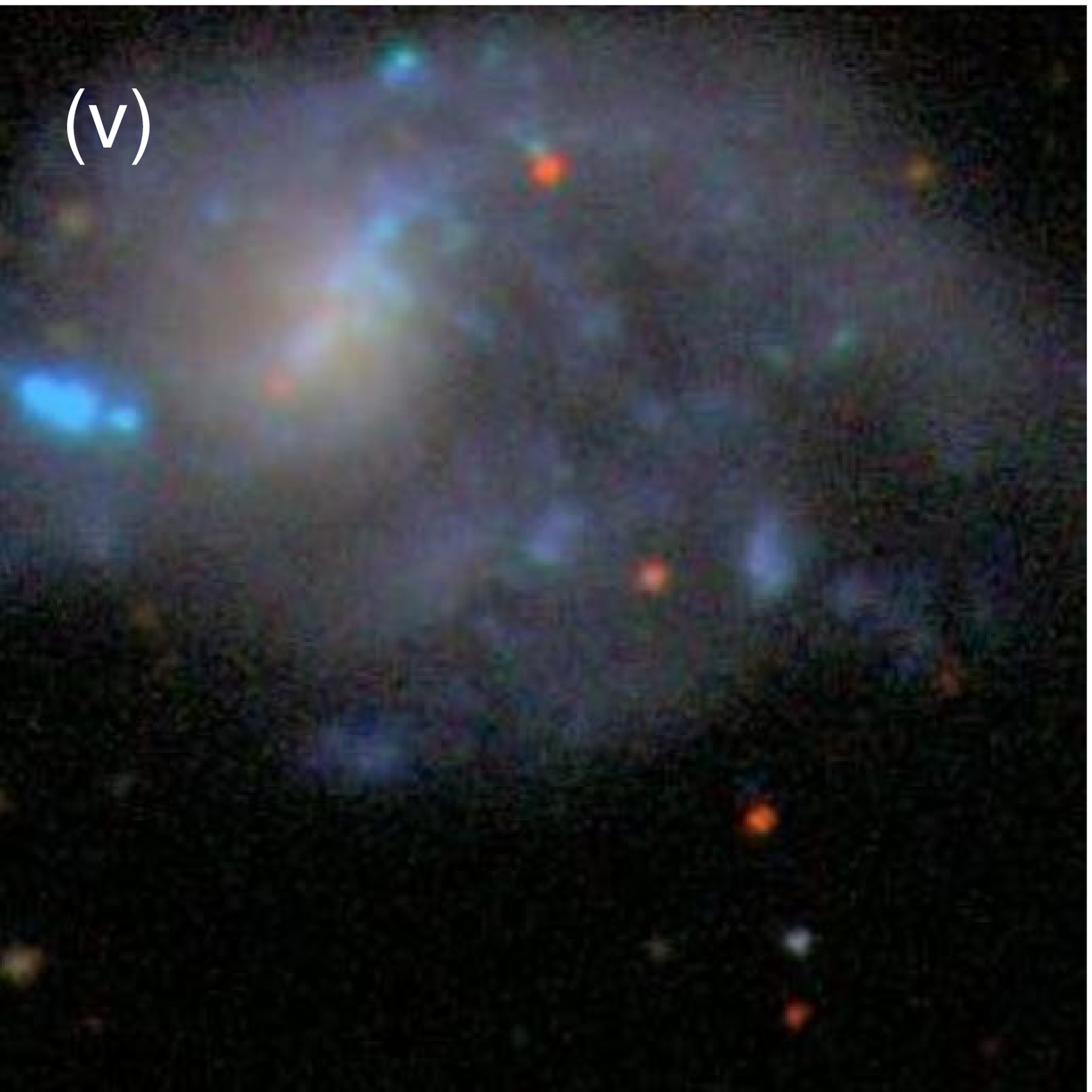}
    \caption{In the centre of each panel is an example of a typical
    source of contamination in the SDSS~DR6 which affects this study:
    (i) a stellar source misclassified as a galaxy; (ii) an object
    classified as a bright ($r \sim 17$) galaxy but for which there is
    no identifiable source/very faint source; (iii) a fragment of a
    satellite trail identified as a chain of several galaxies; (iv) an
    object (perhaps galactic) close to a saturated object; (v) a
    large, extended galaxy which has been de-blended into many
    component objects, many of which have been independently
    classified as galactic. The latter is the primary source of
    contamination in our search for compact groups, particularly at
    bright magnitudes.}
    \label{cont}
  \end{center}
\end{figure*}

Obvious contamination in the compact group catalogues arises as a
result of errors in the photometric galaxy catalogue on which the
search is conducted. In each panel of Figure~\ref{cont} is an example
of a source of contamination for this study: (i) some objects which
are classified as galaxies by the SDSS classification algorithm appear
stellar on visual inspection; (ii) some objects are classified as
reasonably bright sources but on inspection it is obvious that the
magnitude assigned is grossly incorrect; (iii) fragments of satellite
trails are occasionally classified as galaxies; (iv) some objects are
found very close to saturated objects and have unreliable
photometry. The largest source of contamination for this study,
however, comes from (v) incorrect de-blending of extended galaxies
into many smaller sources subsequently identified as galaxies.

Note that when discussing contamination, we do not at this stage
consider redshift information for  galaxies in our groups for which
spectroscopic data is available; we consider only the
photometric properties from which the groups were
identified. Spectroscopic data and the identification of interloping
groups with discordant redshift information are considered in detail
in Section~3.

\subsubsection{Groups in Catalogue A}

To remove the sources of contamination from groups in Catalogue~A,
AWM, DLP and SLE independently visually inspected all 13233 objects
belonging to groups in Catalogue~A and recorded their findings. If
each member of a group appeared galactic in nature, and did not appear
to have any gross errors in its $r-$band photometry, then the group
was marked as ``genuine''; if one or more members of a group appeared
to have been incorrectly classified as galactic, or the $r-$band
magnitude of one or more members appeared grossly incorrect, the group
was marked as ``spurious''; if it was unclear if a group was genuine
or not, it was marked ``for further inspection''. This was done
independently by the three co-authors and, once completed, the three
sets of results were collated.

All groups marked as ``genuine'' by all three co-authors were
classified as genuine and kept in the group catalogue; all groups marked as
``spurious'' by all three co-authors were accepted as spurious and
rejected from the catalogue; any groups where the co-authors were not
unanimous in their independent classifications were re-inspected and
kept or rejected from the catalogue depending on the result of this
re-inspection; any group marked as ``for further inspection'' by {\it
any} co-author was re-inspected, and kept or rejected from the
catalogue depending on the result of this re-inspection. In addition,
a few groups were found to contain a galaxy brighter than the
bright-end limit of the catalogue. These few groups were additionally
removed from the sample.

The final list of groups from Catalogue~A contains 2297 compact groups
(9713 galaxies). Thus 26\% of the groups in Catalogue~A originally
identified by our algorithm as a compact group in SDSS~DR6 were found
to contain a member or member(s) for which the automated photometric
classification procedure was flawed, usually in one of the ways
indicated in Figure~1 (over half of all contamination was due to {\it
(v)} incorrect deblending, approximately one-third was due to {\it
(ii)} grossly incorrect magnitudes, and the remaining fraction was due
to the other types of contamination). However, due to the
independent visual inspection of all members of groups in Catalogue~A
by three different co-authors, the level of contamination in the final
version of the group catalogue is negligible.

\subsubsection{Groups in Catalogue B}

We identified 313508 individual objects belonging to groups found in
Catalogue B. It is impractical to visually inspect each object in the
same way as above to determine if each group is
genuine. Instead, we seek to quantify the contamination fraction by
selecting a random 1500 compact groups and visually inspecting each
member (6252 individual objects).

Of the 1500 compact groups inspected, $\sim 205$ appear to contain a
member or members which had been incorrectly classified by the SDSS
DR6 automated algorithms. Thus, we estimate that the groups identified
in Catalogue~B contain contamination at the 14\,\% level. This is
considerably lower than for Catalogue~A, and reflects the fact that
Catalogue~B contains a smaller fraction of very bright galaxies (for
which the de-blending algorithm runs into difficulty and which
provides the primary source of contamination). Among the groups in
Catalogue~B containing bright galaxies, the contamination level can be
expected to be similar to the initial contamination level of groups in
Catalogue~A.  We have additionally flagged those groups in Catalogue~B
which contain objects which belong to groups originally identified in
Catalogue A and which were subsequently classified as contamination
(Section 2.5).


The final list of groups from Catalogue~B contains 74791 compact
groups (313508 galaxies) and has an estimated contamination fraction
of $14\%$.

\subsubsection*{}

We emphasise that the photometric parameterisation of all the objects
in our published catalogues of compact groups is from the main
SDSS~DR6 reduction. Any errors in the SDSS~DR6 photometric reduction
procedure will therefore propagate through to the compact group
catalogues. Any errors not discussed above which exist in the SDSS DR6
reduction will potentially also exist in the compact group
catalogue. The final list of groups from Catalogue~A only contains
bright galaxies, $r \le 18$, with negligible contamination due to
gross photometric errors. The list of groups from Catalogue~B is much
larger and includes galaxies down to $r = 21$ and has contamination
due to poor photometric classification at the 14\% level. As discussed
in Section 2.2, groups are selected irrespective of the magnitude of
their brightest member galaxy, and for groups where the brightest
member is less than 3 magnitudes brighter than the magnitude limit of
the catalogue, this means that the strict Hickson isolation criteria
may not be met when nearby galaxies fainter than the magnitude limit
of the catalogue are considered. For groups from Catalogue A, the
brightest member galaxy has $r \le 15$ in 189 groups (8\%); for groups
in Catalogue B, the brightest member galaxy has $r \le 18$ in 20057
groups (27\%).

\subsection{Catalogue Format}

\begin{table*}
\begin{tabular*}{0.9\textwidth}{@{\extracolsep{\fill}}crccrccccccccc}
ID &  \multicolumn{3}{c}{$\alpha$ (J2000)} & \multicolumn{3}{c}{$\delta$ (J2000)} & $n_{mem}$ & $\mu$ & $\theta_G$ &  $\theta_N / \theta_G$ & $r_{max}$ & $n_z$ & $z$\\
\hline
SDSSCGA00001 &14 &49 &34.3 &+11 &14 &53.4 &    4 &    20.991 &      0.22 &      4.63 &     15.01 &    1 &     0.055\\
SDSSCGA00002 & 2 &14 & 4.5 &+13 &18 &54.3 &    4 &    21.238 &      0.28 &      3.30 &     14.81 &    1 &     0.060\\
SDSSCGA00003 &23 &54 &13.5 &-10 &23 &17.2 &    4 &    21.279 &      0.16 &      5.96 &     16.42 &    1 &     0.079\\
SDSSCGA00004 &15 &25 &53.7 &+5 &44 &17.8 &    4 &    21.501 &      0.16 &     18.64 &     15.82 &    0 &          \\
SDSSCGA00005 &23 &33 &23.6 &-1 & 8 &43.8 &    4 &    21.519 &      0.29 &      5.19 &     14.53 &    1 &     0.091\\
SDSSCGA00006 &21 &40 &17.4 &-8 & 4 &11.7 &    4 &    21.566 &      0.14 &      4.31 &     16.84 &    0 &          \\
SDSSCGA00007 & 8 &24 &31.6 &+20 &27 &28.5 &    4 &    21.585 &      0.19 &      9.26 &     15.77 &    2 &     0.109\\
SDSSCGA00008 &16 &10 & 2.6 &+5 &54 &53.5 &    4 &    21.747 &      0.31 &      3.74 &     14.89 &    1 &     0.065\\
SDSSCGA00009 &12 & 3 &12.9 &+57 &53 &39.2 &    4 &    21.755 &      0.32 &      7.84 &     14.95 &    2 &     0.034\\
SDSSCGA00010 &16 &26 &50.4 &+25 &53 &34.7 &    4 &    21.913 &      0.20 &      7.58 &     16.01 &    2 &     0.111\\
SDSSCGA00011 &16 &21 &56.5 &+25 &41 &20.1 &    4 &    22.054 &      0.21 &      5.84 &     16.48 &    3 &     0.100\\
SDSSCGA00012 & 7 &44 &42.7 &+16 &55 &21.6 &    4 &    22.130 &      0.29 &      3.40 &     15.36 &    0 &          \\
 \hline

\end{tabular*}
\caption{Compact groups identified in Catalogue A, ranked in order of
decreasing surface brightness. This table has 2297 rows, of which only
the first 12 rows are reproduced here. See Section~2.5 for a
description of each column.}
\end{table*}

\begin{table*}
\begin{tabular*}{0.9\textwidth}{@{\extracolsep{\fill}}crccrccccccccc}
ID &  \multicolumn{3}{c}{$\alpha$ (J2000)} & \multicolumn{3}{c}{$\delta$ (J2000)} & $n_{mem}$ & $\mu$ & $\theta_G$ &  $\theta_N / \theta_G$ & $r_{max}$ & $n_z$ & $z$\\
\hline
SDSSCGB00001 &14 &12 &15.8 &+35 &50 &59.0 &    4 &    19.730 &      0.08 &     14.67 &     16.01 &    1 &     0.059\\
SDSSCGB00002 &16 &15 &45.9 &+54 &40 &19.6 &    4 &    19.791 &      0.10 &     48.11 &     15.35 &    0 &    \\
SDSSCGB00003 &13 &25 &10.4 &+17 & 3 & 8.0 &    4 &    20.333 &      0.13 &     35.94 &     15.05 &    0 &    \\
SDSSCGB00004 &11 &44 &12.1 &+27 & 0 &12.0 &    4 &    20.347 &      0.09 &      7.32 &     16.41 &    1 &     0.093\\
SDSSCGB00005 & 7 &55 &30.4 &+10 &25 &51.8 &    4 &    20.366 &      0.14 &     17.32 &     15.11 &    0 &    \\
SDSSCGB00006 & 9 & 4 &34.9 &+14 &35 &42.4 &    5 &    20.547 &      0.19 &     15.32 &     15.14 &    1 &     0.050\\
SDSSCGB00007 &13 &54 &19.5 &+7 &23 & 8.3 &    4 &    20.577 &      0.12 &     13.62 &     15.62 &    1 &     0.075\\
SDSSCGB00008 &11 & 4 &36.7 &+6 &23 &46.1 &    4 &    20.725 &      0.13 &     24.23 &     15.43 &    1 &     0.032\\
SDSSCGB00009 &16 &28 &28.3 &+41 &13 & 6.2 &    4 &    20.749 &      0.21 &     17.08 &     14.61 &    1 &     0.028\\
SDSSCGB00010 &16 &13 &18.9 &+50 & 2 &12.7 &    4 &    20.805 &      0.04 &      7.93 &     18.78 &    0 &    \\
SDSSCGB00011 &14 &29 &17.4 &-3 & 9 &13.3 &    4 &    20.889 &      0.16 &      3.33 &     16.03 &    1 &     0.082\\
SDSSCGB00012 &13 &39 &44.9 &+45 &39 &58.9 &    4 &    20.966 &      0.05 &     14.28 &     18.08 &    0 &    \\
 \hline

\end{tabular*}
\caption{Compact groups identified in Catalogue B, ranked in order of
decreasing surface brightness. This table has 74791 rows, of which only
the first 12 rows are reproduced here. See Section~2.5 for a description of
each column.}
\end{table*}

\begin{table*}
\begin{tabular*}{0.99\textwidth}{@{\extracolsep{\fill}}ccrccrccccccc}
ID & ObjID (SDSS) &  \multicolumn{3}{c}{$\alpha$ (J2000)} & \multicolumn{3}{c}{$\delta$ (J2000)} & $r$ & $(g - r)$ & SpecObjID (SDSS) & $z_{conf}$ & $z$\\
\hline
SDSSCGA00001.1 &587736807771078936 &14 &49 &34.5 &+11 &14 &53.2 &          15.01 &           0.86 &0                  &        &        \\
SDSSCGA00001.2 &587736807771078937 &14 &49 &34.9 &+11 &14 &55.2 &          15.29 &           0.82 &0                  &        &        \\
SDSSCGA00001.3 &587736807771078935 &14 &49 &34.2 &+11 &14 &44.0 &          16.23 &           1.07 &482677981936877568 &          0.999&          0.055\\
SDSSCGA00001.4 &587736807771078938 &14 &49 &33.6 &+11 &15 & 1.2 &          17.29 &           0.97 &0                  &        &        \\
SDSSCGA00002.1 &587724198822412473 & 2 &14 & 3.9 &+13 &18 &47.2 &          14.81 &           1.08 &120694126130757632 &          0.999&          0.060\\
SDSSCGA00002.2 &587724198822477903 & 2 &14 & 5.1 &+13 &18 &39.5 &          15.07 &           0.87 &0                  &        &        \\
SDSSCGA00002.3 &587724198822477905 & 2 &14 & 5.0 &+13 &19 & 2.3 &          15.52 &           1.12 &0                  &        &        \\
SDSSCGA00002.4 &587724198822412475 & 2 &14 & 4.2 &+13 &19 & 8.1 &          17.71 &           0.68 &0                  &        &        \\
SDSSCGA00003.1 &587727225689538694 &23 &54 &13.2 &-10 &23 &11.0 &          16.42 &           0.85 &182901462030876672 &          0.999&          0.079\\
SDSSCGA00003.2 &587727225689538695 &23 &54 &13.5 &-10 &23 & 8.7 &          16.69 &           1.03 &0                  &        &        \\
SDSSCGA00003.3 &587727225689538696 &23 &54 &13.5 &-10 &23 &23.6 &          16.77 &           0.81 &0                  &        &        \\
SDSSCGA00003.4 &587727225689538697 &23 &54 &13.8 &-10 &23 &25.5 &          16.86 &           0.79 &0                  &        &        \\
SDSSCGA00004.1 &587730023333625957 &15 &25 &53.9 &+5 &44 & 9.7 &          15.82 &           0.99 &0                  &        &        \\
SDSSCGA00004.2 &587730023333625958 &15 &25 &53.7 &+5 &44 &27.7 &          17.20 &           1.09 &0                  &        &        \\
SDSSCGA00004.3 &587730023333625960 &15 &25 &53.4 &+5 &44 &10.2 &          17.57 &           0.84 &0                  &        &        \\
SDSSCGA00004.4 &587730023333625959 &15 &25 &53.8 &+5 &44 &23.7 &          17.83 &           0.95 &0                  &        &        \\
SDSSCGA00005.1 &588015507655819398 &23 &33 &22.7 &-1 & 8 &54.9 &          14.53 &           0.97 &0                  &        &        \\
SDSSCGA00005.2 &588015507655819395 &23 &33 &24.4 &-1 & 8 &54.2 &          16.16 &           1.07 &0                  &        &        \\
SDSSCGA00005.3 &588015507655819396 &23 &33 &23.6 &-1 & 8 &35.0 &          16.33 &           1.05 &108308961144340480 &          1.000&          0.091\\
SDSSCGA00005.4 &588015507655819397 &23 &33 &23.6 &-1 & 8 &31.0 &          17.04 &           0.84 &0                  &        &        \\
SDSSCGA00006.1 &587727213348454817 &21 &40 &17.3 &-8 & 4 &14.8 &          16.84 &           1.36 &0                  &        &        \\
SDSSCGA00006.2 &587727213348454819 &21 &40 &16.9 &-8 & 4 & 8.9 &          16.95 &           0.46 &0                  &        &        \\
SDSSCGA00006.3 &587727213348454818 &21 &40 &17.5 &-8 & 4 &17.4 &          17.22 &           0.83 &0                  &        &        \\
SDSSCGA00006.4 &587727213348454820 &21 &40 &17.8 &-8 & 4 & 5.7 &          17.87 &           0.93 &0                  &        &        \\
SDSSCGA00007.1 &587739407832580481 & 8 &24 &32.1 &+20 &27 &26.2 &          15.77 &           1.22 &586541553271439360 &          1.000&          0.109\\
SDSSCGA00007.2 &587739407832580480 & 8 &24 &31.9 &+20 &27 &39.4 &          16.77 &           1.18 &542631292303310848 &          0.998&          0.110\\
SDSSCGA00007.3 &587739407832580484 & 8 &24 &31.5 &+20 &27 &27.8 &          17.01 &           1.59 &0                  &        &        \\
SDSSCGA00007.4 &587739407832580482 & 8 &24 &31.1 &+20 &27 &20.7 &          17.48 &           0.98 &0                  &        &        \\
SDSSCGA00008.1 &587736543103877672 &16 &10 & 2.5 &+5 &55 & 1.4 &          14.89 &           0.81 &513360322849931264 &          0.999&          0.065\\
SDSSCGA00008.2 &587736543103877675 &16 &10 & 1.5 &+5 &54 &59.2 &          15.74 &           0.96 &0                  &        &        \\
SDSSCGA00008.3 &587736543103877673 &16 &10 & 3.5 &+5 &54 &48.4 &          16.24 &           0.88 &0                  &        &        \\
SDSSCGA00008.4 &587736543103877674 &16 &10 & 3.1 &+5 &54 &45.2 &          16.61 &           1.00 &0                  &        &        \\
SDSSCGA00009.1 &587735696979656747 &12 & 3 &13.8 &+57 &53 &26.0 &          14.95 &           0.56 &370085969178656768 &          1.000&          0.034\\
SDSSCGA00009.2 &587735696979656750 &12 & 3 &12.5 &+57 &53 &36.4 &          15.02 &           0.48 &0                  &        &        \\
SDSSCGA00009.3 &587735696979656749 &12 & 3 &13.4 &+57 &53 &53.0 &          16.64 &           0.47 &369803378328338432 &          0.995&          0.034\\
SDSSCGA00009.4 &587735696979656748 &12 & 3 &11.8 &+57 &53 &41.2 &          17.34 &           1.64 &0                  &        &        \\
SDSSCGA00010.1 &587736920508858742 &16 &26 &50.6 &+25 &53 &28.3 &          16.01 &           0.98 &442988743981268992 &          1.000&          0.110\\
SDSSCGA00010.2 &587736899576725804 &16 &26 &50.4 &+25 &53 &39.6 &          16.97 &           1.09 &0                  &        &        \\
SDSSCGA00010.3 &587736920508858744 &16 &26 &49.7 &+25 &53 &27.8 &          17.32 &           1.09 &443271293421223936 &          1.000&          0.112\\
SDSSCGA00010.4 &587736899576725805 &16 &26 &50.7 &+25 &53 &43.1 &          17.82 &           0.81 &0                  &        &        \\
SDSSCGA00011.1 &587736919434789305 &16 &21 &57.1 &+25 &41 &29.5 &          16.48 &           0.95 &443834327290085376 &          1.000&          0.099\\
SDSSCGA00011.2 &587736919434789307 &16 &21 &56.1 &+25 &41 &19.1 &          16.76 &           1.11 &0                  &        &        \\
SDSSCGA00011.3 &587736919434789306 &16 &21 &56.2 &+25 &41 &21.1 &          16.87 &           0.84 &443271292800466944 &          0.999&          0.100\\
SDSSCGA00011.4 &587736919434789308 &16 &21 &56.5 &+25 &41 &10.9 &          17.26 &           0.56 &443552841290743808 &          0.947&          0.101\\
SDSSCGA00012.1 &587738372204396691 & 7 &44 &43.3 &+16 &55 &21.3 &          15.36 &           0.92 &0                  &        &        \\
SDSSCGA00012.2 &587738372204396692 & 7 &44 &42.4 &+16 &55 &38.5 &          15.95 &           0.85 &0                  &        &        \\
SDSSCGA00012.3 &587738372204396694 & 7 &44 &41.9 &+16 &55 &20.6 &          17.23 &           0.91 &0                  &        &        \\
SDSSCGA00012.4 &587738372204396693 & 7 &44 &43.0 &+16 &55 & 5.8 &          17.73 &           0.86 &0                  &        &        \\
\end{tabular*}
\caption{Individual member galaxies in each compact group found in Catalogue
A. Groups are listed in the same order as Table~1; galaxies in each
group are listed in order of descending $r-$band luminosity. This
table has 9713 rows, of which the first 48 rows are reproduced here
(corresponding to the brightest 12 compact groups). See Section~2.5
for a description of each column.}
\end{table*}

\begin{table*}
\begin{tabular*}{0.99\textwidth}{@{\extracolsep{\fill}}ccrccrccccccc}
ID & ObjID (SDSS) &  \multicolumn{3}{c}{$\alpha$ (J2000)} & \multicolumn{3}{c}{$\delta$ (J2000)} & $r$ & $(g - r)$ & SpecObjID (SDSS) & $z_{conf}$ & $z$\\
\hline
SDSSCGB00001.1 &588017977829097722 &14 &12 &15.9 &+35 &50 &55.3 &          16.01 &           1.18 &0                  &        &        \\
SDSSCGB00001.2 &588017977829097728 &14 &12 &15.8 &+35 &51 & 3.3 &          16.56 &           0.49 &462691635159891968 &          0.989&          0.059\\
SDSSCGB00001.3 &588017977829097727 &14 &12 &15.5 &+35 &50 &59.0 &          16.69 &           0.44 &0                  &        &        \\
SDSSCGB00001.4 &588017977829097726 &14 &12 &16.2 &+35 &50 &58.6 &          17.07 &           0.11 &0                  &        &        \\
SDSSCGB00002.1 &587739849139683368 &16 &15 &45.5 &+54 &40 &22.2 &          15.35 &           0.50 &0                  &        &        \\
SDSSCGB00002.2 &587739849139683374 &16 &15 &46.2 &+54 &40 &17.5 &          16.26 &           0.51 &0                  &        &        \\
SDSSCGB00002.3 &587739849139683371 &16 &15 &46.3 &+54 &40 &22.5 &          16.34 &           0.74 &0                  &        &        \\
SDSSCGB00002.4 &587739849139683372 &16 &15 &45.6 &+54 &40 &16.3 &          17.59 &           0.42 &0                  &        &        \\
SDSSCGB00003.1 &587742903404986565 &13 &25 &10.3 &+17 & 3 &12.0 &          15.05 &           0.44 &0                  &        &        \\
SDSSCGB00003.2 &587742903404986567 &13 &25 &10.4 &+17 & 3 & 0.9 &          16.60 &           0.14 &0                  &        &        \\
SDSSCGB00003.3 &587742903404986570 &13 &25 &10.9 &+17 & 3 & 8.4 &          16.92 &           0.45 &0                  &        &        \\
SDSSCGB00003.4 &587742903404986568 &13 &25 & 9.9 &+17 & 3 &10.8 &          17.32 &           0.45 &0                  &        &        \\
SDSSCGB00004.1 &587741601491714194 &11 &44 &12.2 &+27 & 0 & 8.1 &          16.41 &           0.77 &0                  &        &        \\
SDSSCGB00004.2 &587741708883329169 &11 &44 &12.3 &+27 & 0 &11.8 &          17.12 &           0.78 &0                  &        &        \\
SDSSCGB00004.3 &587741708883329171 &11 &44 &11.8 &+27 & 0 &11.4 &          17.27 &           0.31 &625386958805270528 &          0.978&          0.093\\
SDSSCGB00004.4 &587741708883329170 &11 &44 &12.3 &+27 & 0 &16.6 &          18.02 &           0.14 &0                  &        &        \\
SDSSCGB00005.1 &587741816231690635 & 7 &55 &29.9 &+10 &25 &51.9 &          15.11 &           0.92 &0                  &        &        \\
SDSSCGB00005.2 &587741816231690639 & 7 &55 &30.2 &+10 &25 &46.5 &          15.81 &           0.35 &0                  &        &        \\
SDSSCGB00005.3 &587741816231690637 & 7 &55 &30.7 &+10 &25 &51.4 &          16.57 &           1.11 &0                  &        &        \\
SDSSCGB00005.4 &587741816231690644 & 7 &55 &30.9 &+10 &25 &57.4 &          17.96 &           0.32 &0                  &        &        \\
SDSSCGB00006.1 &587744727686381596 & 9 & 4 &34.8 &+14 &35 &36.3 &          15.14 &           0.59 &0                  &        &        \\
SDSSCGB00006.2 &587744727686381602 & 9 & 4 &34.4 &+14 &35 &39.4 &          15.38 &           0.13 &685341276190539776 &          0.993&          0.050\\
SDSSCGB00006.3 &587744727686381597 & 9 & 4 &34.6 &+14 &35 &52.4 &          15.51 &           0.76 &0                  &        &        \\
SDSSCGB00006.4 &587744727686381607 & 9 & 4 &35.3 &+14 &35 &38.8 &          17.40 &           1.55 &0                  &        &        \\
SDSSCGB00006.5 &587744727686381610 & 9 & 4 &35.5 &+14 &35 &45.1 &          17.49 &           0.16 &0                  &        &        \\
SDSSCGB00007.1 &588017726021369993 &13 &54 &19.7 &+7 &23 &12.4 &          15.62 &           0.89 &508293726485872640 &          0.999&          0.075\\
SDSSCGB00007.2 &588017726021369994 &13 &54 &19.1 &+7 &23 & 3.7 &          16.59 &           0.87 &0                  &        &        \\
SDSSCGB00007.3 &588017726021369995 &13 &54 &19.4 &+7 &23 &15.3 &          17.81 &           0.59 &0                  &        &        \\
SDSSCGB00007.4 &588017726021369996 &13 &54 &19.7 &+7 &23 & 2.1 &          18.51 &           0.67 &0                  &        &        \\
SDSSCGB00008.1 &587732577238712466 &11 & 4 &36.9 &+6 &23 &42.1 &          15.43 &           0.39 &0                  &        &        \\
SDSSCGB00008.2 &587732577238712469 &11 & 4 &36.6 &+6 &23 &51.9 &          17.53 &           0.20 &282545493425258496 &          0.987&          0.032\\
SDSSCGB00008.3 &587732577238712473 &11 & 4 &36.2 &+6 &23 &45.3 &          17.63 &          -0.04 &0                  &        &        \\
SDSSCGB00008.4 &587732577238712467 &11 & 4 &37.2 &+6 &23 &45.2 &          17.70 &           0.07 &0                  &        &        \\
SDSSCGB00009.1 &587729652347961399 &16 &28 &27.9 &+41 &13 & 3.5 &          14.61 &           0.45 &0                  &        &        \\
SDSSCGB00009.2 &587729652347961408 &16 &28 &28.7 &+41 &12 &55.6 &          16.10 &          -0.10 &0                  &        &        \\
SDSSCGB00009.3 &587729652347961407 &16 &28 &28.6 &+41 &13 &11.8 &          16.15 &          -0.00 &0                  &        &        \\
SDSSCGB00009.4 &587729652347961398 &16 &28 &28.0 &+41 &13 &13.7 &          16.26 &           0.08 &229908547784146944 &          0.988&          0.028\\
SDSSCGB00010.1 &587742888898920722 &16 &13 &18.9 &+50 & 2 &15.1 &          18.78 &           1.42 &0                  &        &        \\
SDSSCGB00010.2 &587729226885497049 &16 &13 &18.9 &+50 & 2 &15.2 &          18.80 &           1.60 &0                  &        &        \\
SDSSCGB00010.3 &587742888898920723 &16 &13 &18.9 &+50 & 2 &10.3 &          19.42 &           1.28 &0                  &        &        \\
SDSSCGB00010.4 &587729226885497050 &16 &13 &19.0 &+50 & 2 &10.3 &          19.58 &           1.17 &0                  &        &        \\
SDSSCGB00011.1 &587729776371761364 &14 &29 &17.6 &-3 & 9 & 6.5 &          16.03 &           0.83 &0                  &        &        \\
SDSSCGB00011.2 &587729776371761362 &14 &29 &17.7 &-3 & 9 &21.3 &          16.04 &           0.92 &0                  &        &        \\
SDSSCGB00011.3 &587729776371761366 &14 &29 &17.3 &-3 & 9 & 7.5 &          16.42 &           1.01 &0                  &        &        \\
SDSSCGB00011.4 &587729776371761363 &14 &29 &16.8 &-3 & 9 &17.9 &          16.68 &           1.51 &258619102271111168 &          1.000&          0.082\\
SDSSCGB00012.1 &588298663042941007 &13 &39 &45.1 &+45 &39 &58.8 &          18.08 &           0.51 &0                  &        &        \\
SDSSCGB00012.2 &588298663042941006 &13 &39 &44.8 &+45 &39 &57.8 &          19.16 &           0.70 &0                  &        &        \\
SDSSCGB00012.3 &588298663042941005 &13 &39 &44.7 &+45 &39 &58.7 &          20.15 &           2.18 &0                  &        &        \\
SDSSCGB00012.4 &588298663042941010 &13 &39 &45.0 &+45 &40 & 0.4 &          20.16 &           0.90 &0                  &        &        \\
\end{tabular*}
\caption{Individual member galaxies in each compact group found in Catalogue
B. Groups are listed in the same order as Table~1; galaxies in each
group are listed in order of descending $r-$band luminosity. This table has
313508 rows, of which the first 49 rows are reproduced here (corresponding
to the brightest 12 compact groups). See Section~2.5 for a description
of each column.}
\end{table*}

\begin{figure*}
  \begin{center}
    \includegraphics[angle=0, width=4.cm]{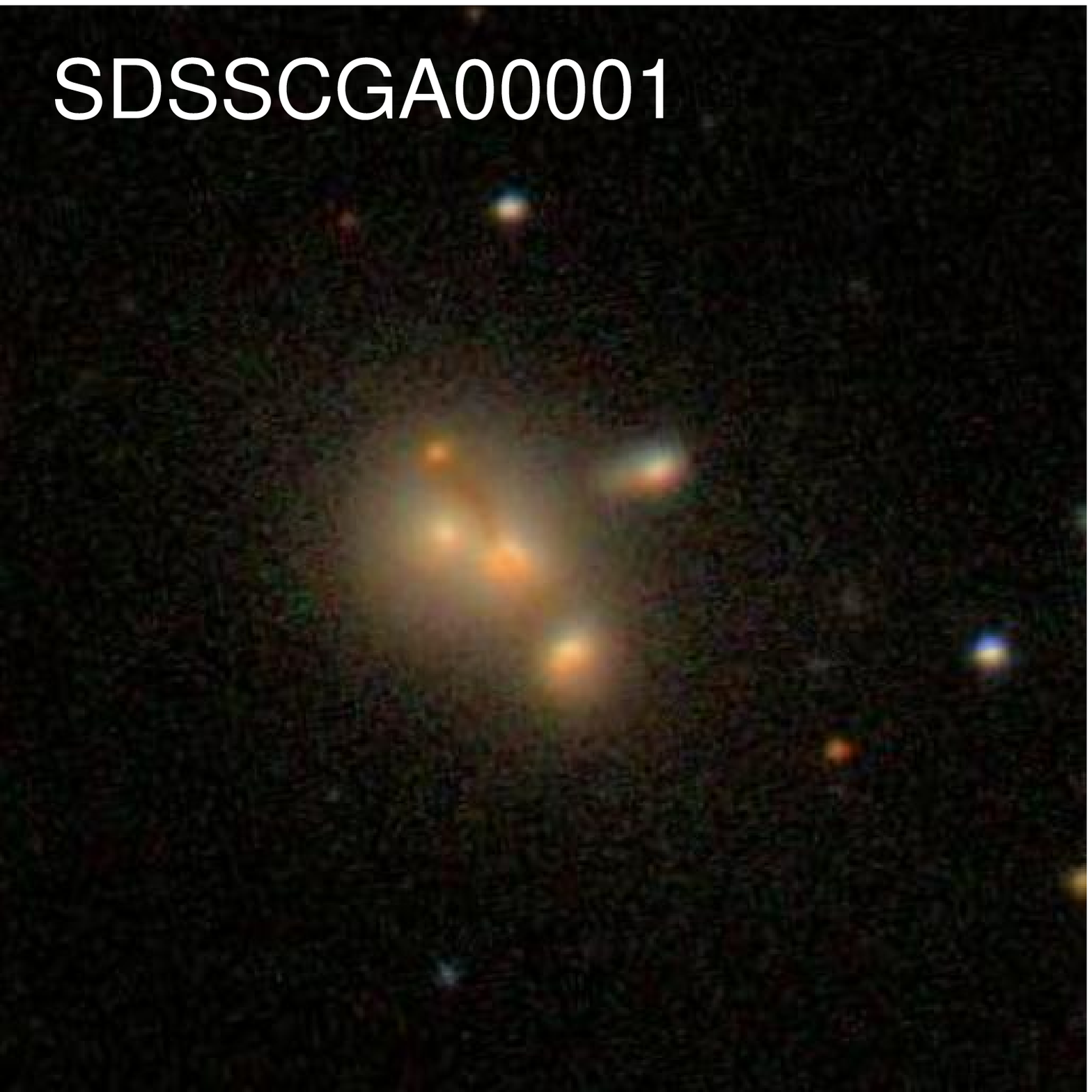}
    \includegraphics[angle=0, width=4.cm]{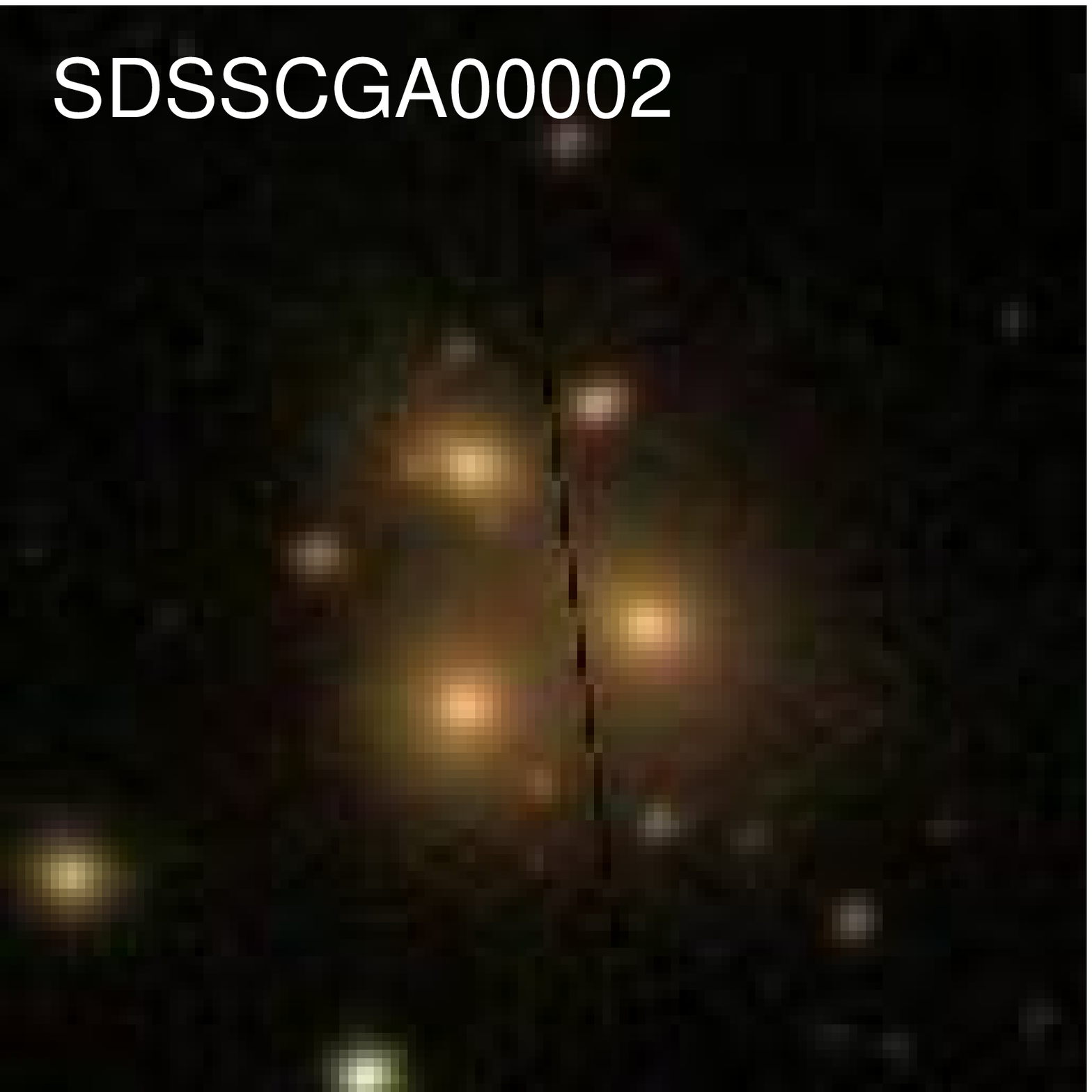}
    \includegraphics[angle=0, width=4.cm]{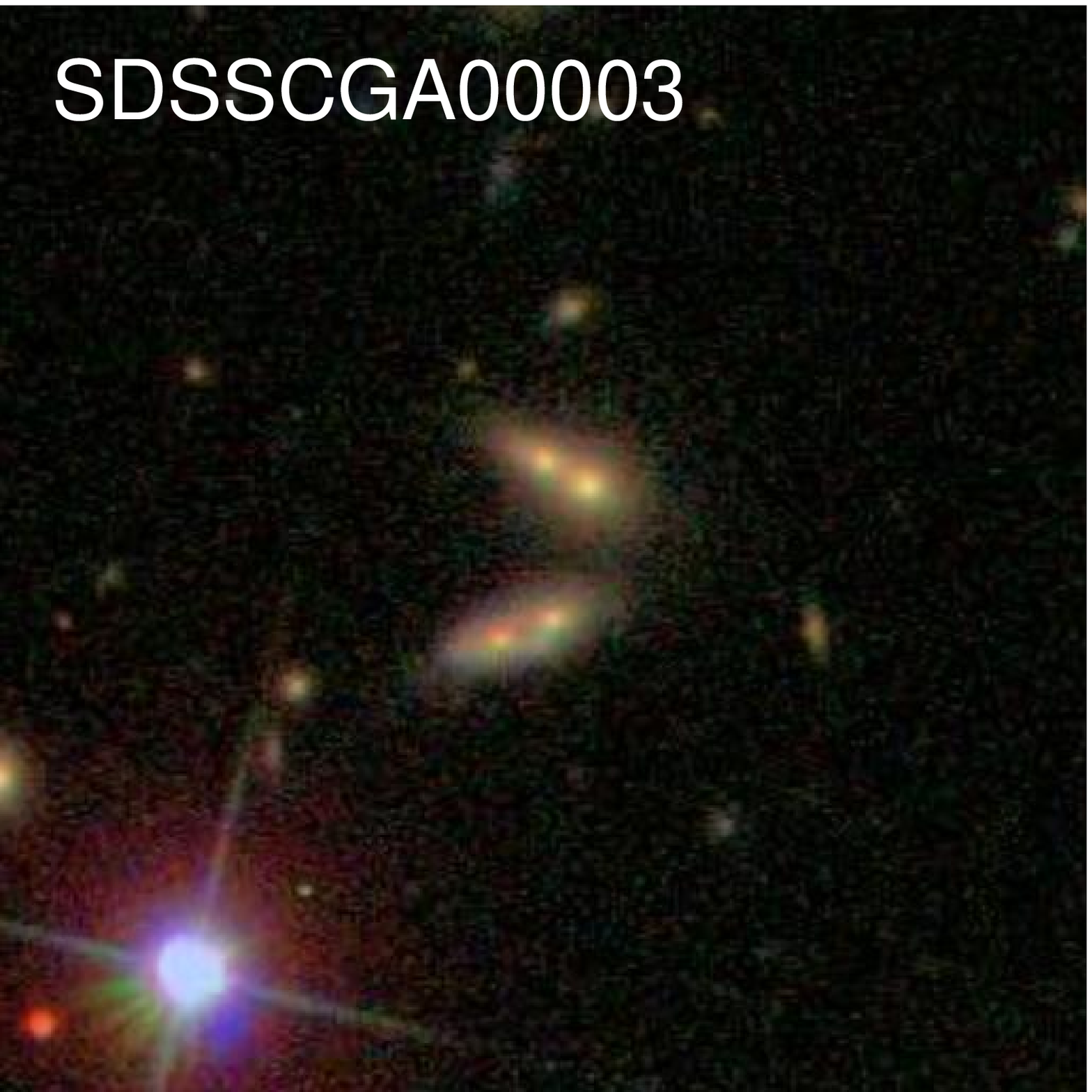}
    \includegraphics[angle=0, width=4.cm]{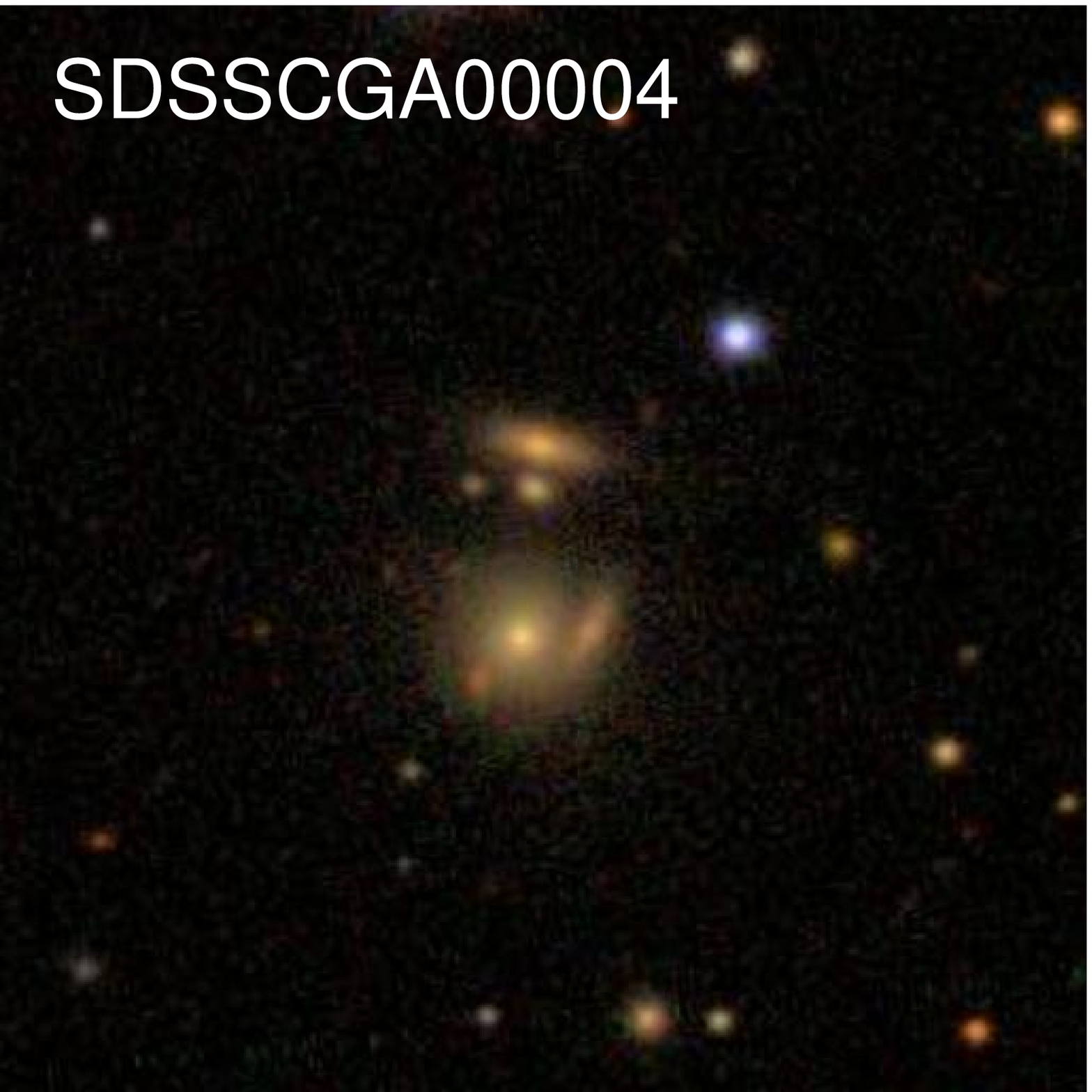}
    \includegraphics[angle=0, width=4.cm]{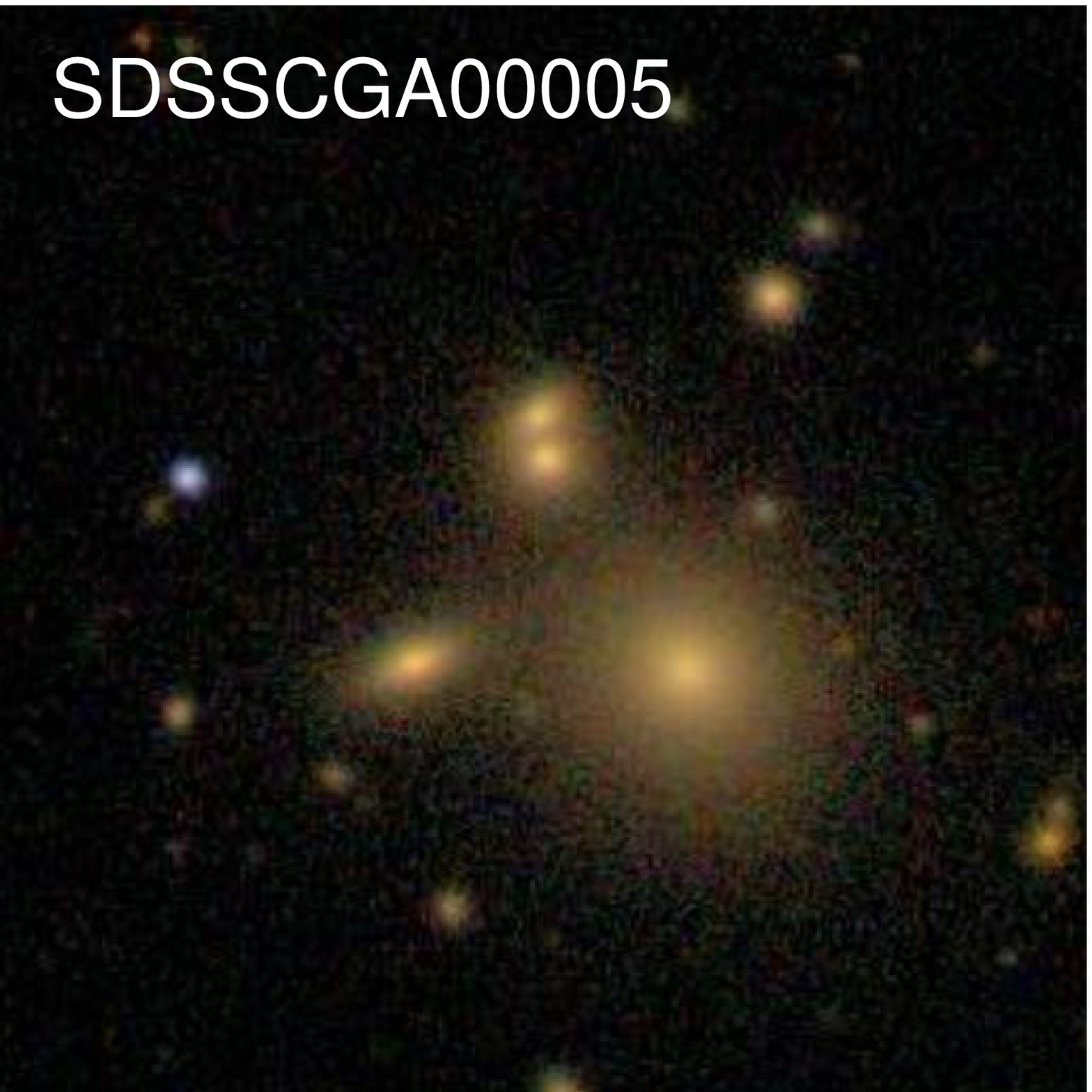}
    \includegraphics[angle=0, width=4.cm]{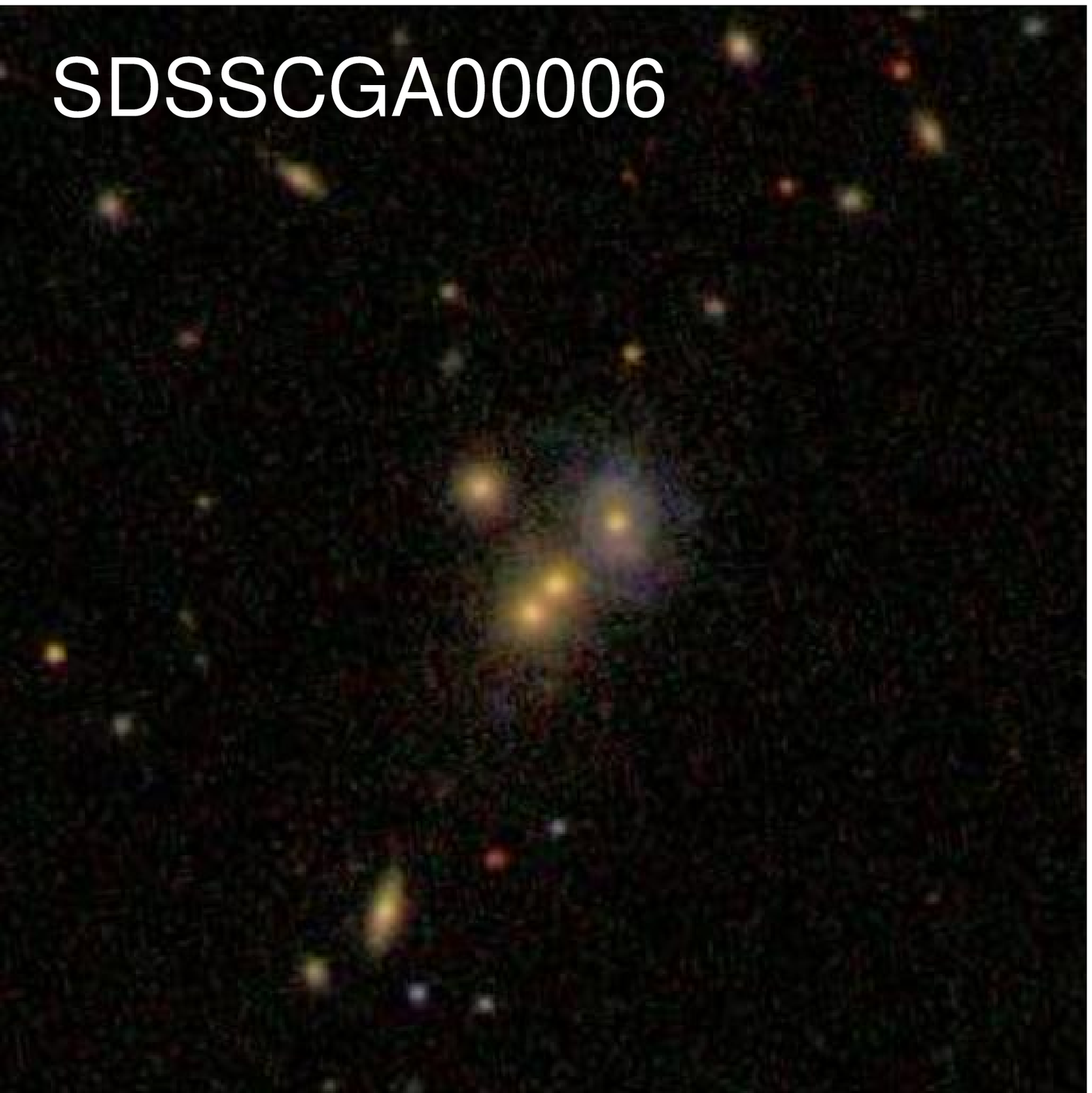}
    \includegraphics[angle=0, width=4.cm]{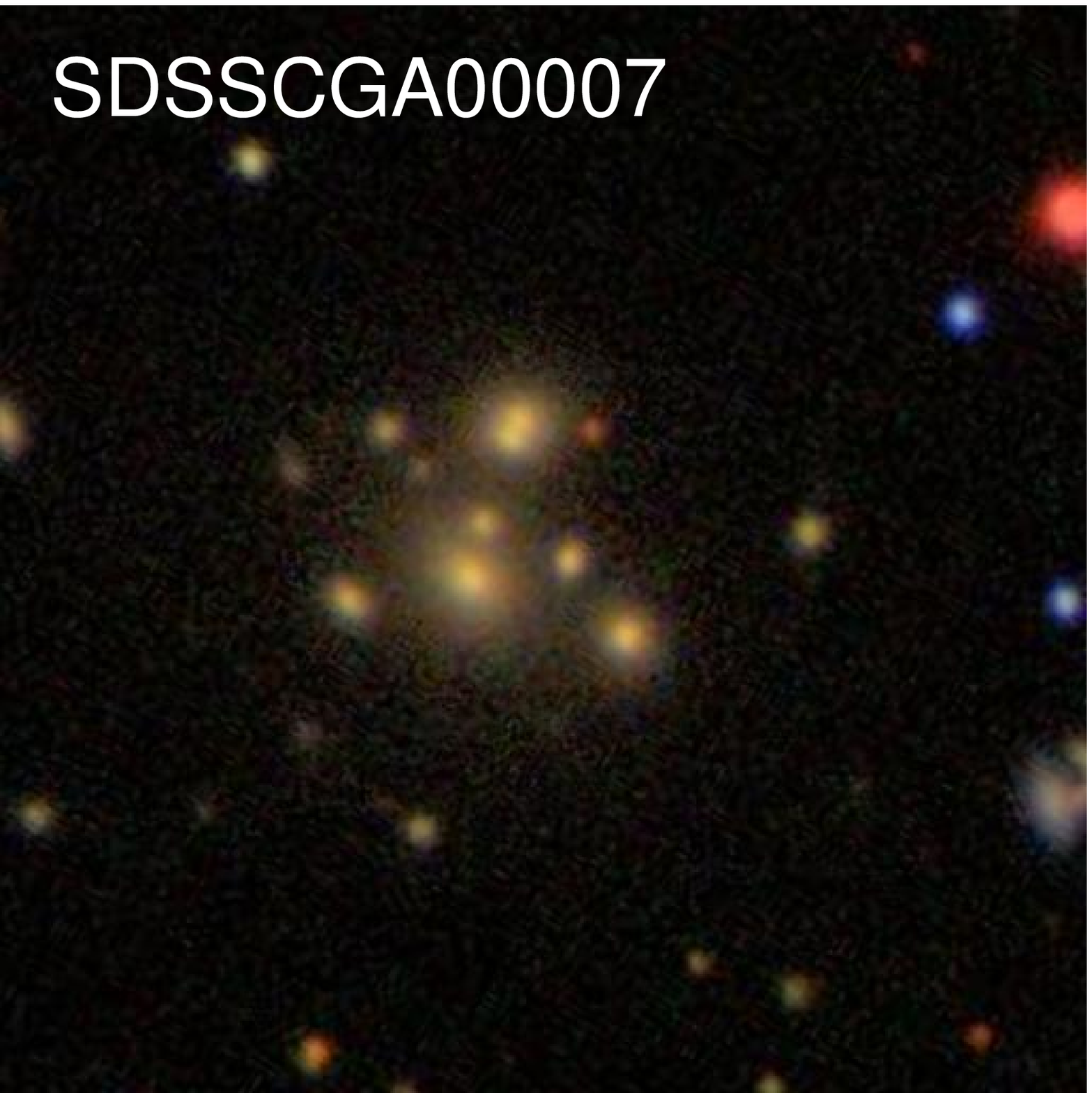}
    \includegraphics[angle=0, width=4.cm]{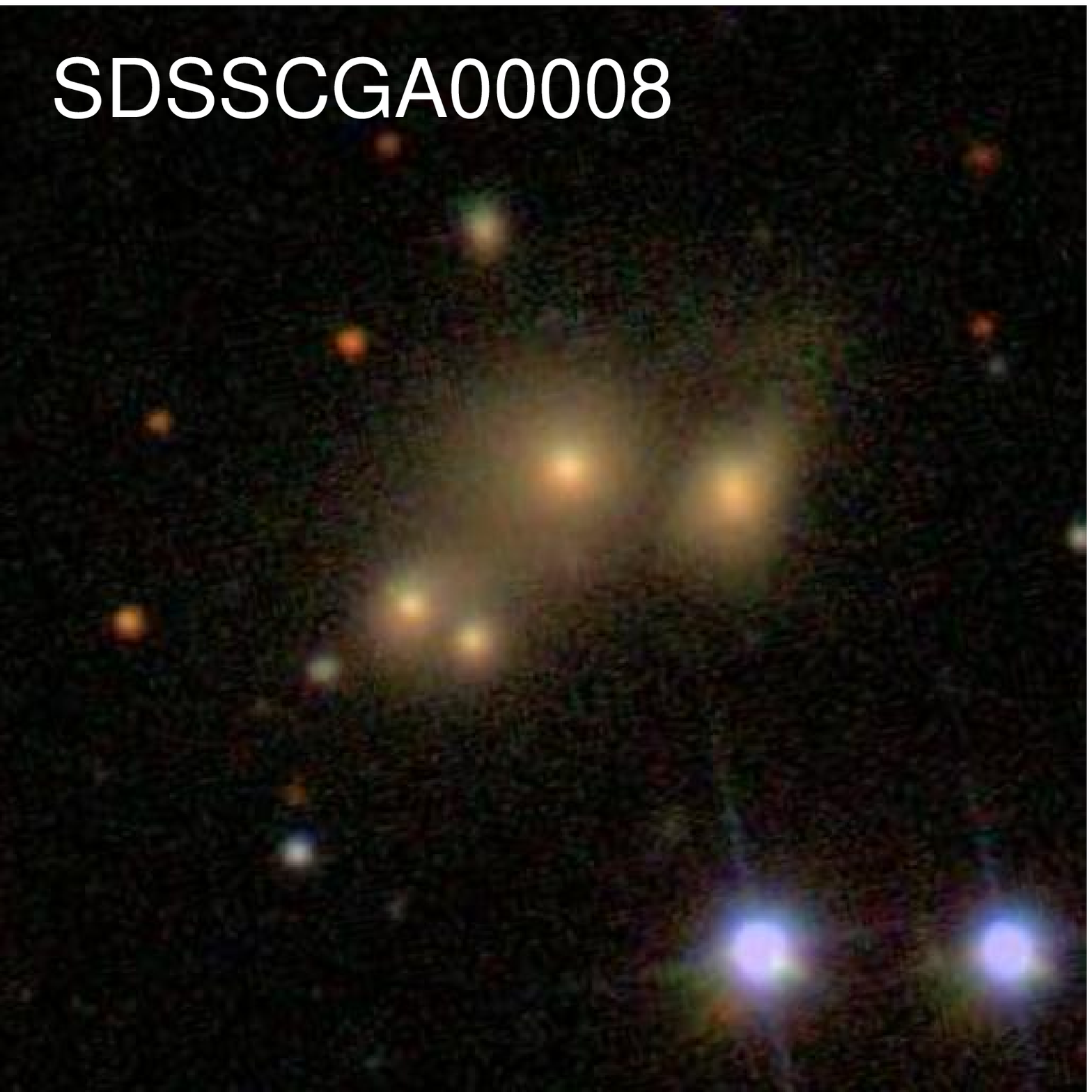}
    \includegraphics[angle=0, width=4.cm]{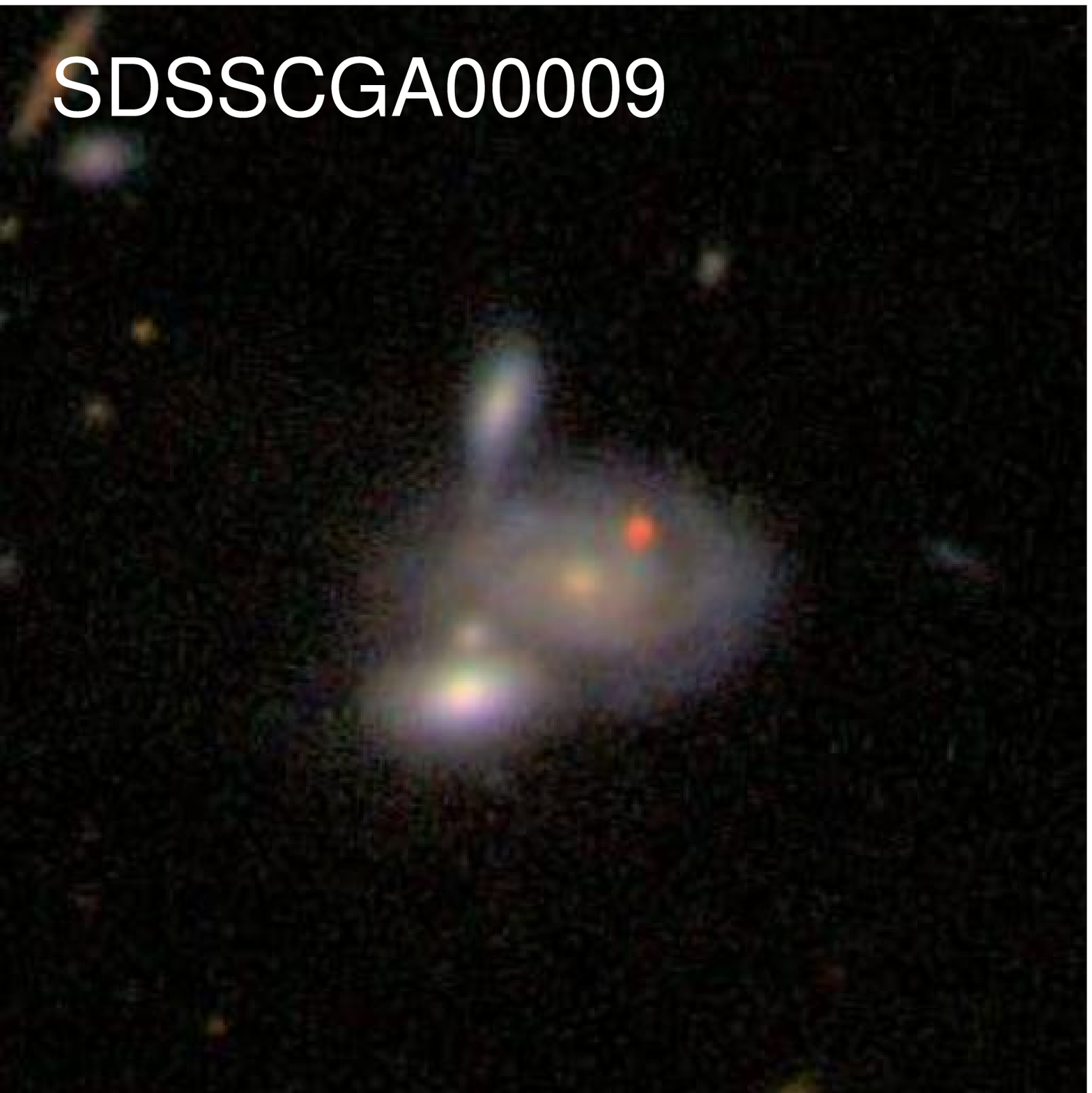}
    \includegraphics[angle=0, width=4.cm]{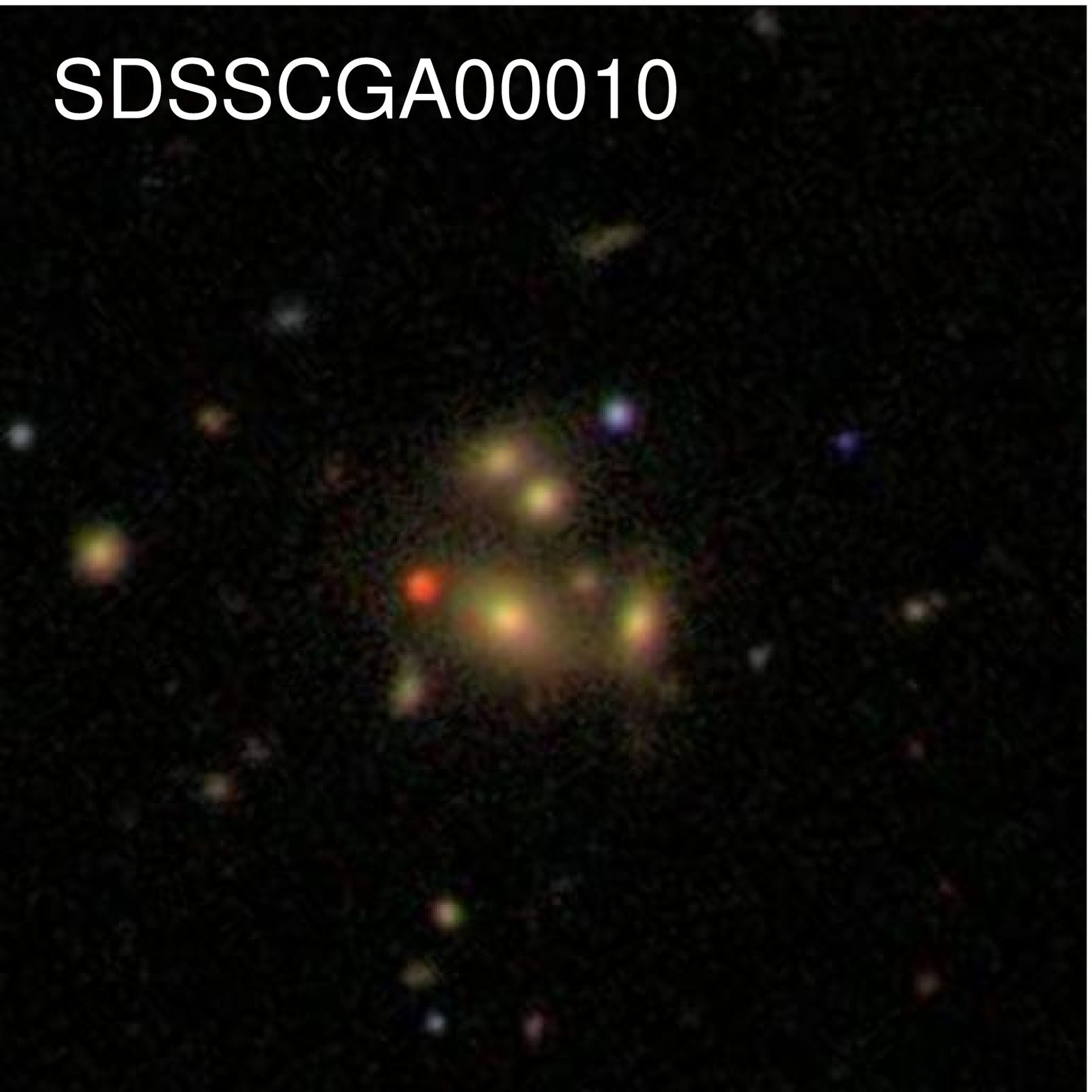}
    \includegraphics[angle=0, width=4.cm]{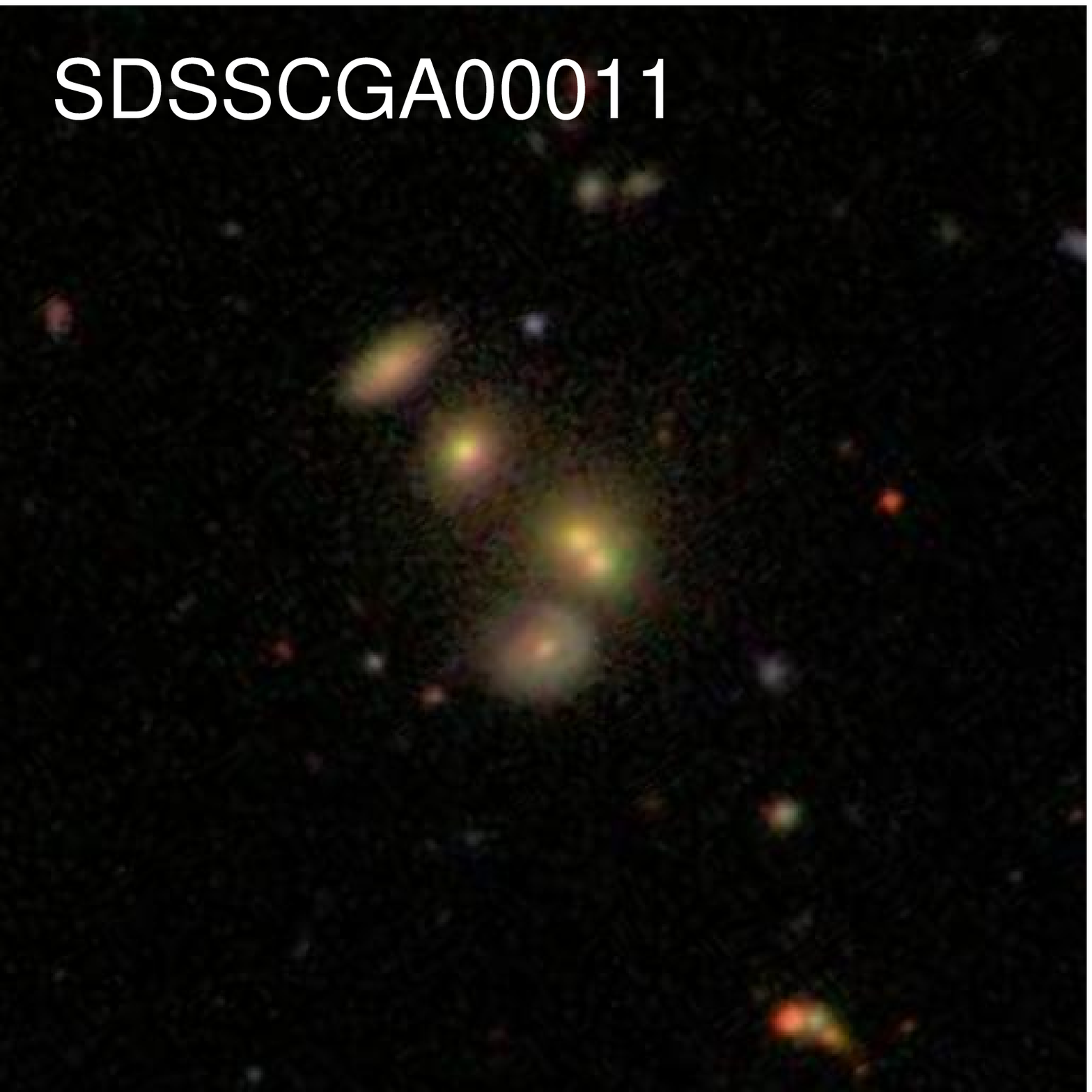}
    \includegraphics[angle=0, width=4.cm]{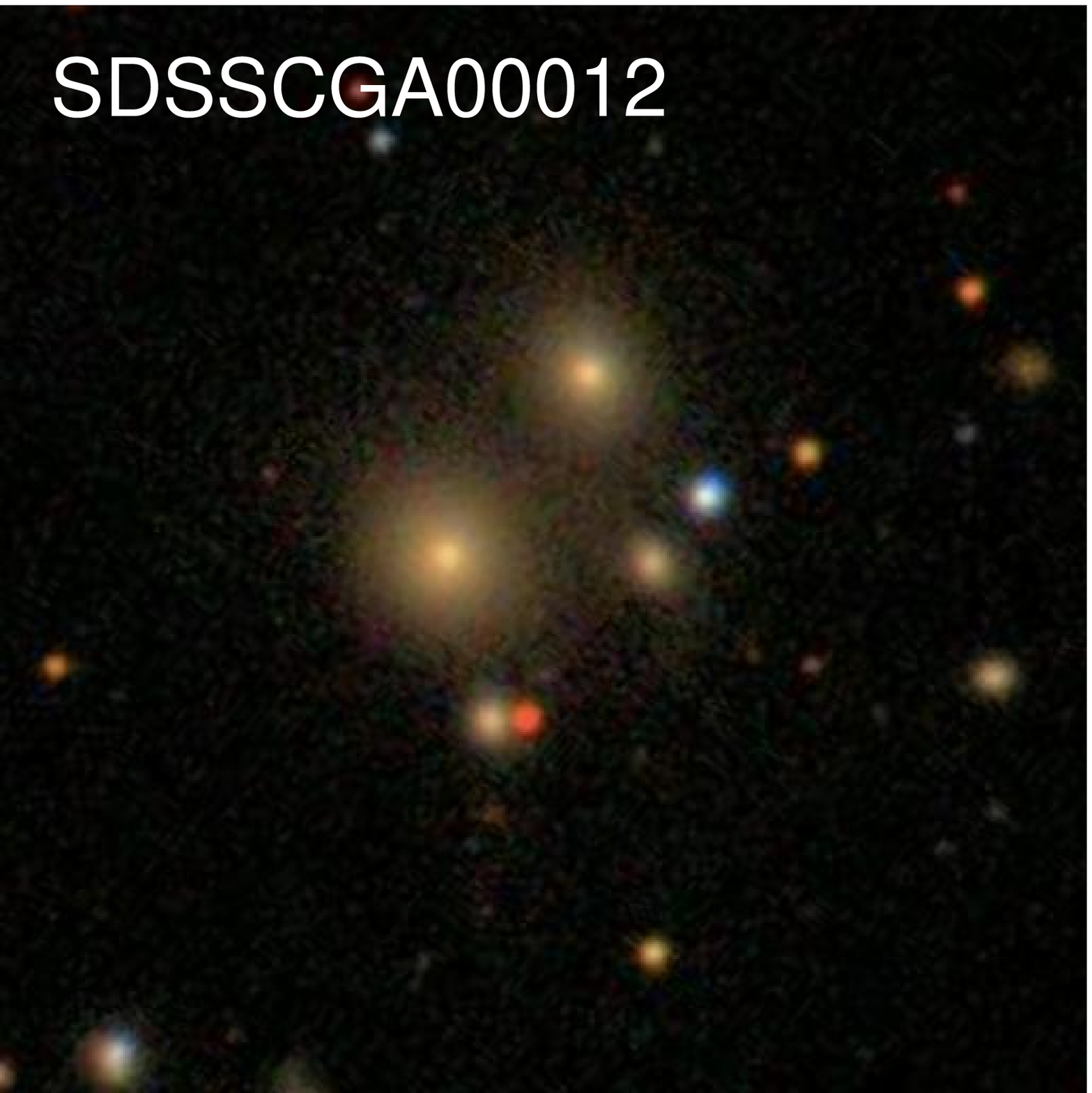}
    \caption{SDSS finding chart images of the twelve compact groups found in
    Catalogue~A with the highest central surface brightnesses which
    are listed in Table~1. Each image is $1.7 \times 1.7$\,arcmins.}
    \label{top12}
  \end{center}
\end{figure*}

\begin{figure*}
  \begin{center}
    \includegraphics[angle=0, width=4.cm]{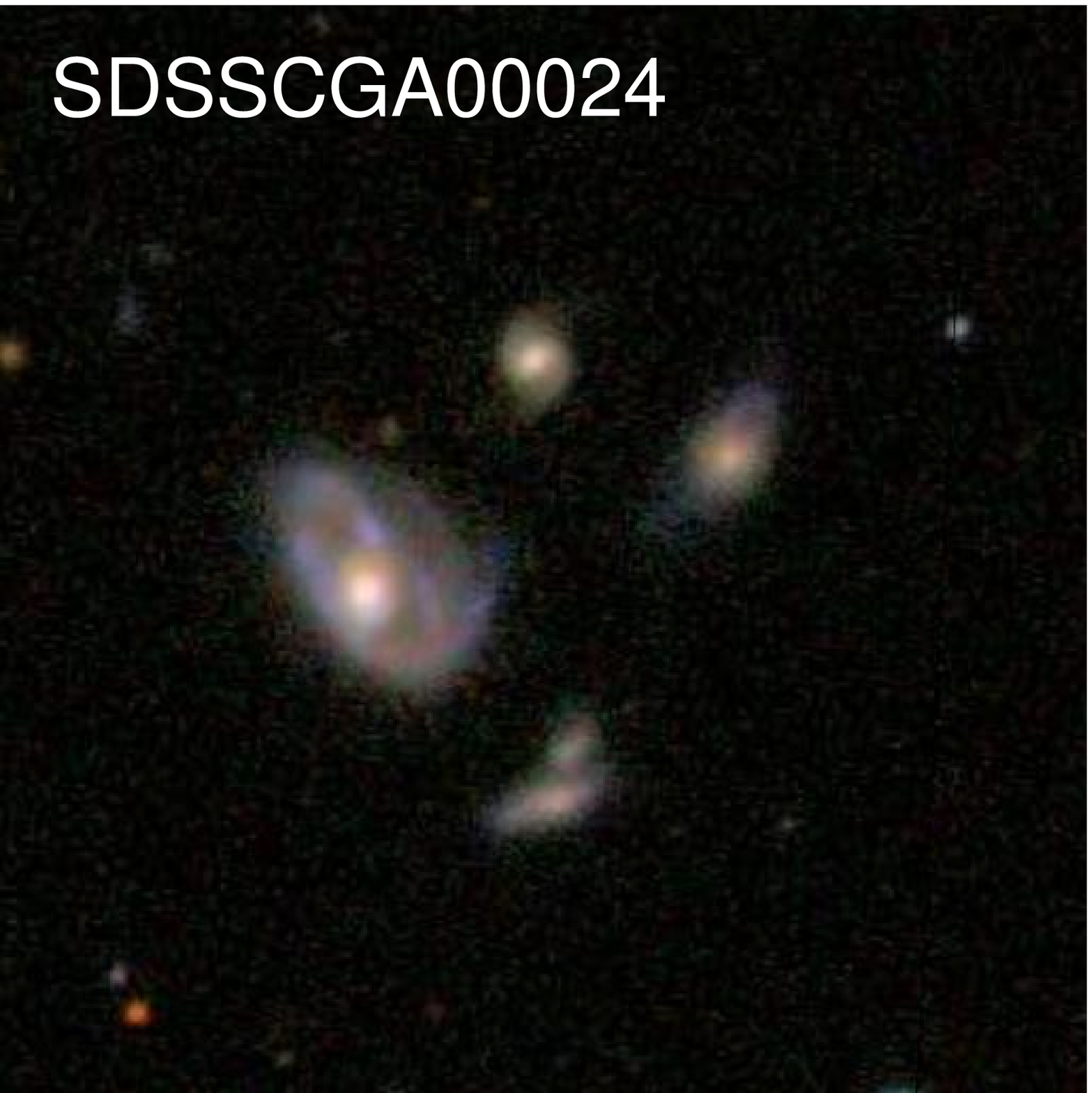}
    \includegraphics[angle=0, width=4.cm]{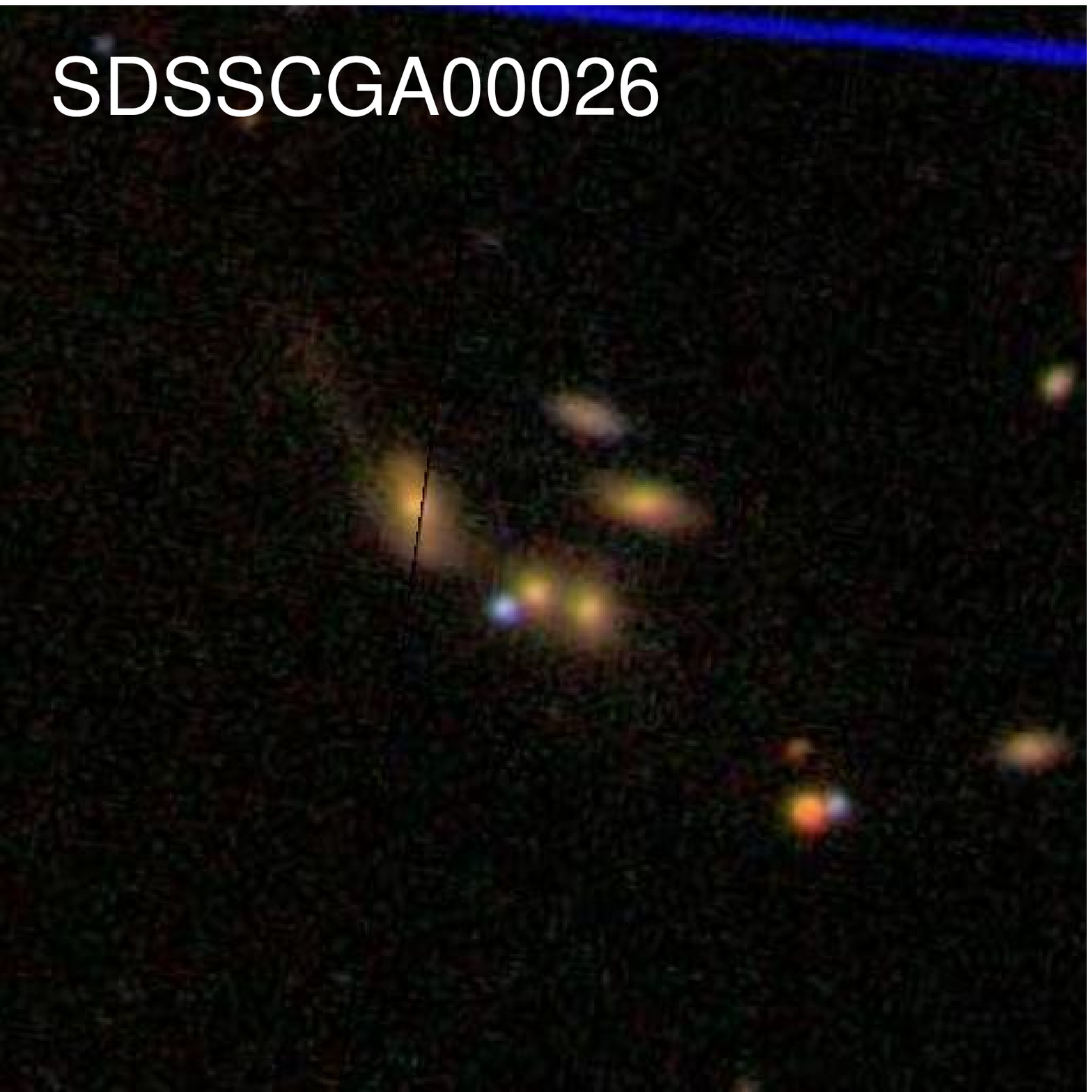}
    \includegraphics[angle=0, width=4.cm]{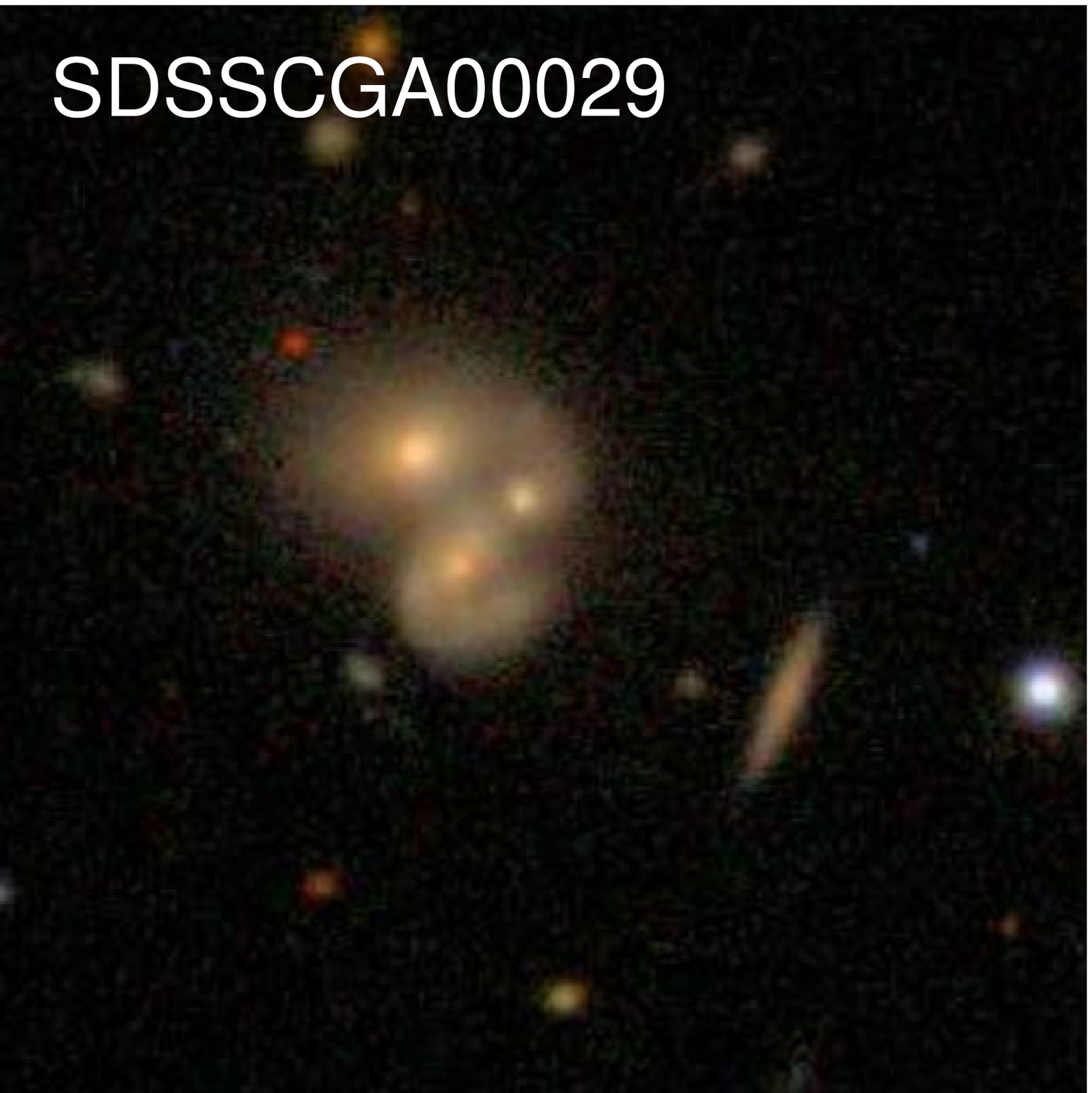}
    \includegraphics[angle=0, width=4.cm]{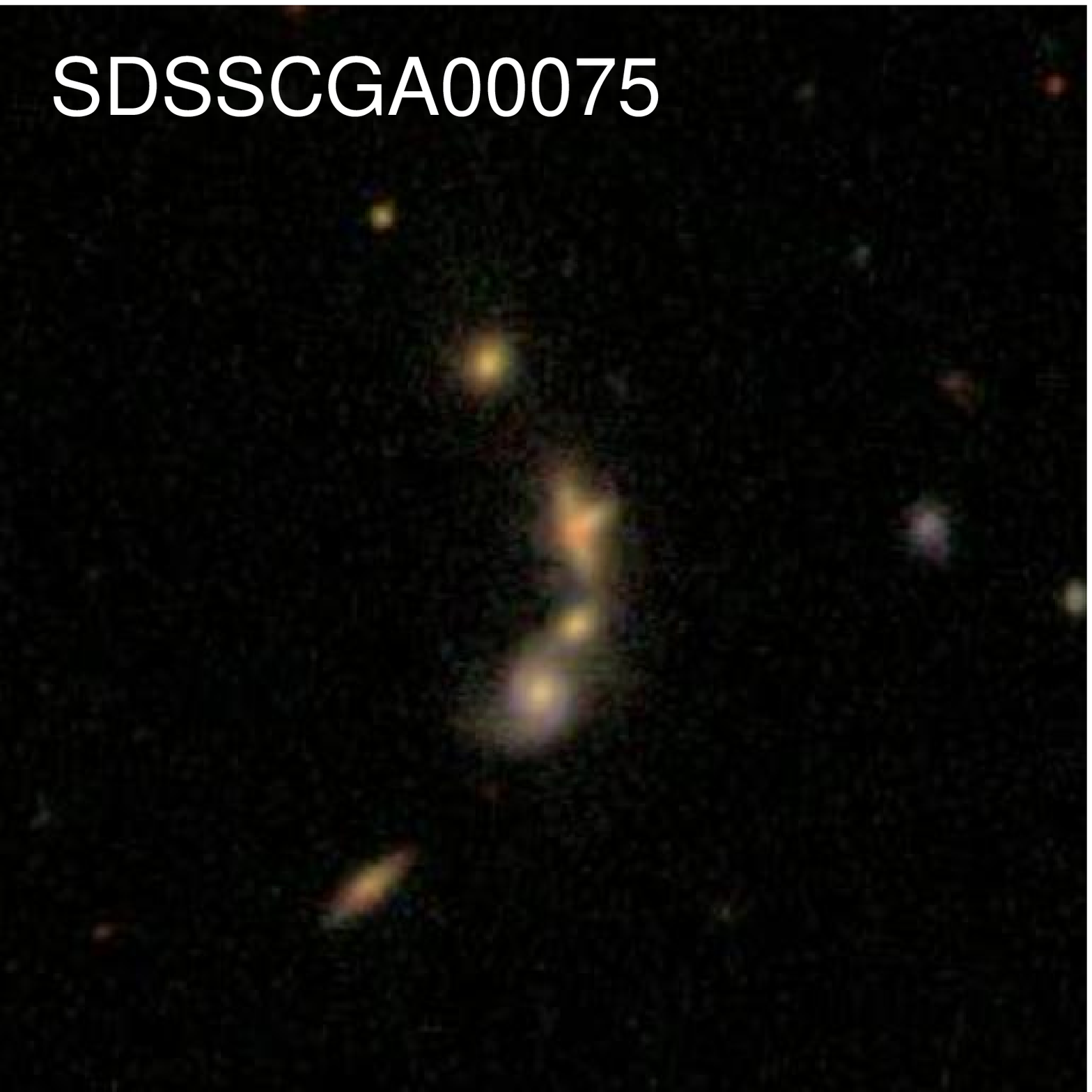}
    \includegraphics[angle=0, width=4.cm]{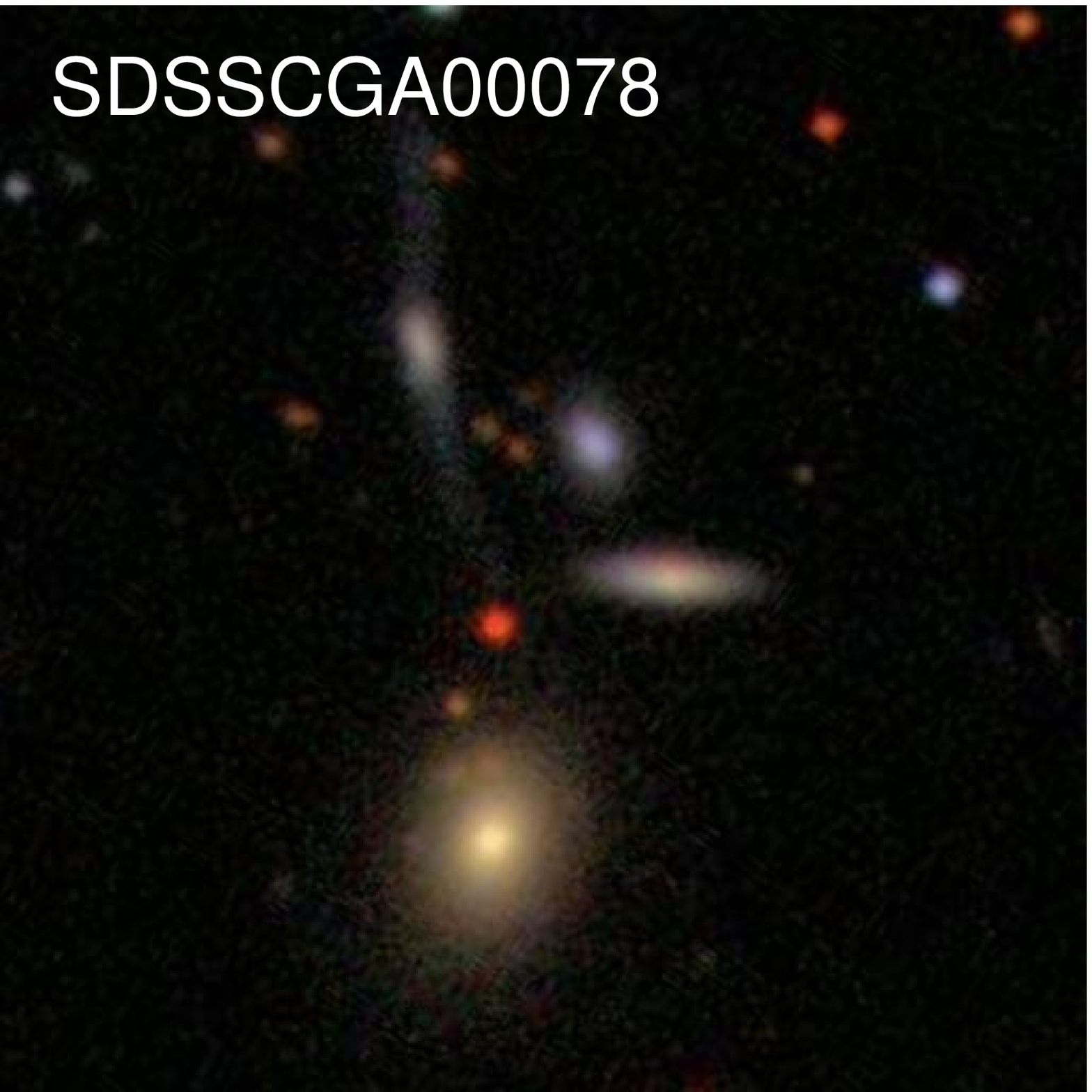}
    \includegraphics[angle=0, width=4.cm]{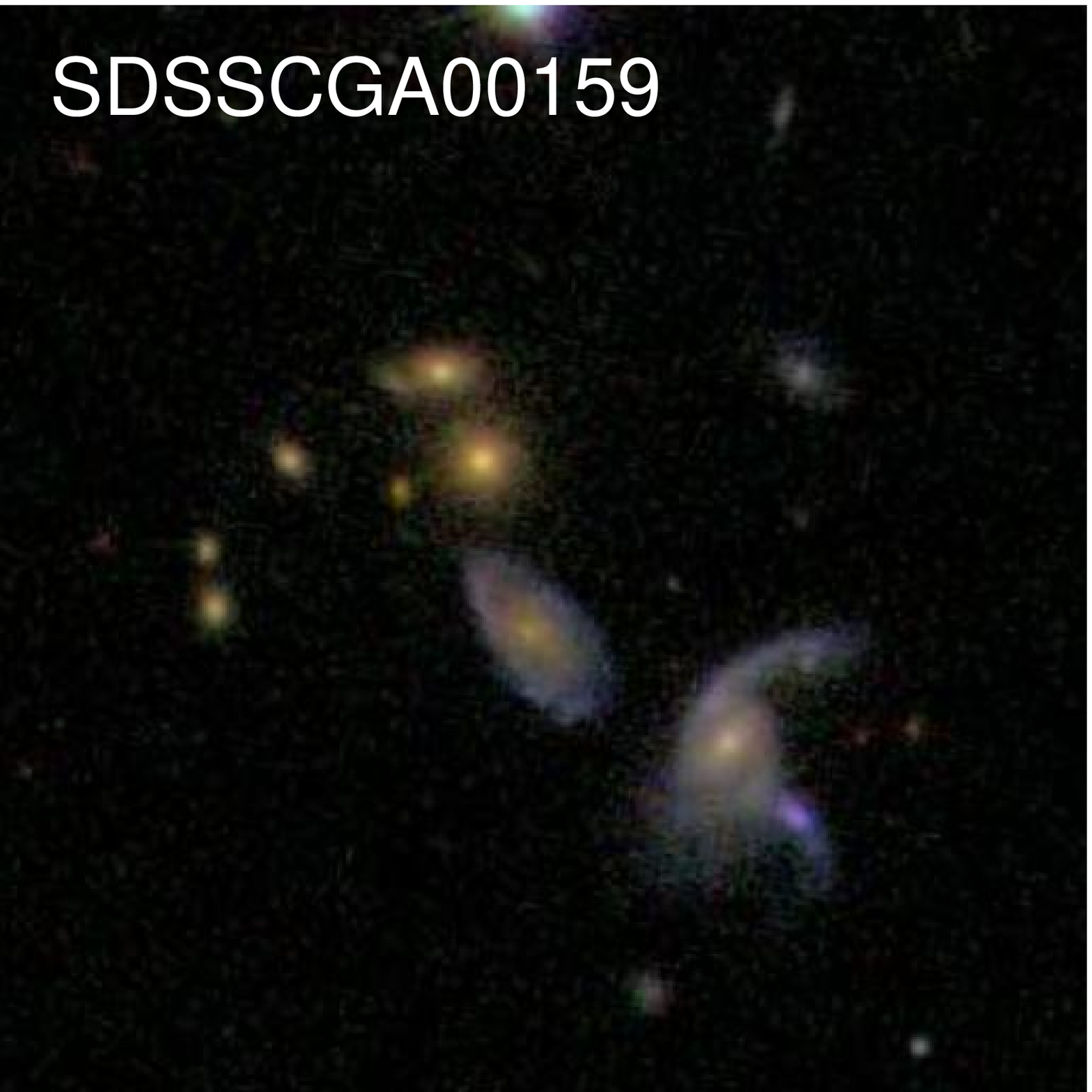}
    \includegraphics[angle=0, width=4.cm]{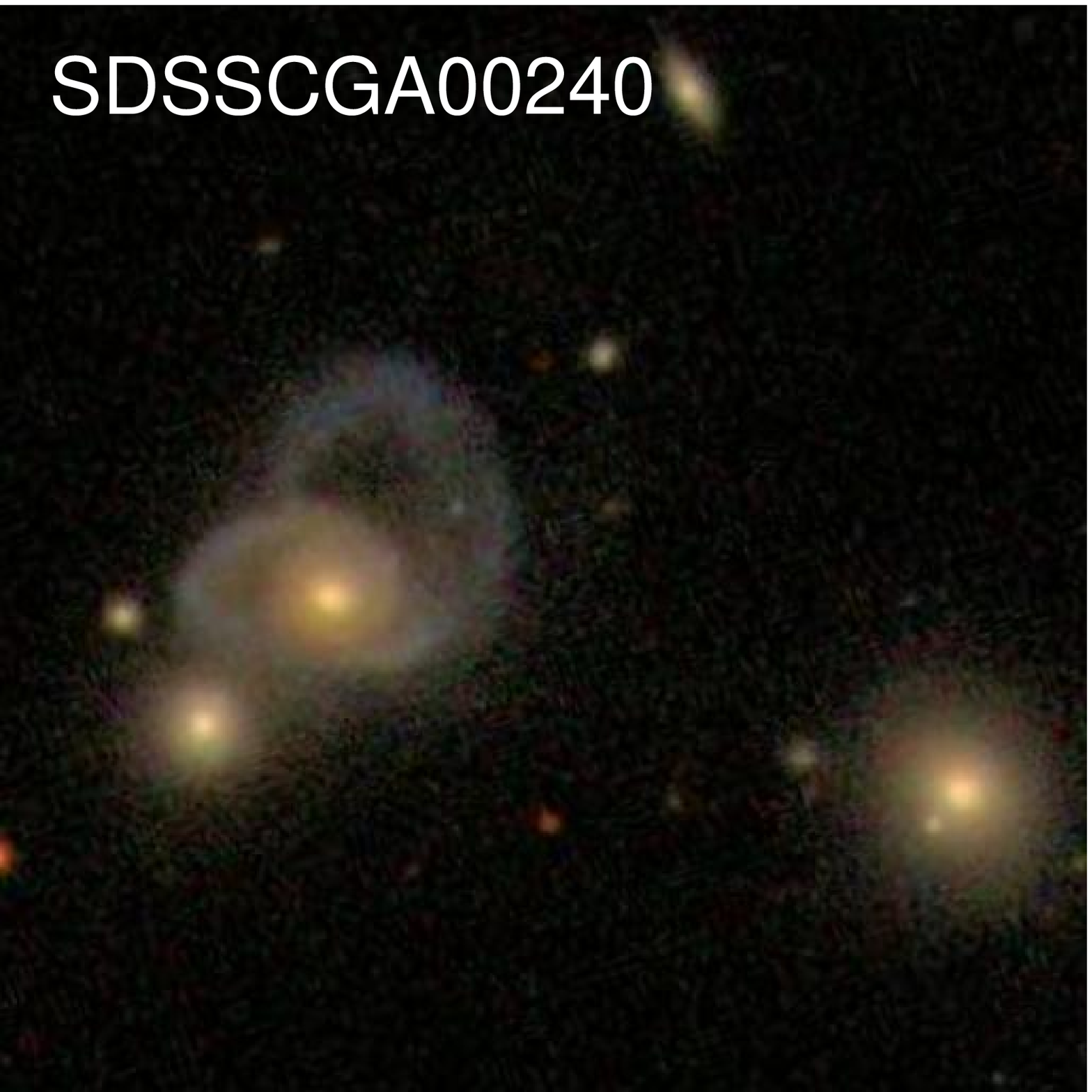}
    \includegraphics[angle=0, width=4.cm]{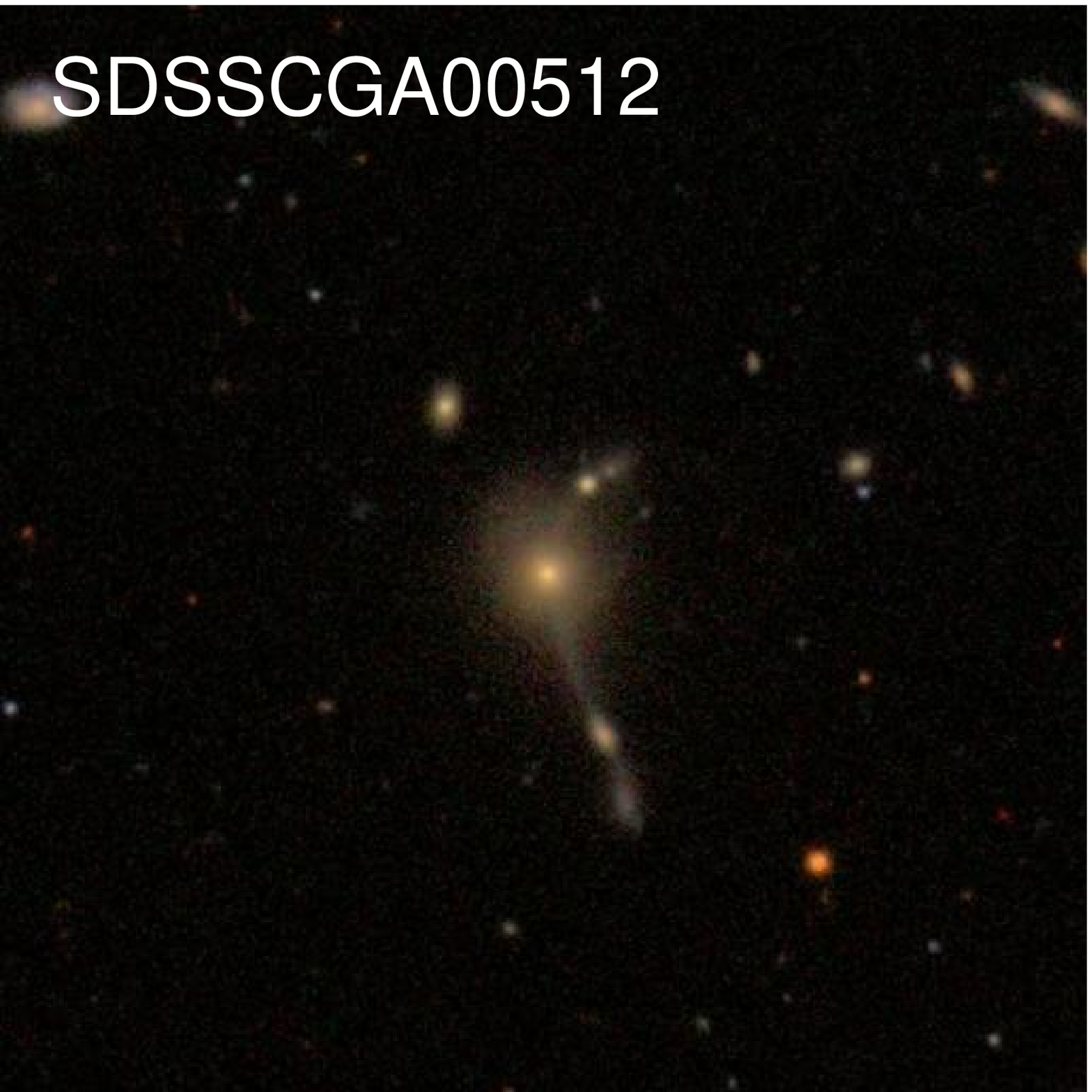}
    \includegraphics[angle=0, width=4.cm]{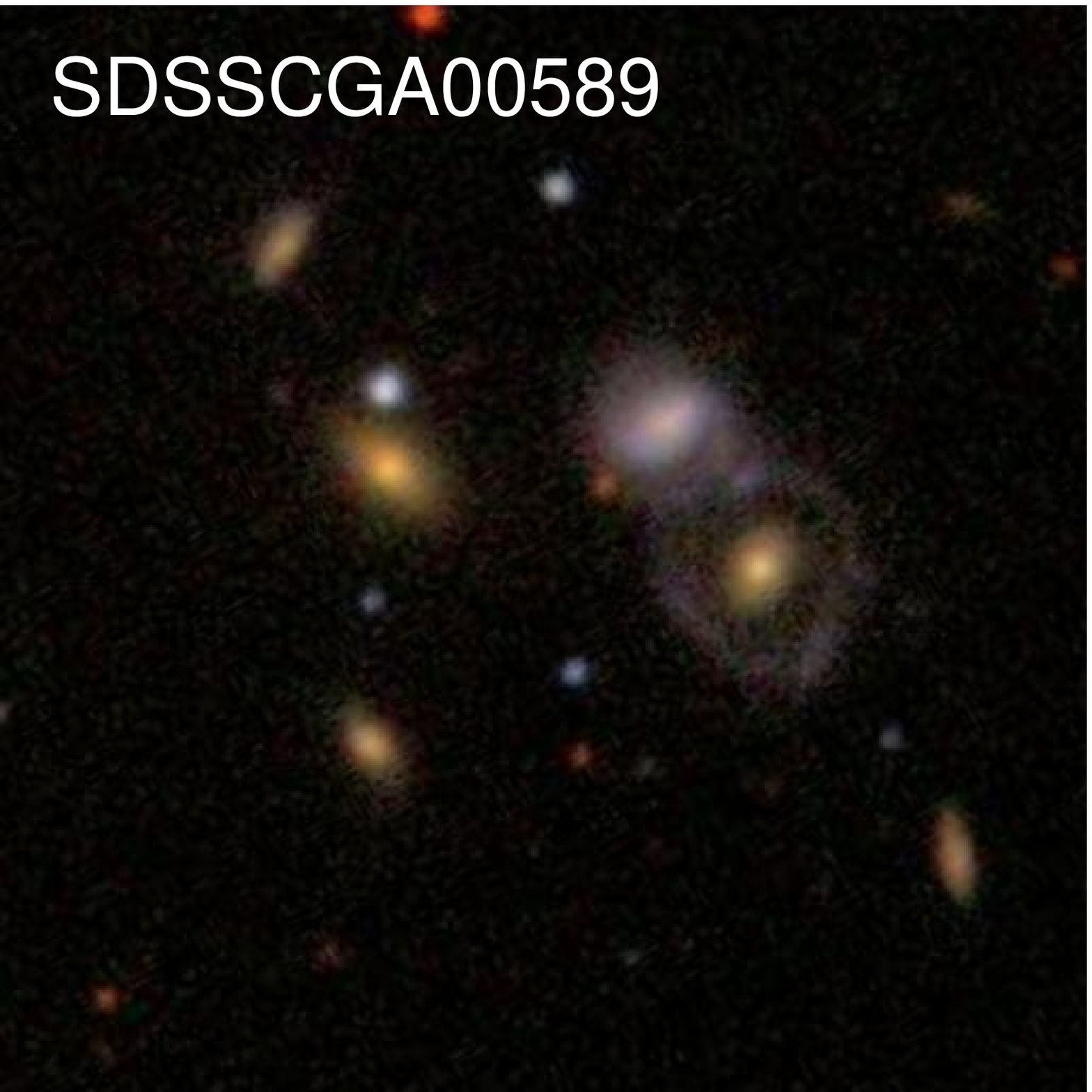}
    \includegraphics[angle=0, width=4.cm]{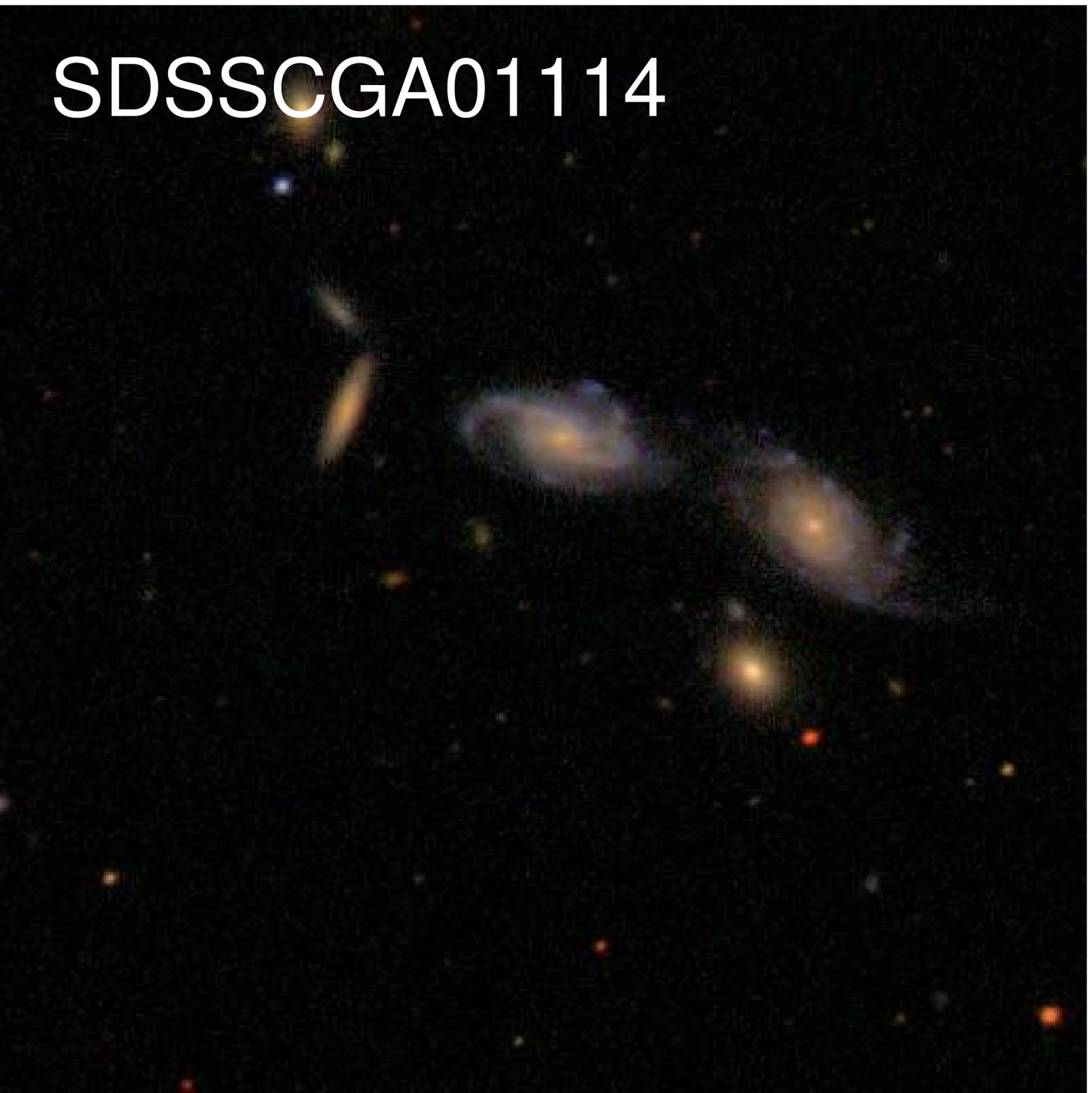}
    \includegraphics[angle=0, width=4.cm]{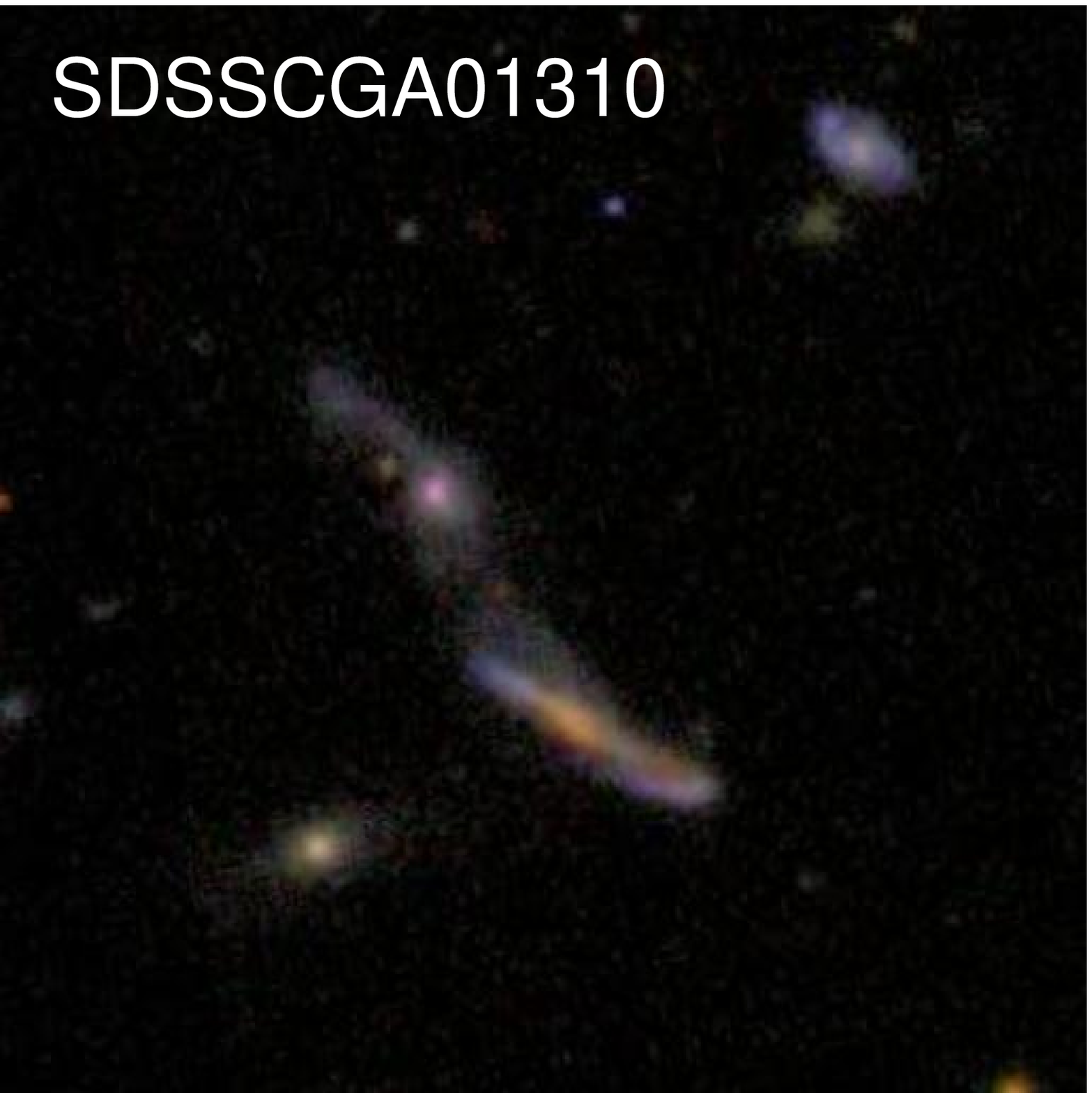}
    \includegraphics[angle=0, width=4.cm]{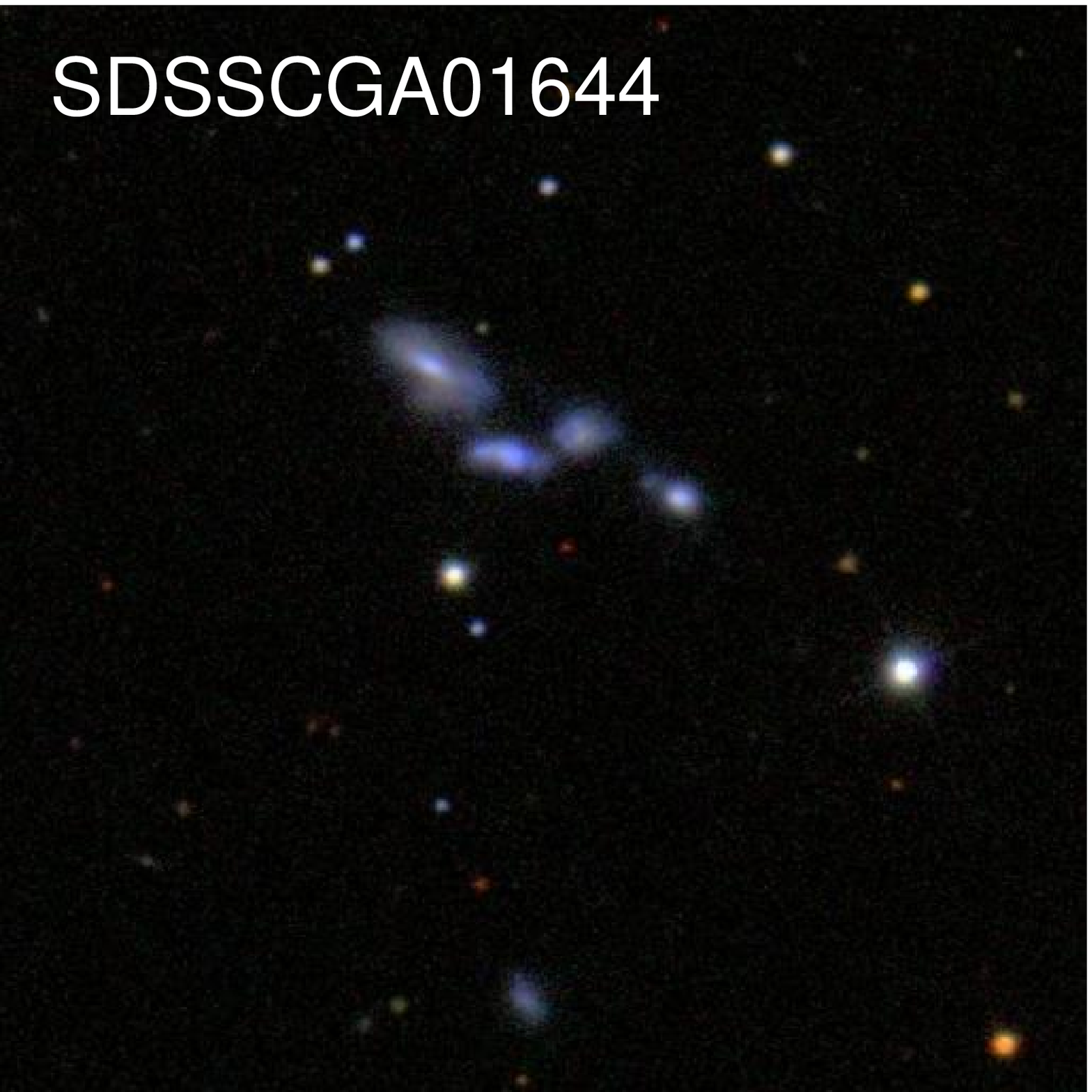}
    \includegraphics[angle=0, width=4.cm]{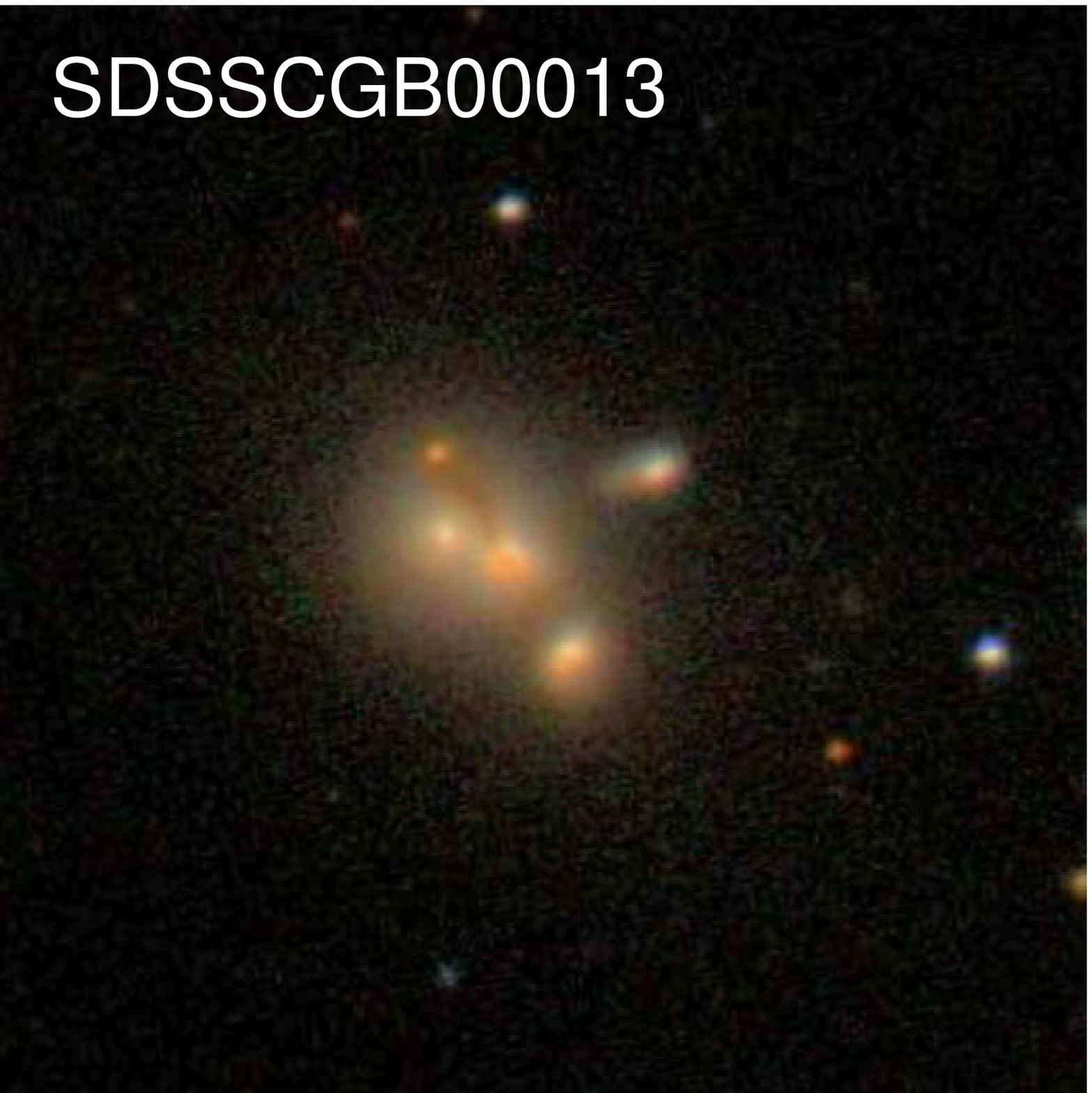}
    \includegraphics[angle=0, width=4.cm]{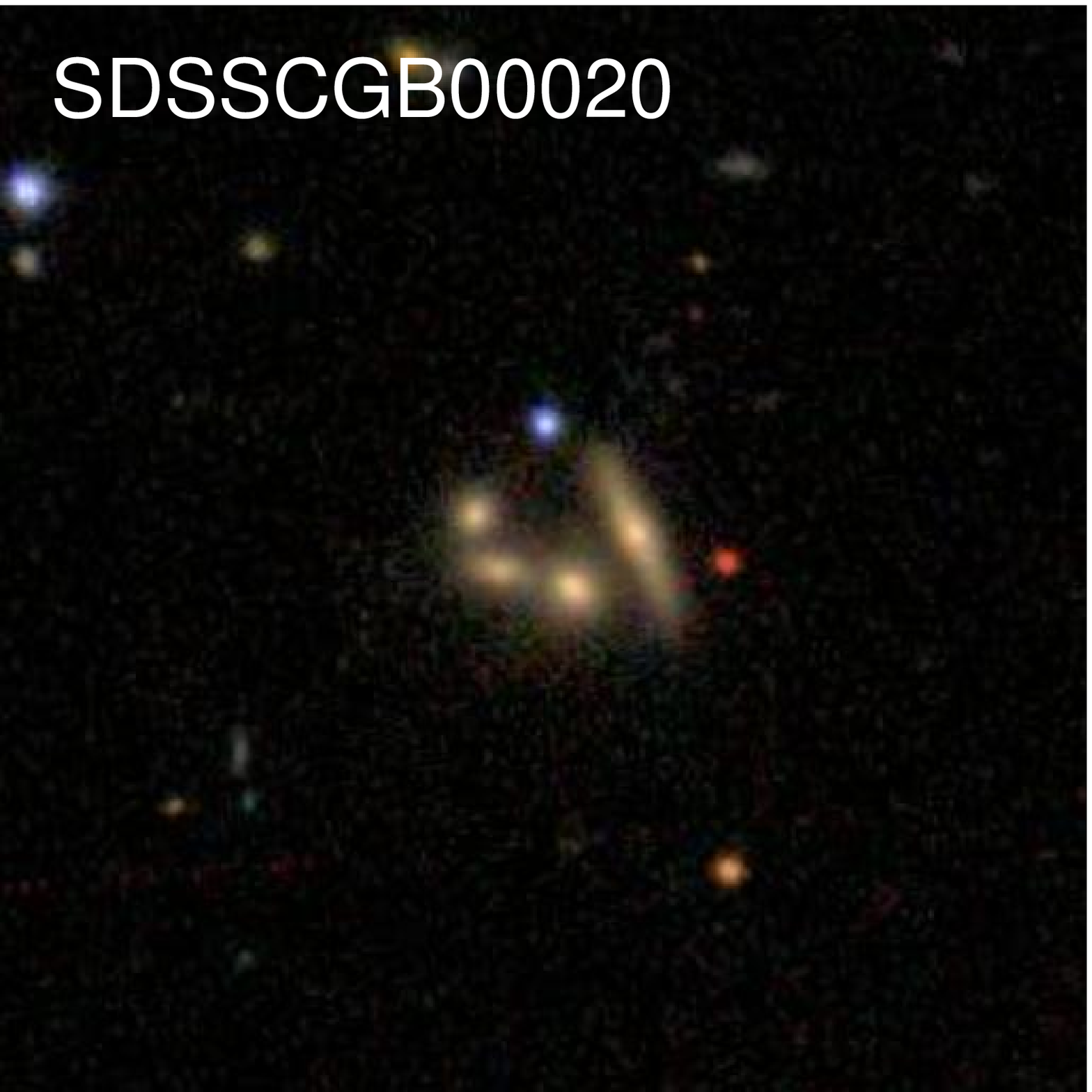}
    \includegraphics[angle=0, width=4.cm]{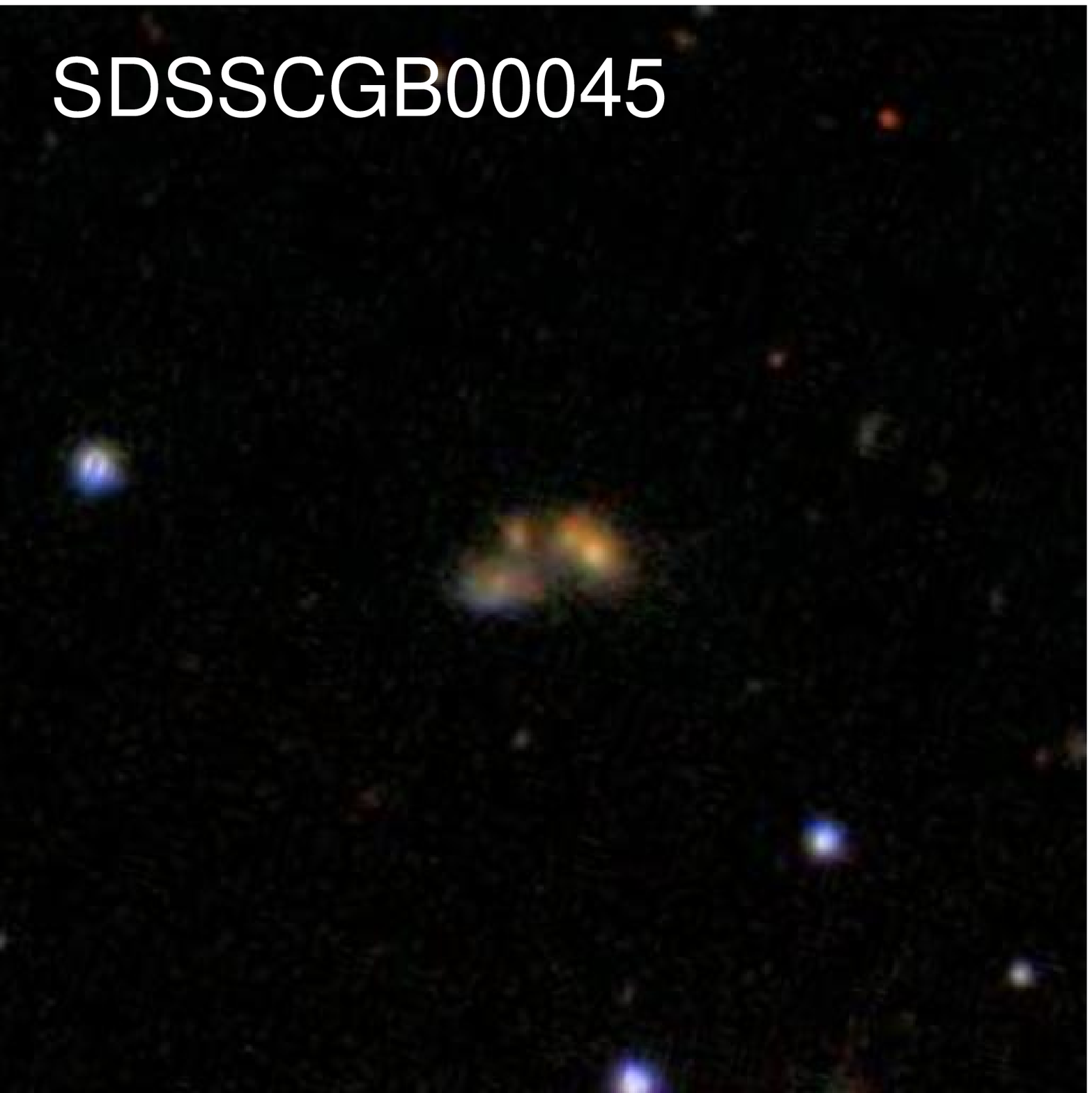}
    \includegraphics[angle=0, width=4.cm]{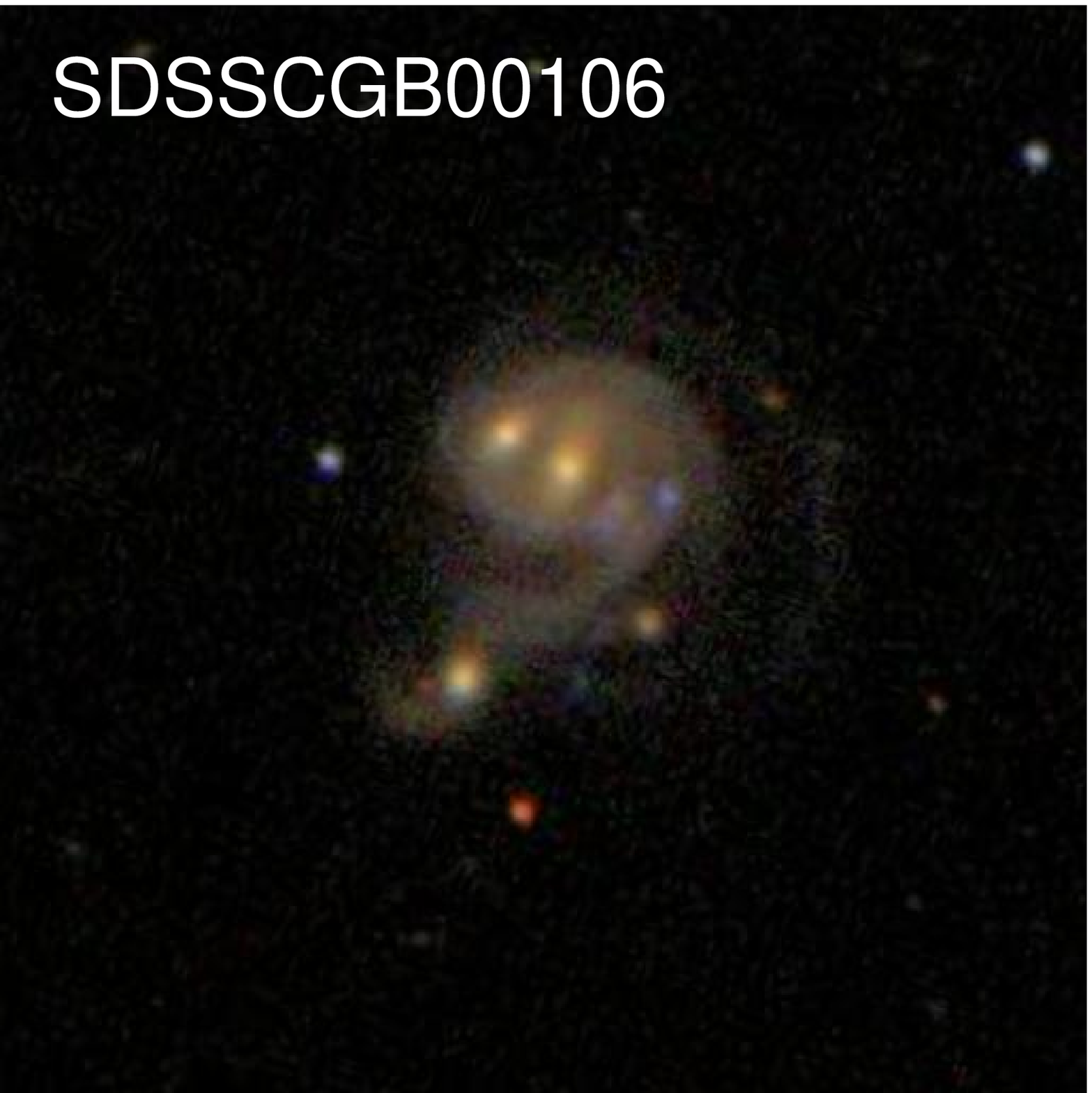}
    \includegraphics[angle=0, width=4.cm]{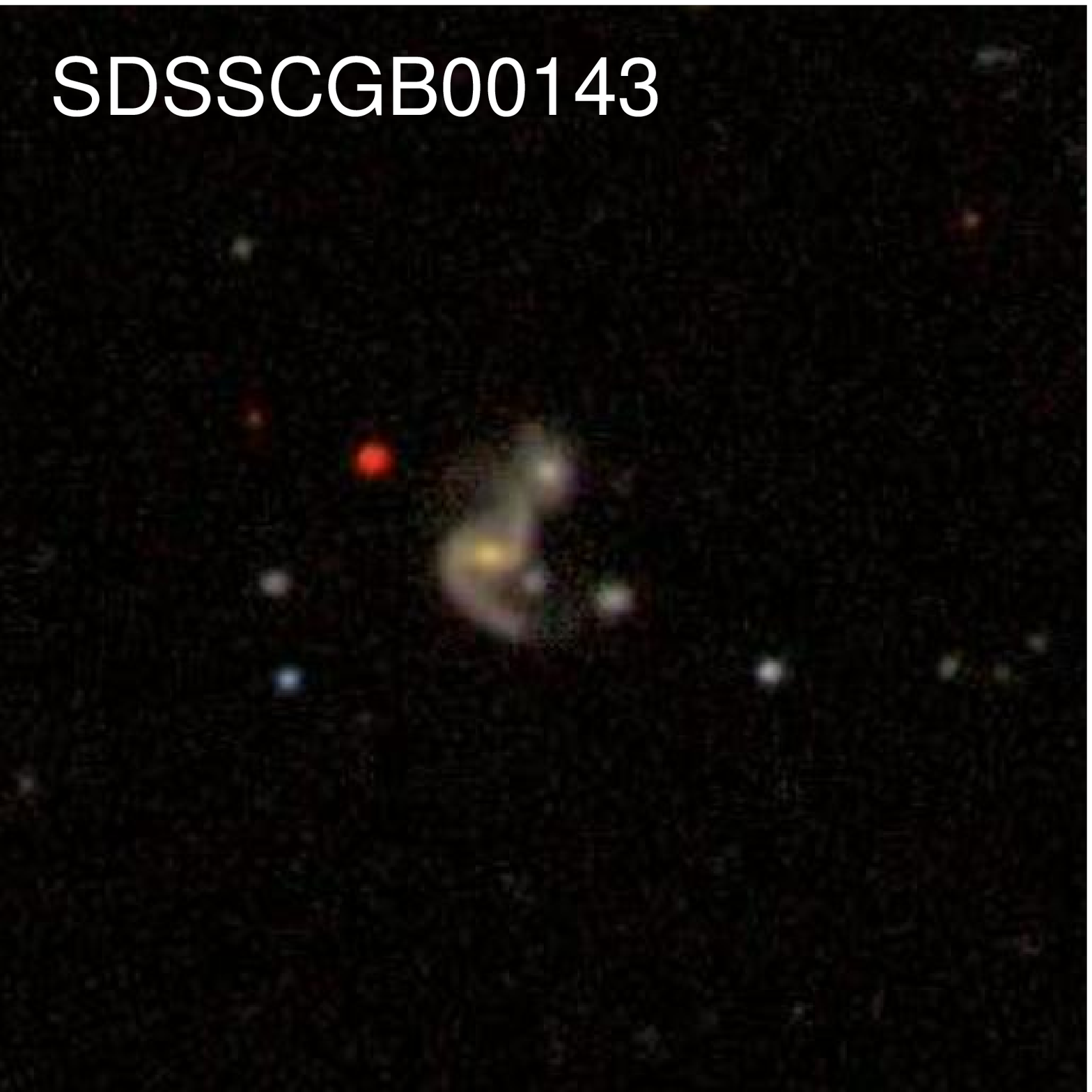}
    \includegraphics[angle=0, width=4.cm]{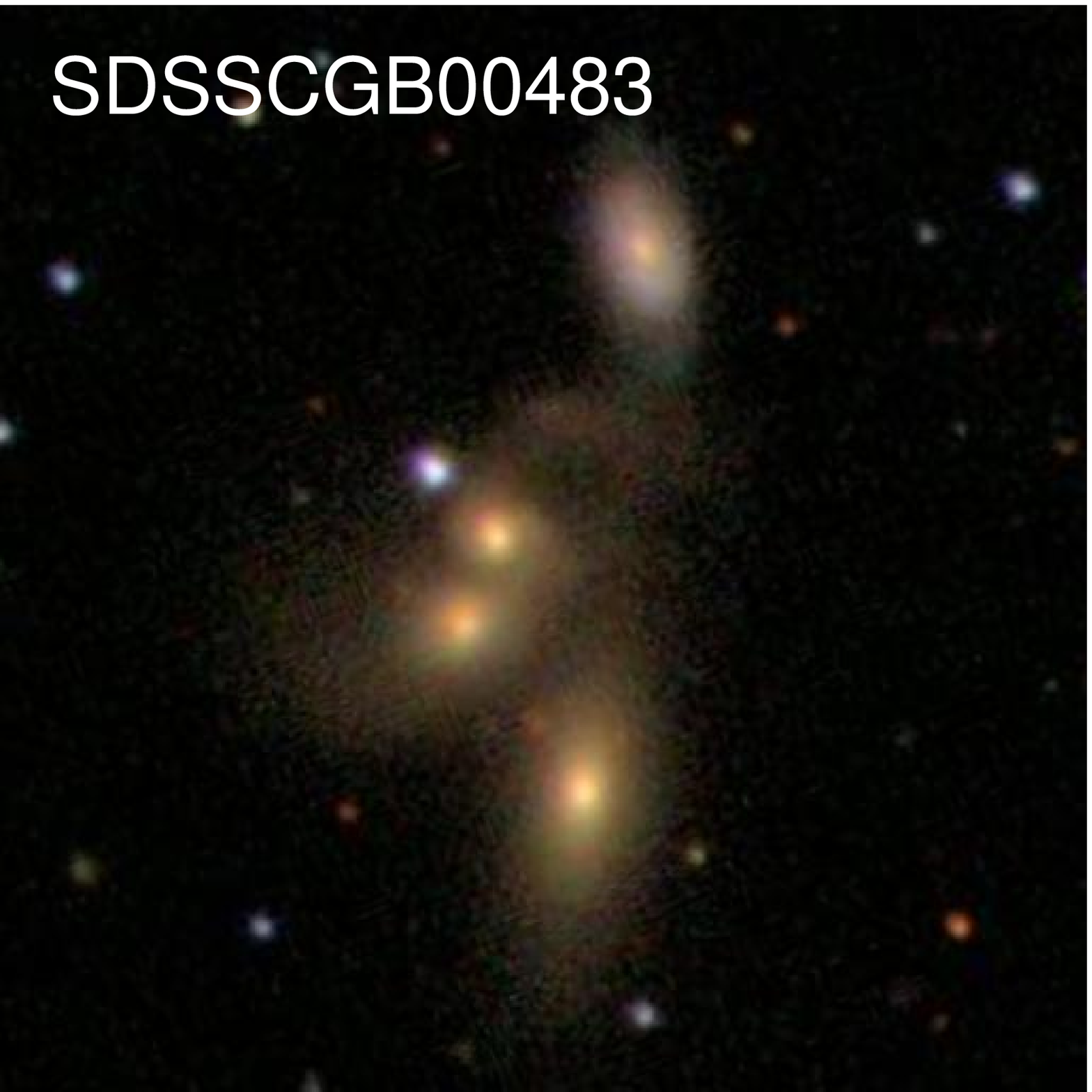}
    \includegraphics[angle=0, width=4.cm]{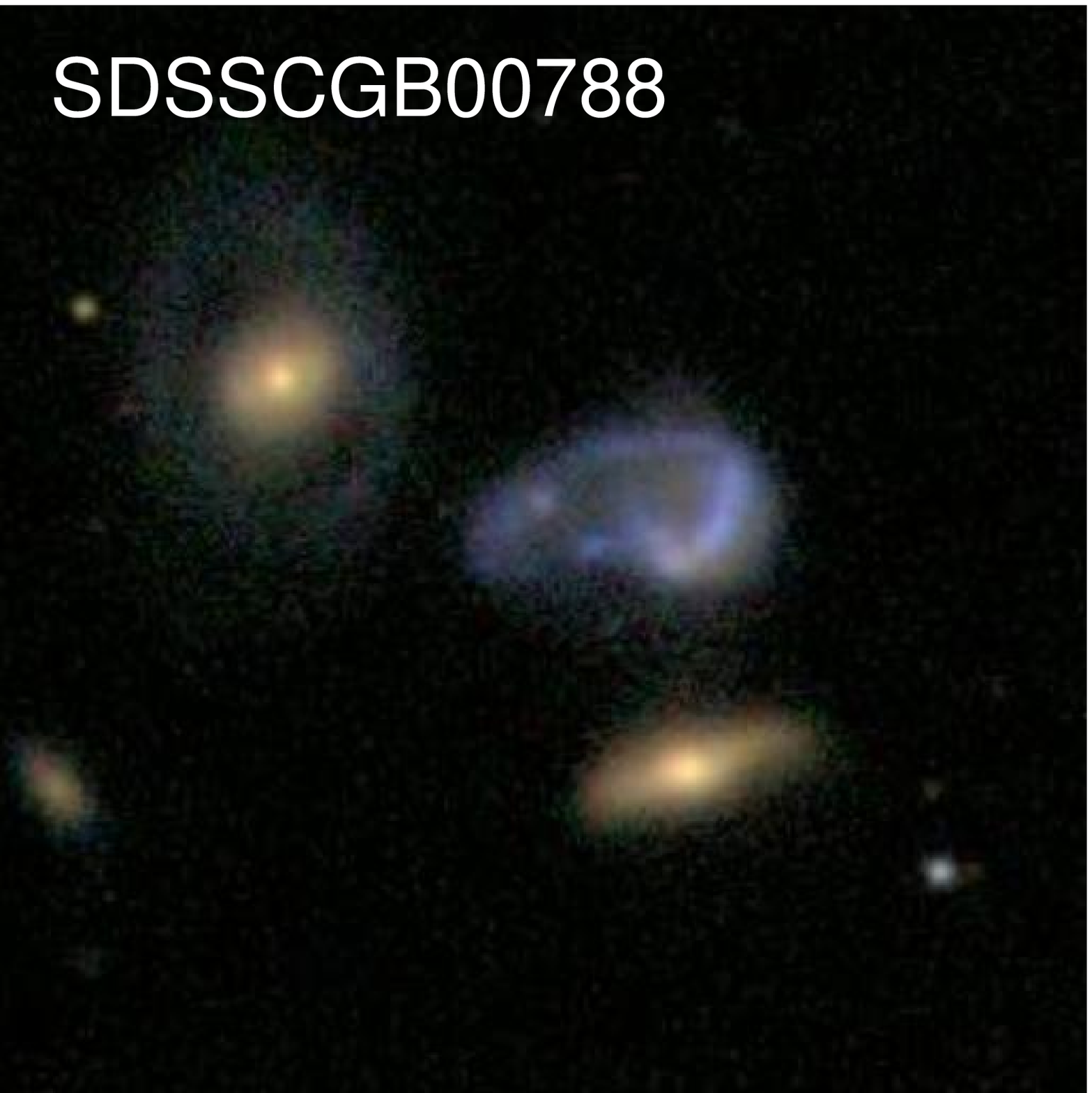}
    \includegraphics[angle=0, width=4.cm]{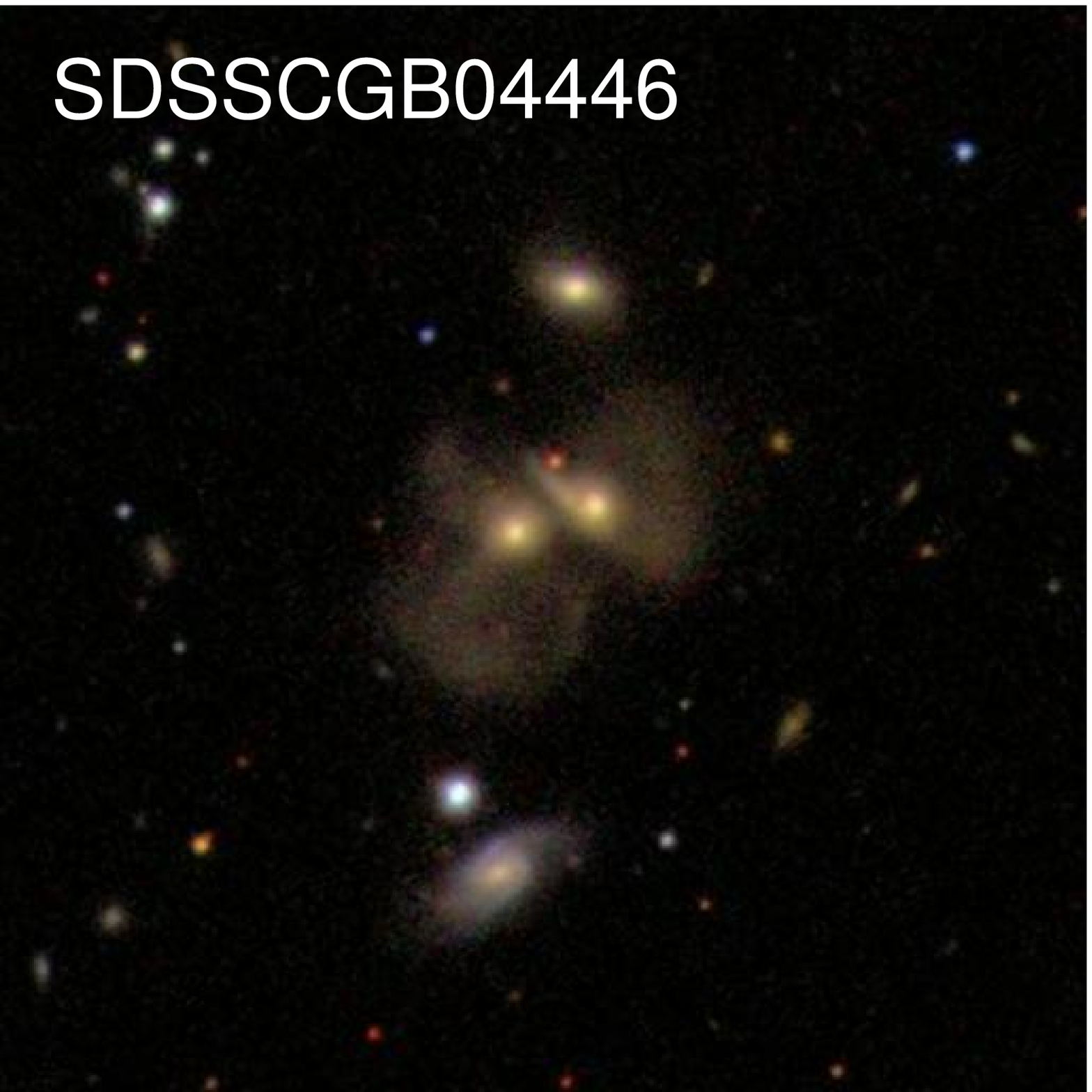}
    \caption{SDSS finding chart images of a selection of the compact
    groups in our catalogues, which show the diversity of the
    identified systems. Each image is $1.7 \times 1.7$\,arcmins,
    except SDSSCGA01114 and SDSSCGA01644, which are $3.4 \times
    3.4$\,arcmins, and SDSSCGB04446, which is $2.6 \times
    2.6$\,arcmins.}
    \label{interest}
  \end{center}
\end{figure*}

Tables~1 and 2 list the properties of compact groups found in Catalogues~A
and B, respectively. Tables~3 and 4 list the properties of the
individual member galaxies for each group in Catalogues A and B,
respectively. In each Table, only lines corresponding to the first 12
compact groups in each catalogue are listed. Full electronic versions
of these Tables can be downloaded through the MNRAS website.

In Tables~1 and 2, compact groups are listed in order of descending
surface brightness (results from Paper~I suggest that groups with a
higher surface brightness are less likely to contain interlopers). The
columns of Tables~1 and 2 are as follows:

\begin{itemize}
\item {\it Column 1} - Group ID (11 characters) Ascending numerical
order (5 digit numbers) prefixed by ``SDSSCGA/B'' depending on the
catalogue;
\item {\it Column 2, 3, 4} - Right Ascension (J2000).  Refers to
geometrical centre of group (hexadecimal format);
\item {\it Column 5, 6, 7} - Declination (J2000). Refers to
geometrical centre of group (hexadecimal format);
\item {\it Column 8} - Number of members in the group, $n_{mem}$;
\item {\it Column 9} - Group surface brightness in the $r-$band, defined by the Hickson
criteria (mags\,arcsec$^{-2}$);
\item {\it Column 10} - Group radius, defined by the Hickson criteria (arcmins);
\item {\it Column 11} - Distance to the next nearest non-member galaxy
in the galaxy catalogue, in the appropriate magnitude range, measured
from the group centre in units of the group radius;
\item {\it Column 12} - $r-$band apparent Petrosian magnitude of
brightest group galaxy, corrected for foreground Galactic extinction;
\item {\it Column 13} - Number of galaxies in the group with a
spectroscopic redshift;
\item {\it Column 14} - Redshift of group (if available). If only one
galaxy has a spectroscopic redshift, then that is adopted as the group
redshift. If multiple members of the group have redshifts, the mean
redshift is listed. Only considers groups with $\Delta v \le
1000$\,km\,s$^{-1}$ (see Section~3.1.2 for definition). Only considers
galaxies with $z_{conf} \ge 0.7$ (see below for definition);
\item {\it Column 15 (online Table 2 only)}. Flag indicating if the
group contains members which were originally identified as belonging
to groups in Catalogue A which were subsequently identified as
containing contamination ($flag = 1$ if true, $0$ if false).
\end {itemize}

The columns of Tables 3 and 4 are as follows:

\begin{itemize}
\item {\it Column 1} - Galaxy ID in the group catalogue. Format is
``$group.gal$'', where ``$group$'' corresponds to Group ID in Tables 1 and
2 (group order is the same as Tables~1 and 2) and ``$gal$'' is a number
from $i=1...n_{mem}$. Galaxies in each group are listed in order of
increasing $r-$band apparent magnitude;
\item {\it Column 2} - SDSS ObjID. Unique identifier of object in the
SDSS database which can be used to retrieve all information stored in
the SDSS on the galaxy;
\item {\it Column 3, 4, 5} - Right Ascension (J2000). Galaxy position
(hexadecimal format);
\item {\it Column 6, 7, 8} - Declination (J2000). Galaxy position
(hexadecimal format);
\item {\it Column 9} - $r-$band apparent Petrosian magnitude of
galaxy, corrected for foreground Galactic extinction;
\item {\it Column 10} - $(g - r)$ colour of galaxy, corrected for
foreground Galactic extinction. No $k-$correction has been applied;
\item {\it Column 11} - SDSS SpecObjID (if available). Unique
identifier of the SDSS spectrum for this galaxy;
\item {\it Column 12} - SDSS $z_{conf}$ (if applicable). If a spectrum
exists for the galaxy, this parameter indicates how reliable the
derived redshift is. 0 (poor) $\le z_{conf} \le$ 1 (excellent);
\item {\it Column 13} - Redshift (if available). 
\end{itemize}

Figure~\ref{top12} shows images of the twelve compact groups found in
Catalogue A with the highest central surface brightnesses. It is
striking that these twelve groups are generally dominated by red
galaxies. In Paper~II, we found that compact group galaxies identified
in the mock catalogue were predominantly ``red and dead''
galaxies. While these brightest groups appear to be consistent with
this prediction, it is important to emphasise that bluer, later-type
galaxies are present in significant numbers in these
catalogues. Figure~\ref{interest} shows a selection of compact groups
which were selected for their visual impact and to show the range and
variety of galaxy types that are present in these catalogues.

\subsection{Comparison with Lee et al. (2004)}

\cite{lee2004} searched for compact groups of galaxies in the Early
Data Release (EDR) of the SDSS (\citealt{stoughton2002}) using all
galaxies with $14.0 \le r \le 21.0$. They used the Hickson criteria,
modified so that only groups with $\mu \le 24$\,mags\,arcsec$^{-2}$
were considered. As an independent check on the success of our search
algorithm and the reproducibility of our results, we now determine if
Catalogue~B contains those groups identified by
\cite{lee2004}. However, not all of the Lee et al. groups will be
identified in the SDSS~DR6 due to differences in the underlying
catalogues. In particular,

\begin{itemize}
\item the photometric calibration for the SDSS $ugriz$ system has been
updated between the SDSS EDR and SDSS DR6 (see
\citealt{adelmanmccarthy2008} and references therein).
\item \cite{lee2004} do not use the SDSS star-galaxy classification
algorithm, and instead use a different classification scheme from
\cite{scranton2002}, which is itself based upon an earlier processing
of the SDSS EDR than the published version.
\end{itemize}

To take the above considerations taken into account, we search the
SDSS~DR6 for those objects with the best positional coincidence to the
individual member galaxies of the \cite{lee2004} groups, as given in
their Table~2. This allows us to derive the photometric properties of
the \cite{lee2004} groups using the SDSS~DR6 catalogue. We find that

\begin{enumerate}
\item Of the 175 groups (more than 3 members) identified in
\cite{lee2004}, good positional matches to objects in the SDSS~DR6
(better than 0.5 arcsecs) are found for all members of 164 groups. We
therefore do not consider the 11 remaining groups further, since it is
not possible for us to have found these groups using the SDSS~DR6;

\item Of the remaining 164 groups, 13 groups have at least one member
which is best matched to an object flagged as stellar in DR6.  Thus we
do not examine these 13 groups further;

\item Of the remaining 151 groups, 2 groups have members which are
flagged as ``DEBLENDED\_AS\_PSF''. Thus these 2 groups will not be
identified in our catalogue (Section~2.2) and we do not consider them
further;

\item Of the remaining 149 groups, 1 group has a member with $r =
14.41$, brighter than our bright-end limit. Thus this group will not
be identified in our catalogue and we do not consider it further;

\item Of the remaining 148 groups, 11 have a magnitude range in the
$r-$band ($\Delta\,m$) of greater than 3 magnitudes in the
SDSS~DR6. Thus these groups will not be identified in our catalogue
using the Hickson criteria, and we do not consider them further;

\item Of the remaining 137 groups, 58 do not match the isolation
criteria for selection as a compact group using the SDSS DR6. Thus
these groups will not be identified in our catalogue and we do not
consider them further.
\end{enumerate}

We conclude that, of the original 175 groups identified by
\cite{lee2004}, 96 are not identified in SDSS~DR6. This is due to a
different photometric classification of member objects (and nearby
neighbours), which results in a failure of the group to meet the
Hickson criteria in our SDSS~DR6 galaxy catalogue.

Finally, we cross-correlate the remaining 79 compact groups from
\cite{lee2004} with the groups identified in Catalogue~B (Tables 2 and
4). We identify all 79 groups, demonstrating that our procedure for
identifying compact groups in SDSS~DR6 is consistent with the earlier
study using the SDSS~EDR.

\section{Basic properties of the SDSS DR6 Compact Group Catalogue}

\begin{table*}
\begin{tabular*}{0.65\textwidth}{c|rr|rr|rr|rr}
& \multicolumn{4}{|c}{Catalogue A} & \multicolumn{4}{|c}{Catalogue B}\\
& \multicolumn{2}{|c|}{$\Delta v < \infty$}& \multicolumn{2}{|c}{$\Delta v \le 1000$ km\,s$^{-1}$}& \multicolumn{2}{|c|}{$\Delta v < \infty$}& \multicolumn{2}{|c}{$\Delta v \le 1000$ km\,s$^{-1}$}\\
$n_z$  & $n_{grps}$ & $n_{gals}$& $n_{grps}$ & $n_{gals}$ & $n_{grps}$ & $n_{gals}$& $n_{grps}$ & $n_{gals}$\\
\hline
&&&&&&&&\\
$0$       & 500  & 2115 & 500  & 2115  & 60516 & 252805 & 60516 & 252805\\
$> 0$     & 1797 & 4108 & 1008 & 1915  & 14275 & 16405  & 13414 & 14414 \\
&&&&&&&&\\
1         & 373  & 373   & 373  & 373  & 12537 & 12537  & 12537 & 12537 \\
2         & 776  & 1552  & 422  & 844  & 1419  & 2838   & 772   & 1544  \\
3         & 439  & 1317  & 159  & 477  & 251   & 753    & 87    & 261  \\
4         & 183  & 732   & 49   & 196  & 63    & 252    & 18    & 72   \\
5         & 23   & 115   & 5    & 25   & 5     & 25     & 0     & 0    \\
6         & 2    & 12    & 0    & 0    & 0     & 0      & 0     & 0    \\
7         & 1    & 7     & 0    & 0    & 0     & 0      & 0     & 0    \\
&&&&&&&&\\
$n_{mem}$ & 148  & 613   & 34   & 139  & 54    & 220    & 13    & 52 \\
\end{tabular*}
\caption{The number of groups, $n_{grps}$, with $n_z$ redshifts per
group ($z_{conf} \ge 0.7$; also listed is the number of individual
galaxies this represents in each case, $n_{gal}$). The first two
columns for each catalogue gives these numbers irrespective of whether
multiple redshifts per group agree with each other or not ($\Delta v <
\infty$); the last two columns require $\Delta v \le 1000$\kms.}
\end{table*}

\subsection{Number of galaxies per group}

\begin{figure*}
  \begin{center}
    \includegraphics[angle=270, width=14.cm]{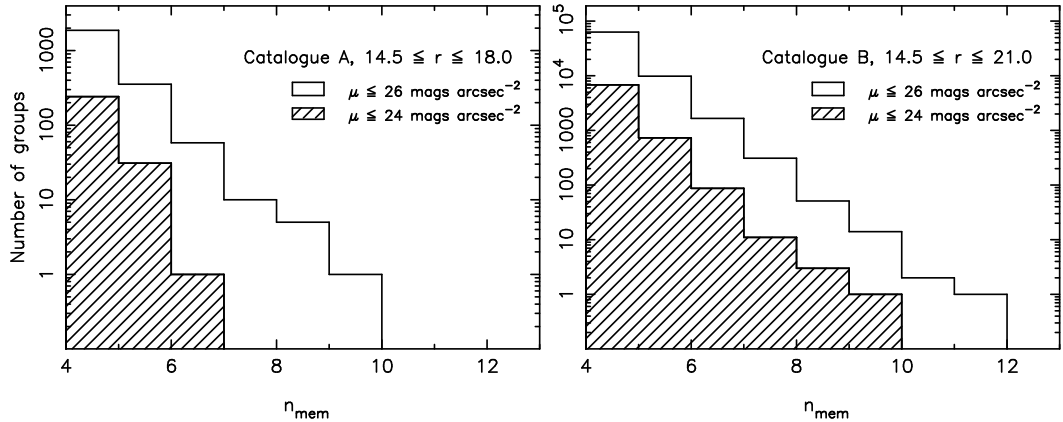}
    \caption{Distribution of number of galaxies in each group
    identified in Catalogues~A and B (left and right panels,
    respectively) for two different cuts in group surface brightness,
    $\mu \le 26$\,mags\,arcsec$^{-2}$ (blank histogram) and $\mu \le
    24$\,mags\,arcsec$^{-2}$ (hatched histogram).}
    \label{nmem}
  \end{center}
\end{figure*}

Figure~\ref{nmem} shows the distribution of the number of galaxies per
group for Catalogue~A (left panel) and Catalogue~B (right
panel). Blank histograms show the complete samples, whereas the
hatched histograms show the distribution for those groups where $\mu
\le 24$\,mags\,arcsec$^{-2}$. The fraction of groups which contain
galaxies at discordant redshifts (that is, chance line-of-sight
alignments rather than physically dense groups) is expected to be
greatly reduced for this latter sample. Clearly, the overwhelming
majority of groups from both Catalogues~A and B have 6 members or less;
very few groups in the brighter subsamples have more than 5 members,
strongly suggesting that groups more numerous than this contain
interlopers, at least in part. If groups with more than 6 genuine
members exist, they are intrinsically very rare.

\subsection{Group surface brightness}

\begin{figure*}
  \begin{center}
    \includegraphics[angle=270, width=14.cm]{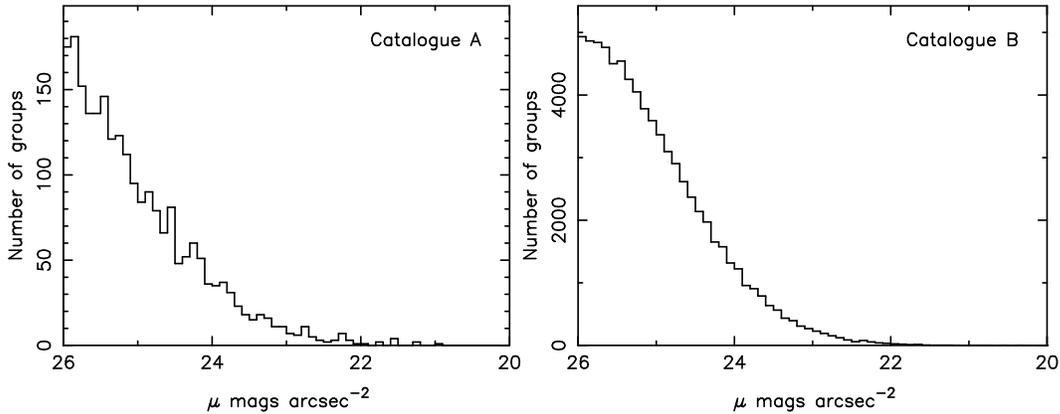}
    \caption{The distribution of group surface brightness in the
    $r-$band for all groups identified in Catalogue~A (left panel) and
    Catalogue B (right panel).}
    \label{sbdens}
  \end{center}
\end{figure*}

The left panel of Figure~\ref{sbdens} shows the $r-$band group surface
brightness distribution for all groups identified in Catalogue~A, and
the right panel shows the same for all groups identified in
Catalogue~B. More groups are found at lower surface brightness;
however, as shown in Paper~I, brighter groups are more likely to be
genuine and not contain any interlopers (see Section 3.3.2 for a more
detailed investigation of this point). 

\subsection{Spectroscopic Information}

\subsubsection{Spectral Completeness}

The compact groups and their member galaxies presented in Tables 1 --
4 were selected based on their photometric properties, in particular
their projected positions and $r-$band apparent magnitudes. However,
the SDSS DR6 has spectroscopic information available for 1.27 million
objects, of which 679733 objects are galaxies which satisfy our
criteria for inclusion in Catalogue~B. Thus, many of the galaxies we
identify as being a member of a compact group will have spectroscopic
data available.

For the groups identified in Catalogue A, 4131 out of 9713 member
galaxies (43\,\%) have spectroscopic information available, 4108 of
which yield a reliable redshift ($z_{conf} \ge 0.7$). For groups
identified in Catalogue B, 16566 out of 313508 member galaxies (5\,\%)
have spectroscopic information available, 16405 of which yield a
reliable redshift. The fraction of galaxies with spectroscopic data
available in Catalogue~B is significantly less than in Catalogue~A,
since Catalogue~A contains a higher fraction of brighter galaxies
(which are preferentially selected as spectroscopic targets).

\subsubsection{Estimate of interloper fractions}

\begin{figure}
  \begin{center}
    \includegraphics[angle=270, width=8.cm]{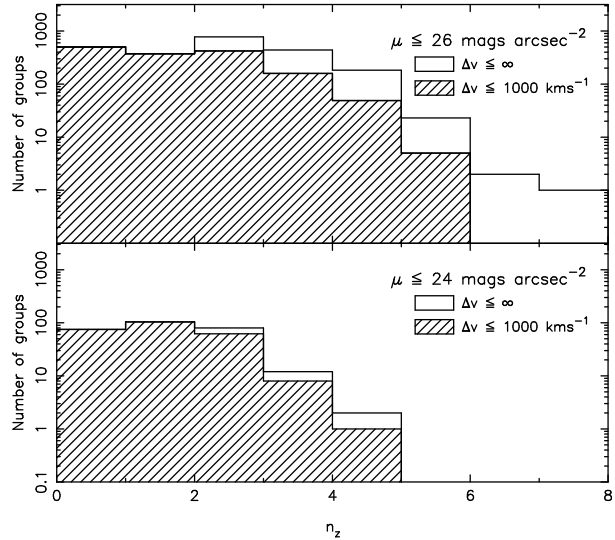}
    \caption{Top panel: The blank histogram shows the distribution of
    the number of galaxies with reliable redshifts per group
    ($z_{conf} \ge 0.7$), for compact groups identified in
    Catalogue~A. The hatched histogram shows the same distribution but
    with the additional constraint that $\Delta\,v \le 1000$\,\kms
    (ie., we ignore groups with discordant velocity data). Bottom
    panel: Same as top panel, except only using groups where $\mu \le
    24$\,mags\,arcsec$^{-2}$.}
    \label{nz}
  \end{center}
\end{figure}

\begin{figure}
  \begin{center}
    \includegraphics[angle=270, width=8.cm]{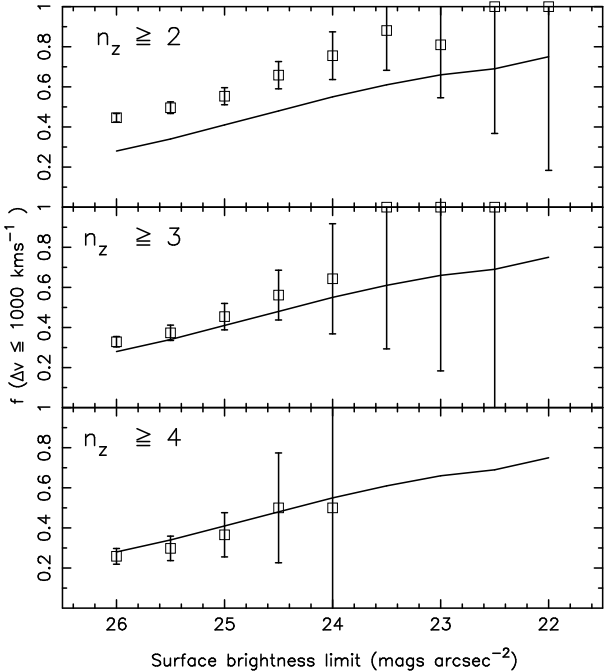}
    \caption{For groups identified in Catalogue A with multiple
    redshifts, points show how the fraction of groups with concordant
    redshifts ($\Delta v \le 1000$\,\kms) varies as a function of the
    surface brightness detection limit. Vertical error bars are
    Poisson uncertainties. The solid lines show the {\it predicted}
    fraction of groups with concordant redshifts (ie. no interlopers)
    from our mock catalogue study in Paper~I. The fraction of groups
    with concordant redshifts decreases for groups with higher $n_z$
    since it is more likely that an interloping galaxy will be
    targeted. For groups with $n_z \ge 3, 4$, the observed and
    predicted number of groups with interlopers agree remarkably
    well.}
    \label{sb}
  \end{center}
\end{figure}

Table~5 lists the number of groups, $n_{grps}$, with $n_z$ redshifts
per group, as well as the number of individual galaxies this
represents in each case, $n_{gals}$. We only consider galaxies where
$z_{conf} \ge 0.7$. This information is listed for two different cuts
in $\Delta v$, where we define

\begin{eqnarray}
\Delta v & = & 0~,~~~n_z \le 1 \nonumber\\
\Delta v & = & max \left[ \left| c z_i - c z_j \right| / (1 + z_{ij}) \right]~,~~~n_z \ge 2.
\end{eqnarray}

\noindent $z_{ij} = (z_i+z_j)/2$,~$i,j = 1...n_z$,~$i \ne j$, and so
$\Delta v$ is a measure of the maximum line-of-sight velocity
difference between group members for groups where $n_z \ge 2$. The cut
$\Delta v \le 1000$\,\kms selects only those groups where multiple
redshifts (if available) are concordant. This removes groups which
contain interloping galaxies which happen to lie along the same
line-of-sight. As was demonstrated in Paper~II (and assumed in most
previous studies of compact groups; e.g. \citealt{hickson1992}), few,
if any, genuine compact groups are expected to possess members with
line-of-sight velocity differences in excess of $1000$\kms.

The blank histogram in the top panel of Figure~\ref{nz} shows the
distribution of $n_z$ for compact groups identified in
Catalogue~A. The hatched histogram shows the same distribution but
with the additional constraint that $\Delta v \le 1000$\,\kms. The
bottom panel of Figure~\ref{nz} shows the same distributions but only
considers those groups with $\mu \le 24$\,mags\,arcsec$^{-2}$. Results
from Paper~I suggests that the fraction of identified groups which
contain interlopers decreases from 71\% for $\mu \le
26$\,mags\,arcsec$^{-2}$ to 44\% for $\mu \le
24$\,mags\,arcsec$^{-2}$.

Table~5 and Figure~\ref{nz} demonstrate that the majority of groups
which have redshift measurements for multiple members have discordant
redshifts ($\sim 55\%$ in Catalogue~A), implying that {\it at least}
55\,\% of identified groups in Catalogue~A contain interlopers.  For
the brighter sub-sample of groups with $\mu \le
24$\,mags\,arcsec$^{-2}$, a smaller fraction of groups with multiple
redshifts has discordant redshifts ($\sim 24\%$). These results are
qualitatively consistent with the expected contamination rates
discussed in Paper~I.

The solid lines in Figure~\ref{sb} show the {\it predicted} fraction
of interloper-free compact groups recovered as a function of the
limiting surface brightness of the groups (calculated in Paper~I and
listed in Table~1 of that paper). For example, when considering all
groups brighter than $\mu = 24$\,mags\,arcsec$^{-2}$, we expect that
44\% of those groups will contain interlopers. For groups from
Catalogue~A with $n_z \ge 2$, the points in Figure~\ref{sb} shows the
fraction with multiple {\it concordant} redshifts as a function of the
limiting group surface brightness. Vertical error bars show the
Poisson uncertainties. The three panels show results for different
values of $n_z$.

Despite the fact that full redshift information (ie., $n_z = n_{mem}$)
is not available for most groups, Figure~\ref{sb} demonstrates that
the predicted interloper fractions from Paper~I appear in good
agreement with the best estimates for interloper fractions for groups
identified in Catalogue A. The observed interloper fraction and the
predicted one are in best agreement for higher $n_z$, since it is the
more likely the interloper(s), if present, will have been targeted for
spectroscopy. Figure~\ref{sb} thus presents strong empirical evidence
that the selection of groups by their surface brightness can greatly
reduce the fraction of interloping galaxies/groups present.

\subsubsection{Redshift distribution of groups identified in Catalogue A}

\begin{figure}
  \begin{center}
    \includegraphics[angle=270, width=8.cm]{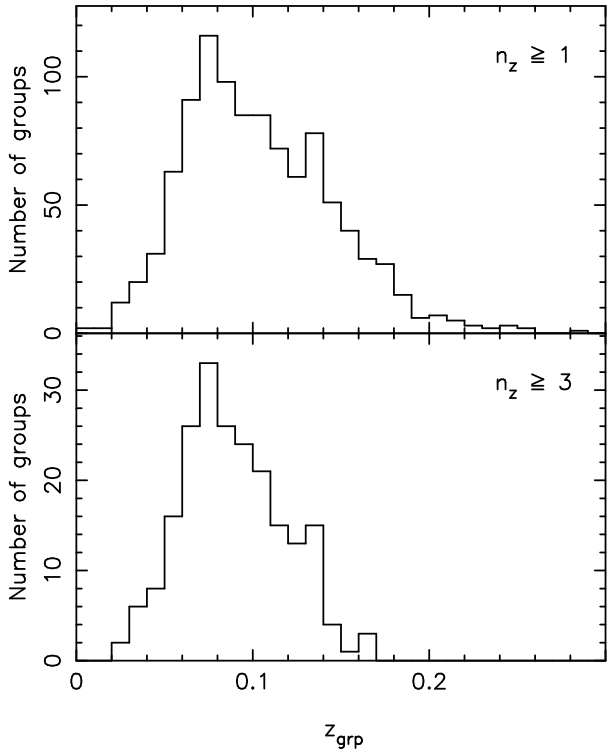}
    \caption{Mean redshift distribution of compact groups from
    Catalogue~A, for those groups with redshift measurements where
    $z_{conf} \ge 0.7$. Only those groups with $\Delta v \le 1000$\kms
    are shown. The top panel shows those groups for which there exists
    a redshift measurement for at least one galaxy. The lower panel
    shows those groups for which at least 3 galaxies per group have a
    redshift measurement. The latter distribution is least likely to
    contain groups with interloping galaxies.}
    \label{zdist}
  \end{center}
\end{figure}

Figure~\ref{zdist} shows the mean redshift distribution for all
compact groups from Catalogue~A with redshift measurements where
$z_{conf} \ge 0.7$. Only groups for which $\Delta v \le 1000$\kms are
shown. The top panel shows those groups for which there exists a
redshift measurement for at least one galaxy. The lower panel shows
those groups for which at least 3 galaxies per group have a redshift
measurement. The lower distribution is less likely to contain
redshifts of groups with interloping galaxies due to its higher
spectroscopic completeness.

The peak (modal) redshift of the groups in each panel of
Figure~\ref{zdist} is $z_{peak} = 0.07 - 0.08$. The median redshift of
all groups with at least one redshift is ${z_{med}} \simeq 0.1$, while
for those with at least 3 concordant redshifts it is ${z_{med}} \simeq
0.09$ (for comparison, the median redshift of the \cite{lee2004}
sample is $z_{med} \approx 0.13$). We note that the classical sample
of Hickson compact groups (\citealt{hickson1982}) have a considerably
lower median redshift than these samples, with a median redshift of
$z_{med} \sim 0.03$ (\citealt{hickson1992}).

A K-S test shows that the distributions in the two panels of
Figure~\ref{zdist} are unlikely to have been drawn from the same
underlying distribution at the $>99.9\%$ level. In particular, the
fraction of groups with $z > 0.1$ in the $n_z \ge 3$ sample is much
less than in the $n_z \ge 1$ sample.  It is therefore probable that
all genuine groups found in Catalogue A have $z \lesssim 0.17$, the
maximum redshift of groups in the $n_z \ge 3$ sample.

\subsubsection{Velocity dispersions of groups identified in Catalogue A}

\begin{figure}
  \begin{center}
    \includegraphics[angle=270, width=8.cm]{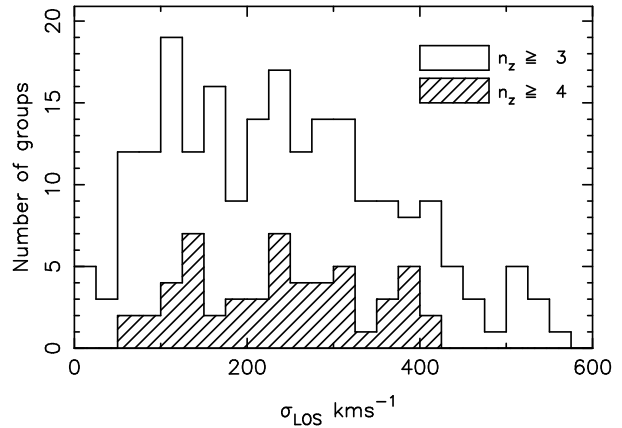}
    \caption{Line-of-sight velocity dispersions for all groups from
    Catalogue~A with $n_z \ge 3, 4$ (blank and hatched histograms,
    respectively) with concordant redshift information. A K-S test
    reveals that the two distributions are statistically similar.}
    \label{vdisp}
  \end{center}
\end{figure}

Figure~\ref{vdisp} shows the physical line-of-sight velocity dispersions
($\sigma_{LOS}$) for all groups from Catalogue~A with $n_z \ge 3, 4$
(blank and hatched histograms, respectively) with concordant redshift
information. A K-S test shows that the two distributions are
statistically similar. Although $\Delta v \le 1000$\kms is required for a group
to be included in these histograms, most groups have velocity
dispersions significantly smaller than this; the maximum velocity
dispersion of groups with $n_z = 3 (4)$ is $\sigma_{LOS} = 546\,$\kms
(395\,\kms). The median velocity dispersion of groups with
$n_z = 3 (4)$ is $\sigma_{LOS} = 244\,$\kms (227\,\kms). These numbers
compare favourably to the classical Hickson compact groups, which have
a median velocity dispersion of $\sigma_{LOS} \simeq 200$\kms
(\citealt{hickson1992}).

\subsection{Number Density of Compact Groups}

We identify 2297 compact groups in the SDSS DR6 down to a limiting
magnitude of $r = 18$, corresponding to $\sim
0.24$\,groups\,degree$^{-2}$. 0.9\% of galaxies down to this limiting
magnitude are identified as a member of a compact group. From the
redshift analysis in the previous section, we can expect $\sim 71\%$
of these compact groups to contain interlopers, and so the density of
interloper-free compact groups is probably nearer $\sim
0.07$\,groups\,degree$^{-2}$. However, most of the systems containing
interlopers can still be expected to contain a close pair, triplet or
higher-order multiple of galaxies (Paper~I).

In Paper~I, we identify 15122 compact groups in an all-sky mock
catalogue with a limiting magnitude of $r = 18$ mags, the same as
Catalogue~A. The predicted density of identified groups in this
catalogue is therefore $\sim 0.37$\,groups\,degree$^{-2}$,
approximately 50\% higher than the observed density in the
SDSS~DR6. Given that both the observed and mock catalogues were
examined using an identical procedure, we surmise that this disparity
is a result of inaccuracies in the modeling on which the mock
catalogue is based. For example, it may be that the limited resolution
of the Millennium simulation causes the merger time of close systems
to be incorrectly estimated (see a discussion of these issues in
Paper~I and \citealt{springel2005}). It should be noted that our
investigation of the effect of interlopers (Section 3.2.2) and the
results from Paper~II suggest that many of the properties of the
galaxies and groups identified in the mock catalogue are close to the
observed properties. Future contributions dealing with the observed
properties of the identified groups in SDSS~DR6 will include
comparisons to the mock catalogue, and may reveal where the source of
the discrepancy lies.

Considering all galaxies down to a limiting magnitude of $r = 21$, we
identify 74791 compact groups. Accounting for contamination from
incorrect photometric classification (Section~2.4.2), this corresponds
to $\sim 6.7$\,groups\,degree$^{-2}$, with 0.9\% of all galaxies down
to this limiting magnitude identified as a member of a compact
group. As the redshift study in the previous section makes clear, the
majority of these groups will consist in part or in full of
interloping galaxies along the same line-of-sight. While the fraction
of groups with interlopers is predicted to be as high as 71\% for a
catalogue with a limiting magnitude of $r = 18$, we can expect that
the interloper fraction will significantly increase for a catalogue
with a fainter limiting magnitude (since the probability that a given
line-of-sight will pass close to an unrelated galaxy will increase).

\subsection{Density of the compact group environment}

\begin{figure*}
  \begin{center}
    \includegraphics[angle=270, width=14.cm]{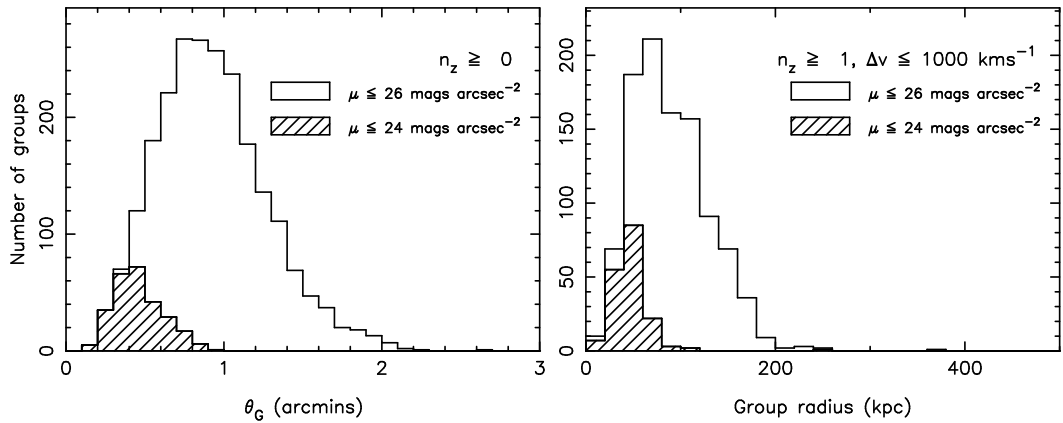}
    \caption{Left panel: the blank histogram shows the distribution of
    the angular projected radius ($\theta_G$) of all identified groups
    in Catalogue~A. The hatched histogram shows the distribution of
    $\theta_G$ for all groups from Catalogue~A with $\mu \le
    24$\,mags\,arcsec$^{-2}$. Right panel: for groups from Catalogue~A
    with concordant redshift information, the blank histogram shows
    the distribution of the physical group radius in kpc. The
    hatched histogram shows the same distribution for groups with $\mu
    \le 24$\,mags\,arcsec$^{-2}$.}
    \label{size}
  \end{center}
\end{figure*}

\subsubsection{Angular size on sky}

The left panel of Figure~\ref{size} shows the distribution of the
projected radii ($\theta_G$) of groups from Catalogue~A. The blank
histogram shows the distribution for all groups, whereas the hatched
histogram shows the distribution for those groups with $\mu \le
24$\,mags\,arcsec$^{-2}$. Virtually all of these compact groups have
$\theta_G < 2$\,arcmins, and all groups with $\mu \le
24$\,mags\,arcsec$^{-2}$ have $\theta_G < 1$\,arcmin. This reflects
the fact that $\mu \propto \theta_G^{-2}$, as expected, and emphasizes
the high apparent density of these groups.

\subsubsection{Physical group radii and intergalactic separation}

For those groups with concordant redshift information, we can
calculate the de-projected physical radius of each group, $r_G =
(D_M \theta_G)/(1+z_{grp})$, where $D_M$ is the transverse comoving distance
(following \citealt{hogg1999}). We assume $H_o =
70$\,\kms\,Mpc$^{-1}$, $\Omega_\Lambda = 0.7$ and $\Omega_M = 0.3$.
The blank histogram in the right panel of Figure~\ref{size} shows the
distribution of group radii for all groups with the
necessary redshift information identified in Catalogue~A, and the hatched
histogram shows the corresponding distribution for those groups with
$\mu \le 24$\,mags\,arcsec$^{-2}$.

Groups detected in Catalogue~A typically have $r_G \lesssim 200$\,kpc,
and the sub-set with $\mu \le 24$\,mags\,arcsec$^{-2}$ typically have
$r_G \lesssim 100$\,kpc. The mean radii of all the compact
groups is ${\bar r}_G = 89.4 \pm 1.3$\,kpc ( ${\bar r}_G = 47.1 \pm
1.8$\,kpc for the bright sub-set), where the uncertainty is the random
error in the mean. We estimate the typical inter-galactic separation
of individual galaxies within groups to be ${\bar r}_{sep} \simeq (
\frac{4}{3} \pi {\bar r}_G^{~3} / n_{mem})^{\frac{1}{3}}$. For
$n_{mem} = 4$, this corresponds to ${\bar r}_{sep} \simeq 91$\,kpc
(${\bar r}_{sep} \simeq 48$\,kpc) for the main (bright) sample of
groups from Catalogue~A. This is significantly less than the expected
virial radii of typical $L_\star-$type galaxies, and confirms the high
spatial density of these environments.

\section{Summary}

In this paper we have presented a publically available catalogue of
compact groups of galaxies identified in the SDSS~DR6
(\citealt{adelmanmccarthy2008}) using the original Hickson criteria
(\citealt{hickson1982}).  We identify 2297 (74791) compact groups down
to a limiting magnitude of $r = 18$ ($21$). 0.9\% of all galaxies at
both magnitude limits are identified as members of compact groups,
although once interlopers are accounted for this fraction will
decrease.  Figures~\ref{top12} and \ref{interest} show a selection of
compact groups from our catalogue and illustrates the diversity of the
groups in this sample. Many galaxies in these groups appear to be
early-type, in agreement with previous studies
(e.g. \citealt{hickson1988a,palumbo1995}, Paper~II), although there
are many examples of groups with late-type members (many of them
visually spectacular). A later paper in this series will investigate
the detailed morphology of galaxies in these compact groups.

Spectroscopic information is available for 43\,\% (5\,\%) of compact
group galaxies to a limiting magnitude of $r = 18$ (21); we find that
the median redshift of the compact groups identified in Catalogue A is
${z_{med}} = 0.09$. The median line-of-sight velocity dispersions
within groups from Catalogue A are $\sigma_{LOS} \simeq 220 -
250$\,\kms. and the typical inter-galactic separations are of order
$50 - 100$\,kpc. The fraction of groups with interloping galaxies as
members is significant, and is shown to be in good agreement with the
predictions from the mock galaxy catalogue from Paper~I, despite the
latter over-predicting the number density of compact groups by $\sim
50\%$. We empirically show that the selection of groups by group
surface brightness can reduce the interloper fraction significantly;
this will be a powerful tool for future studies of these catalogues.

The catalogue of compact groups derived in this paper is publically
available: Tables~1 -- 4 contain basic information for all the groups
and member galaxies we identify, including the identifiers necessary
to extract full information on each galaxy from the main SDSS~DR6
database. Full versions of these tables are available for download
from the MNRAS website.

\section*{Acknowledgments}
 
AWM acknowledges support from a Research Fellowship from the Royal
Commission for the Exhibition of 1851. He also thanks Sara Ellison and
Julio Navarro for additional financial assistance. SLE and DRP
acknowledge the receipt of NSERC Discovery Grants which funded some of
this research. We thank the anonymous referee for thoughtful and
useful suggestions which helped improve the clarity of this paper.

Funding for the SDSS and SDSS-II has been provided by the Alfred
P. Sloan Foundation, the Participating Institutions, the National
Science Foundation, the U.S. Department of Energy, the National
Aeronautics and Space Administration, the Japanese Monbukagakusho, the
Max Planck Society, and the Higher Education Funding Council for
England. The SDSS Web Site is http://www.sdss.org/.

The SDSS is managed by the Astrophysical Research Consortium for the
Participating Institutions. The Participating Institutions are the
American Museum of Natural History, Astrophysical Institute Potsdam,
University of Basel, University of Cambridge, Case Western Reserve
University, University of Chicago, Drexel University, Fermilab, the
Institute for Advanced Study, the Japan Participation Group, Johns
Hopkins University, the Joint Institute for Nuclear Astrophysics, the
Kavli Institute for Particle Astrophysics and Cosmology, the Korean
Scientist Group, the Chinese Academy of Sciences (LAMOST), Los Alamos
National Laboratory, the Max-Planck-Institute for Astronomy (MPIA),
the Max-Planck-Institute for Astrophysics (MPA), New Mexico State
University, Ohio State University, University of Pittsburgh,
University of Portsmouth, Princeton University, the United States
Naval Observatory, and the University of Washington.

\end{document}